\documentclass[11pt,a4paper]{article}
\pdfoutput=1
\usepackage{jheppub}

\usepackage{etoolbox}
\makeatletter
\patchcmd{\maketitle}{\@fpheader}{\vspace{-15mm}}{}{}
\makeatother

\usepackage[normalem]{ulem}
\usepackage{amsmath}
\usepackage{amssymb}
\usepackage{amscd}
\usepackage{enumerate}
\usepackage{amsfonts}
\usepackage{epsfig}
\usepackage{mathtools}
\usepackage{yfonts}
\usepackage{bbold}
\usepackage{breqn}
\usepackage{multirow}
\usepackage[export]{adjustbox}
\definecolor{davecolor}{rgb}{0.95,  0.5,  0.2}

\definecolor{darkgreen}{rgb}{0,0.5,0}
\definecolor{darkblue}{rgb}{0,0,0.6}
\definecolor{purple}{rgb}{0.4,0.15,0.21}
\definecolor{black}{rgb}{.2,.2,.2}

\usepackage{dsfont}

\usepackage{graphicx}
\usepackage{bm}
\usepackage{xcolor}

\definecolor{davecolor}{rgb}{0.95,  0.5,  0.2}

\def\({\left(}
\def\){\right)}
\def\[{\left[}
\def\]{\right]}
\def\<{\langle}
\def\>{\rangle}

\def\CM{{\cal M}}

\def\CO{{\cal O}}

\def\CL{{\cal L}}

\newcommand\half{{\ensuremath{\frac{1}{2}}}}
\newcommand\p{\ensuremath{\partial}}

\newcommand\ket[1]{\ensuremath{\lvert{#1}\rangle}}
\newcommand\bra[1]{\ensuremath{\langle{#1}\rvert}}

\newcommand{\be}{\begin{equation}}
\newcommand{\ee}{\end{equation}}
\newcommand{\bea}{\begin{eqnarray}}
\newcommand{\eea}{\end{eqnarray}}
\newcommand{\benn}{\begin{equation*}}
\newcommand{\eenn}{\end{equation*}}
\newcommand{\bwt}{\begin{widetext}}
\newcommand{\ewt}{\end{widetext}}

\newcommand{\bi}{\begin{itemize}}
\newcommand{\ei}{\end{itemize}}
\newcommand{\ben}{\begin{enumerate}}
\newcommand{\een}{\end{enumerate}}
\newcommand{\bca}{\begin{cases}}
\newcommand{\eca}{\end{cases}}
\newcommand{\bln}{\begin{align}}
\newcommand{\eln}{\end{align}}
\newcommand{\bst}{\begin{split}}
\newcommand{\est}{\end{split}}
\newcommand{\ba}{\begin{aligned}}
\newcommand{\ea}{\end{aligned}}

\newcommand\de{{\ensuremath{{\delta}}}}
\newcommand\De{{\ensuremath{{\Delta}}}}

\newcommand{\fr}{\frac}
\newcommand\ra{{\rightarrow}}

\usepackage{empheq}
\usepackage{enumitem}

\newcommand{\Mcal}{{\cal M}}
\newcommand{\Ocal}{{\cal O}}
\newcommand{\Dmax}{\Delta_{\max}}

\newcommand\bk{{\bf k}}
\newcommand\bkp{{\bf k}^\prime}

\newcommand\UV{\textrm{UV}}
\newcommand{\IR}{\textrm{IR}}
\newcommand\Loop{\textrm{loop}}
\newcommand\tot{\textrm{tot}}
\newcommand{\eff}{\textrm{eff}}

\newcommand\LUV{\Lambda_{\mathrm{UV}}}
\newcommand\LIR{\Lambda_{\mathrm{IR}}}
\newcommand\imax{\mathfrak{i}_{\mathrm{max}}}

\newcommand{\gbar}{\bar{g}}
\newcommand{\mgap}{m_{\mathrm{gap}}}

\newcommand{\cO}{{\cal O}}
\newcommand{\corr}[1]{\left\langle #1 \right\rangle}
\newcommand{\fraki}{\mathfrak{i}}
\newcommand{\frakj}{\mathfrak{j}}
\newcommand{\ifrak}{\mathfrak{i}}
\newcommand{\cI}{\mathcal{I}}
\newcommand{\cIless}{\cI_{\, \fraki \, < \, \frakj}}
\newcommand{\cIeq}{\cI_{\, \fraki \, = \, \frakj}}

\newcommand{\atanh}{\tanh^{-1}}
\newcommand{\threeFtwo}{{_3 F_2}}
\newcommand{\fourFthree}{{_4 F_3}}
\newcommand{\pth}[1]{\left(#1 \right)}

\newcommand{\Fbar}{\overline{F}_{\cO}}
\newcommand{\Fbarp}{\overline{F}_{\cO'}}
\newcommand{\Fnobar}{F_{\cO}(p_1, \dots, p_n)}
\newcommand{\KL}{K\"{a}ll\'{e}n-Lehmann }

\title{Nonperturbative dynamics of (2+1)d \boldmath $\phi^4$-theory from Hamiltonian truncation}
\author[a]{Nikhil Anand,} 
\author[b]{Emanuel Katz,} 
\author[b]{Zuhair U. Khandker,}
\author[c,d]{Matthew T. Walters}
\affiliation[a]{Department of Physics, McGill University, Montr\'{e}al, QC H3A 2T8, Canada} 
\affiliation[b]{Department of Physics, Boston University, Boston, MA 02215, U.S.A.} 
\affiliation[c]{Theoretical Physics Department, CERN, 1211 Geneva 23, Switzerland}
\affiliation[d]{Institute of Physics, \'{E}cole Polytechnique F\'{e}d\'{e}rale de Lausanne (EPFL), CH-1015 Lausanne, Switzerland}

\abstract{We use Lightcone Conformal Truncation (LCT)---a version of Hamiltonian truncation---to study the nonperturbative, real-time dynamics of $\phi^4$-theory in 2+1 dimensions. This theory has UV divergences that need to be regulated. We review how, in a Hamiltonian framework with a total energy cutoff, renormalization is necessarily \emph{state-dependent}, and UV sensitivity cannot be canceled with standard local operator counter-terms. To overcome this problem, we present a prescription for constructing the appropriate state-dependent counterterms for (2+1)d $\phi^4$-theory in lightcone quantization. We then use LCT with this counterterm prescription to study $\phi^4$-theory, focusing on the $\mathbb{Z}_2$ symmetry-preserving phase. Specifically, we compute the spectrum as a function of the coupling and demonstrate the closing of the mass gap at a (scheme-dependent) critical coupling. We also compute Lorentz-invariant two-point functions, both at generic strong coupling and near the critical point, where we demonstrate IR universality and the vanishing of the trace of the stress tensor.}

\begin{document}
\maketitle


\section{Introduction and Summary}
\label{sec:Intro}

In this work, we study the nonperturbative, real-time dynamics of $\phi^4$-theory in 2+1 dimensions (focusing on the $\mathbb{Z}_2$ symmetry-preserving phase). The Lagrangian of the theory is\footnote{Note that in this work all local operators are normal-ordered.}
\be
\CL = \frac{1}{2}\p_\mu\phi\p^\mu\phi - \frac{1}{2}m^2\phi^2 - \frac{1}{4!}g\phi^4.
\label{eq:L}
\ee
Specifically, we numerically compute the spectrum as a function of the dimensionless coupling $\gbar\equiv\fr{g}{4\pi m}$ and demonstrate the closing of the mass gap at a (scheme-dependent) critical coupling. We also compute two-point functions of local operators, both at generic strong coupling and near the critical point. 

The method we use to perform these computations is Lightcone Conformal Truncation (LCT), which is a version of Hamiltonian truncation. Hamiltonian truncation is a powerful framework for studying QFTs nonperturbatively. The basic idea of these methods is to first express the QFT Hamiltonian in a well-chosen basis, then truncate the basis to a finite size using some prescription, and numerically diagonalize the finite-dimensional Hamiltonian in order to obtain an approximation to the physical spectrum and eigenstates of the QFT. Finally, one looks for convergence in physical observables as the truncation threshold is increased. 

The use of Hamiltonian truncation methods in studying strongly-coupled QFTs was pioneered by Yurov and Zamolodchikov~\cite{Yurov:1989yu,Yurov:1991my} and subsequently by L\"{a}ssig, Mussardo, and Cardy~\cite{Lassig:1990xy}. Since then, there has been tremendous progress, and Hamiltonian truncation has been applied to a wide array of QFTs. To mention a few recent examples, truncation has been used to study spontaneous symmetry breaking~\cite{Coser:2014lla,Rychkov:2014eea,Rychkov:2015vap}, scattering~\cite{Bajnok:2015bgw,Gabai:2019ryw}, and quench dynamics~\cite{Rakovszky:2016ugs,Hodsagi:2018sul} in strongly-coupled 2d systems. For a recent overview with a comprehensive list of references, see~\cite{2018RPPh...81d6002J}. At the same time, there have been significant conceptual and technical advancements in the overall framework of Hamiltonian truncation that have greatly enhanced the applicability and precision of these methods as a whole. These include a systematic Wilsonian renormalization framework for including the effects of high-energy states discarded by truncation, significantly improving the convergence of such methods~\cite{Feverati:2006ni,Watts:2011cr,Giokas:2011ix,Elias-Miro:2015bqk,Elias-Miro:2017xxf,Elias-Miro:2017tup}, as well as advances in the numerical diagonalization of large matrices for truncation applications~\cite{Lee:2000ac,Lee:2000xna}.

LCT is a relatively more recent version of Hamiltonian truncation that is formulated in lightcone quantization, instead of the usual equal-time quantization. One of the main motivations for working in lightcone quantization is that it allows for LCT to be formulated in infinite volume (at least formally, as we will see), which facilitates the computation of physical observables like correlation functions. In this way, LCT provides access to different types of dynamical observables and nicely complements other Hamiltonian truncation methods. This particular truncation method involves using a basis of low-dimension primary operators of some UV CFT (in this case, free field theory with a single massless scalar) to study the full RG flow resulting from relevant deformations (here, the mass term and quartic interaction). Recent progress in both the formulation and application of LCT can be found in~\cite{Katz:2013qua,Katz:2014uoa,Katz:2016hxp,Anand:2017yij,Fitzpatrick:2018ttk,Delacretaz:2018xbn,Fitzpatrick:2018xlz,Anand:2019lkt,Fitzpatrick:2019cif}, and a pedagogical introduction to the method can be found in~\cite{Anand:2020gnn}.

Despite the successes of Hamiltonian truncation, there has been a persistent barrier to further progress: QFTs with UV divergences. The problem is well-known in the literature (\emph{e.g.},~\cite{Hogervorst:2014rta,Rutter:2018aog,EliasMiro:2020uvk}). As we will review, the basic problem is that in a Hamiltonian framework, one places an energy cutoff on a QFT---instead of a loop-momentum cutoff typical of Feynman diagram calculations---and this requires counterterms that are \emph{state-dependent}. In other words, the UV sensitive contributions encountered in any process depend on the details of the incoming and outgoing states. The upshot is that the usual local operator counterterms (which are \emph{state-independent} by construction) no longer suffice to renormalize the theory, and instead there is a potential proliferation of counterterms due to the fact that UV sensitivity needs to be canceled state-by-state. This has been a significant barrier to progress, and indeed most applications of Hamiltonian truncation, including LCT, have been restricted to UV finite theories in low spacetime dimensions.

In this work, we present a solution to the problem of state-dependent counterterms for (2+1)d $\phi^4$-theory in infinite volume within lightcone quantization. Our solution is simple in that it only requires introducing state-dependent counterterms at fixed $O(g^2)$ in perturbation theory, and the necessary counterterms are easy to evaluate numerically (albeit in a brute-force way). Despite its simplicity, the counterterm prescription is crucial for performing reliable computations at strong coupling and especially near the critical point.

There has been much recent interest in applying Hamiltonian truncation methods to theories in $d>2$. In particular, ref.~\cite{Hogervorst:2014rta} presented a generalization of the Truncated Conformal Space Approach (TCSA), which was initially formulated in 2d in~\cite{Yurov:1989yu}, to theories in arbitrary spacetime dimension. They then used this approach to compute the spectrum of $\phi^4$-theory in $d=2.5$, which does not have UV divergences. Ref.~\cite{Rutter:2018aog} determined the allowed structure of state-dependent counterterms to second order in the energy cutoff for theories in general $d$, focusing in particular on the case of $\phi^4$-theory. Ref.~\cite{Hogervorst:2018otc} presented a general formulation of Hamiltonian truncation for theories on the $d$-dimensional sphere, and computed the partition function and correlation functions for $i\phi^3$-theory in $d=3$, which contains logarithmic divergences.

Recently, Elias-Mir\'{o} and Hardy presented a solution to state-dependent counterterms for (2+1)d $\phi^4$-theory in finite volume within equal-time quantization~\cite{EliasMiro:2020uvk}. They were able to use their prescription to compute the spectrum of the theory and check the predictions of a weak/strong-coupling self-duality. Our counterterm solution and that of Elias-Mir\'{o} and Hardy work in complementary settings: infinite- versus finite-volume and lightcone- versus equal-time quantization. At the same time, in their respective settings, both solutions are very general and should be applicable to many other QFTs. We are hopeful that these works will open the door to applying Hamiltonian truncation to many new classes of QFTs. 

Returning to the model at hand, we use LCT along with our counterterm prescription to study the dynamics of (2+1)d $\phi^4$-theory, focusing on the $\mathbb{Z}_2$ symmetry-preserving phase. We first test our approach with the following consistency checks:
\begin{itemize}
\item Figure~\ref{fig:Spectrum}: We compute the spectrum as a function of the dimensionless coupling $\gbar$ and demonstrate the closing of the mass gap as $\gbar$ is dialed up from zero. 
\item Figure~\ref{fig:SpectrumRatios}: We show that higher eigenvalues approach zero consistently with the mass gap and that our state-dependent counterterm is crucial for ensuring that eigenvalue ratios match theoretical predictions as we approach the critical point.
\end{itemize}
We then obtain the following new, nonperturbative results for (2+1)d $\phi^4$-theory:
\begin{itemize}
\item Figure~\ref{fig:SDStrong}: We compute the \KL spectral densities of the operators $\phi^2$ and $\phi^4$ at strong coupling in order to illustrate the types of observables one can compute using LCT. 
\item Figure~\ref{fig:Universality}: Close to the critical point, we demonstrate universality in the spectral densities of the operators $\phi^n$ for $n=1,\ldots,6$, which is a prediction of criticality.
\item Figure~\ref{fig:CriticalCouplingCorrelators}: We compute the position space correlators of the $\mathbb{Z}_2$-even operators $\phi^2$, $\phi^4$, and $\phi^6$ near the critical point and demonstrate that the universal IR behavior is well-fit by a simple power law, as expected for an IR fixed point.
\item Figure~\ref{fig:StressTensorTrace}: We compute the spectral density of the trace of the stress tensor $T^\mu_{\,\,\mu}$ and demonstrate that it vanishes near criticality, providing evidence that the critical point is described by a CFT.
\item Figure~\ref{fig:TraceCorrelator}: We compute the position space correlation function of $T^\mu_{\,\,\mu}$ close to the critical point and show that the deviation from zero in the IR is well-fit by a power law consistent with the universality seen in $\phi^{2n}$.
\end{itemize}

The critical point where the mass gap closes should be in the same universality class as the 3d Ising CFT. In this work, we do not have sufficient IR precision in our results to reliably extract 3d Ising critical exponents. Nevertheless, our demonstration of universality in correlation functions and the vanishing of the stress tensor trace is strong evidence that we are probing physics governed by the critical CFT. It would be exciting if LCT can be harnessed at higher truncation levels in the future as a new tool for studying 3d Ising physics. Also, it is worth emphasizing that at generic strong coupling (\emph{i.e.}, not too close to the critical point), our results are less limited by our finite IR resolution and the two-point functions we compute are new results for the dynamics of (2+1)d $\phi^4$-theory. 

The calculations in this work were performed using minimal computational resources. At maximum truncation level, our basis consists of 35,425 states, and all of our calculations were performed on personal laptops with runtimes on the order of several days. It is encouraging that even at modest basis sizes, we see convergence in many observables as well as the onset of critical behavior near the critical point. We expect that the size of the basis and the corresponding precision of the method can be substantially increased in future efforts. 

While our counterterm prescription for removing UV divergences is new to this work, the general LCT setup and the interpretation of the results rely crucially on previous works. In particular, ref.~\cite{Katz:2016hxp} first formulated LCT in general dimensions. Subsequently, ref.~\cite{Anand:2017yij} studied (1+1)d $\phi^4$-theory, including the RG flow to the 2d Ising model, and demonstrated the computation of spectral densities and the onset of critical phenomena. Finally, ref.~\cite{Anand:2019lkt} developed an efficient method for computing LCT matrix elements for theories in (2+1)d, including $\phi^4$-theory. 

This paper is organized as follows. In section~\ref{sec:UV}, we review the problem of state-dependent counterterms and present our solution. Our counterterm prescription is stated in section~\ref{subsec:Counterterm}. In section~\ref{sec:Implementation}, we briefly review the basic setup of LCT, including a description of the truncation parameters involved. In section~\ref{sec:Checks}, we perform several consistency checks of our method in the free massive theory and in perturbation theory. In section~\ref{sec:Results}, we present our strong-coupling results. We conclude in section~\ref{sec:Discussion}. 

Several appendices supplement the main text. For the interested reader, appendix~\ref{sec:Overview} provides a self-contained overview of LCT in 3d and the calculations needed for this work, while appendices \ref{sec:Basis}-\ref{sec:MatrixElements} detail several new techniques for constructing the LCT basis and evaluating Hamiltonian matrix elements. Appendix~\ref{sec:FockSpace} contains some details for using Fock space methods to compute matrix elements, which are generally less efficient and hence not used in this work, but which often come in handy in other contexts. Appendix~\ref{app:StateDependence} discusses more details on the structure of state-dependent cutoffs in ET and LC quantization, and appendix~\ref{app:LambdaEffective} discusses the connection between UV and IR cutoffs in LCT. Appendix~\ref{app:CountertermExample} provides a low-truncation example of the construction of our state-dependent counterterm. Finally, appendix~\ref{app:VaryParameters} supplements the discussion in section~\ref{sec:Results}.

\section{Hamiltonian Truncation and UV Divergences}
\label{sec:UV}

\subsection{The Problem of State-Dependent Counterterms}

\begin{figure}[t!]
\begin{center}
\includegraphics[width=0.75\textwidth]{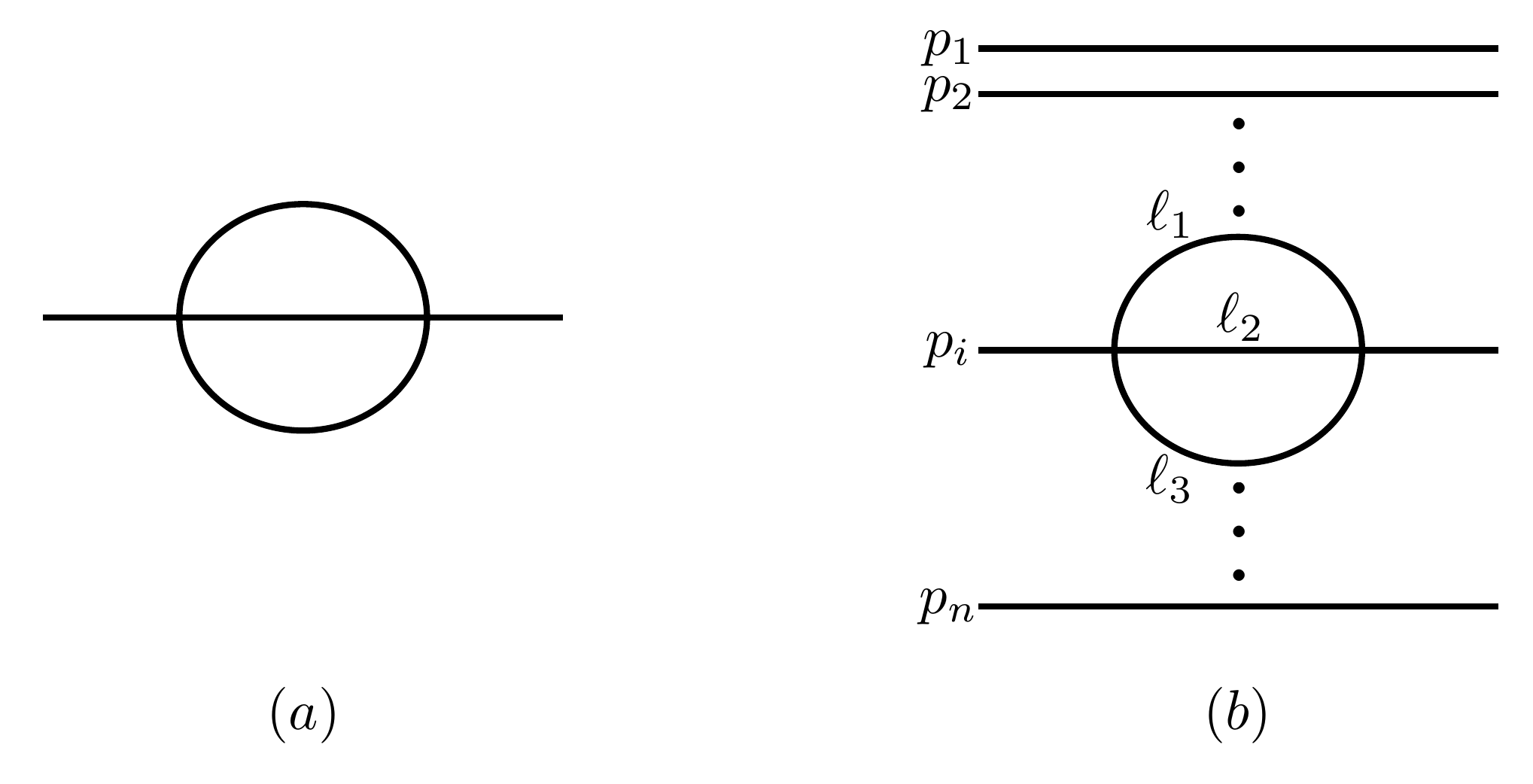}
\caption{(a) Sunset diagram; (b) Sunset diagram with spectators.}
\label{fig:Sunset} 
\end{center}
\end{figure}

In 3d $\phi^4$-theory, given by the Lagrangian in (\ref{eq:L}), the leading perturbative correction to the bare mass $m^2$ is logarithmically divergent, \be
\left. \delta m^2 \right|_{O(g^2)} = -\frac{g^2}{96\pi^2} \log\left( \frac{\left(\Lambda_\Loop/m+1\right)^2}{8\left(\Lambda_\Loop/m-1\right)} \right) \approx -\frac{g^2}{96\pi^2} \log\fr{\Lambda_\Loop}{8m}.
\label{eq:1pMassShift}
\ee
Here, $\Lambda_\Loop$ is a UV cutoff on loop momenta, and the correction comes from the ``sunset'' diagram shown in figure~\ref{fig:Sunset}(a). The standard renormalization procedure is to introduce a mass counterterm that cancels the dependence on $\Lambda_\Loop$, yielding UV-insensitive results for physical observables. For this particular divergence, the counterterm one adds to the Lagrangian is
\be
\delta \CL_{\textrm{c.t.}} = \frac{1}{2}  c \, \phi^2, 
\label{eq:MassCT} 
\ee
where the coefficient $c$ is chosen to cancel the UV sensitivity in (\ref{eq:1pMassShift}) and may have additional finite terms depending on the renormalization scheme. 

A crucial point for the present discussion is that the counterterm $\delta \CL_{\textrm{c.t.}}$ in (\ref{eq:MassCT}) is \emph{state-independent}. By this, we mean that the counterterm does not depend on the external states of any given process. This is something we usually take for granted. Indeed, the fact that $\delta \CL_{\textrm{c.t.}}$ is state-independent is immediately evident from the fact that we can express it in terms of the local operator $\phi^2$, which makes no reference to external states. In a Feynman diagram language, this state-independence is a consequence of the fact that $\Lambda_\Loop$ is a cutoff on local loop momenta, which is agnostic about the details of the diagram's external legs.  

In a Hamiltonian framework, the situation is starkly different. To regulate divergences one places a UV cutoff on the total energy of intermediate states instead of a cutoff on local loop momenta. An immediate consequence of doing this is that UV sensitivities, and the counterterms needed to remove them, necessarily become \emph{state-dependent}.\footnote{These counterterms are often referred to in the literature as ``nonlocal'', in contrast with typical counterterms such as eq.~\eqref{eq:MassCT}, which can be written in terms of local operators.} This state dependence is essentially a consequence of energy positivity. As we will now discuss, when summing over intermediate states, the ``energy budget'' available to loop momenta depends on the division of the total momentum among the particles in the state. This momentum distribution varies from state to state, leading to state-dependent UV sensitivities. 

Let us begin with a simple conceptual picture. The sunset diagram in figure~\ref{fig:Sunset}(a) represents a sum over three-particle intermediate states. In practice, we impose a UV cutoff $E_{\UV}$ on the total energy allowed for the intermediate states, which regulates the divergence from the sunset diagram. However, let's now consider the same diagram, but in the presence of spectator particles, as shown in figure~\ref{fig:Sunset}(b). This second diagram represents a sum over $(n+2)$-particle states. While $E_{\UV}$ again sets the cutoff on the total energy of the $(n+2)$-particle states, the UV divergence is controlled specifically by the maximum energy available to the three particles in the sunset part of this diagram. This maximum energy is strictly less than $E_{\UV}$, as some of the energy budget is taken by the relative momentum of the spectators with respect to the particles in the loops. The loop momenta  in figure~\ref{fig:Sunset}(b) thus see a \emph{lower cutoff} than those in figure~\ref{fig:Sunset}(a), due to the additional particles in the external state.

More concretely, consider a general $n$-particle Fock space state $|p_1,\dots,p_n\>$. Figure~\ref{fig:Sunset}(b) shows a leading correction to the mass of this $n$-particle state due to the $\phi^4$ interaction, where one of the external particles (labeled by $p_i$) splits into three particles, which then recombine, while the remaining $n-1$ external particles are spectators. Let's define $\mu_\tot^2$ as the invariant mass-squared of the full $(n+2)$-particle intermediate state, while $\mu_\Loop^2$ is the invariant mass-squared of only the three particles that participate in the interaction.

In a Hamiltonian framework, we place a UV cutoff on the total energy, which in a fixed momentum frame is equivalent to placing a cutoff on $\mu_{\text{tot}}^2$. We would like to know what a cutoff on $\mu_{\text{tot}}^2$ implies for $\mu_{\Loop}^2$, which is the source of the UV divergence. As we discuss in more detail in appendix~\ref{app:StateDependence}, the answer to this question depends on whether we are working in equal-time or lightcone quantization.

In equal-time (ET) quantization, we have $p^\mu = (E,\vec{p}\,)$, where the spatial momenta $\vec{p}$ are conserved in interactions, and $\mu^2 = E^2 - |\vec{p} \,|^2$. In lightcone (LC) quantization, we instead have $p_\mu = (p_+,p_-,p_\perp)$, with the components $p_-$ and $p_\perp$ conserved in interactions, and $\mu^2 = 2p_+p_- - p_\perp^2$. If we introduce a UV cutoff $\mu_{\text{tot}}^2 \leq \Lambda^2$, then in these two quantization schemes we obtain the resulting loop momentum cutoff (\emph{i.e.}, the maximum energy running through the sunset diagram):\footnote{See appendix~\ref{app:StateDependence} for a derivation and further explanation of these two expressions.}
\be
\begin{aligned}
&\text{ET}: \hspace{5mm} \mu_{\Loop}^2 \leq \left(  \Lambda -  \sum_{j\neq i} \sqrt{|\vec{p}_j|^2 + m^2}   \right)^2 - |\vec{p}_i|^2, \\
&\text{LC}: \hspace{5mm}  \mu_{\Loop}^2 \leq x_i \Lambda^2 - x_i \sum_{j\neq i} \frac{p_{j\perp}^2 + m^2}{x_j} - p_{i\perp}^2,
\label{eq:MuIntCutoff}
\end{aligned}
\ee
where $x_i \equiv \fr{p_{i-}}{p_{\tot-}}$ is the fraction of the total lightcone momentum carried by $p_i$.

In LC quantization, a cutoff $\Lambda$ on the total energy of intermediate states thus leads to the state-dependent shift in the bare mass:
\be
\includegraphics[width=0.33\textwidth,valign=c]{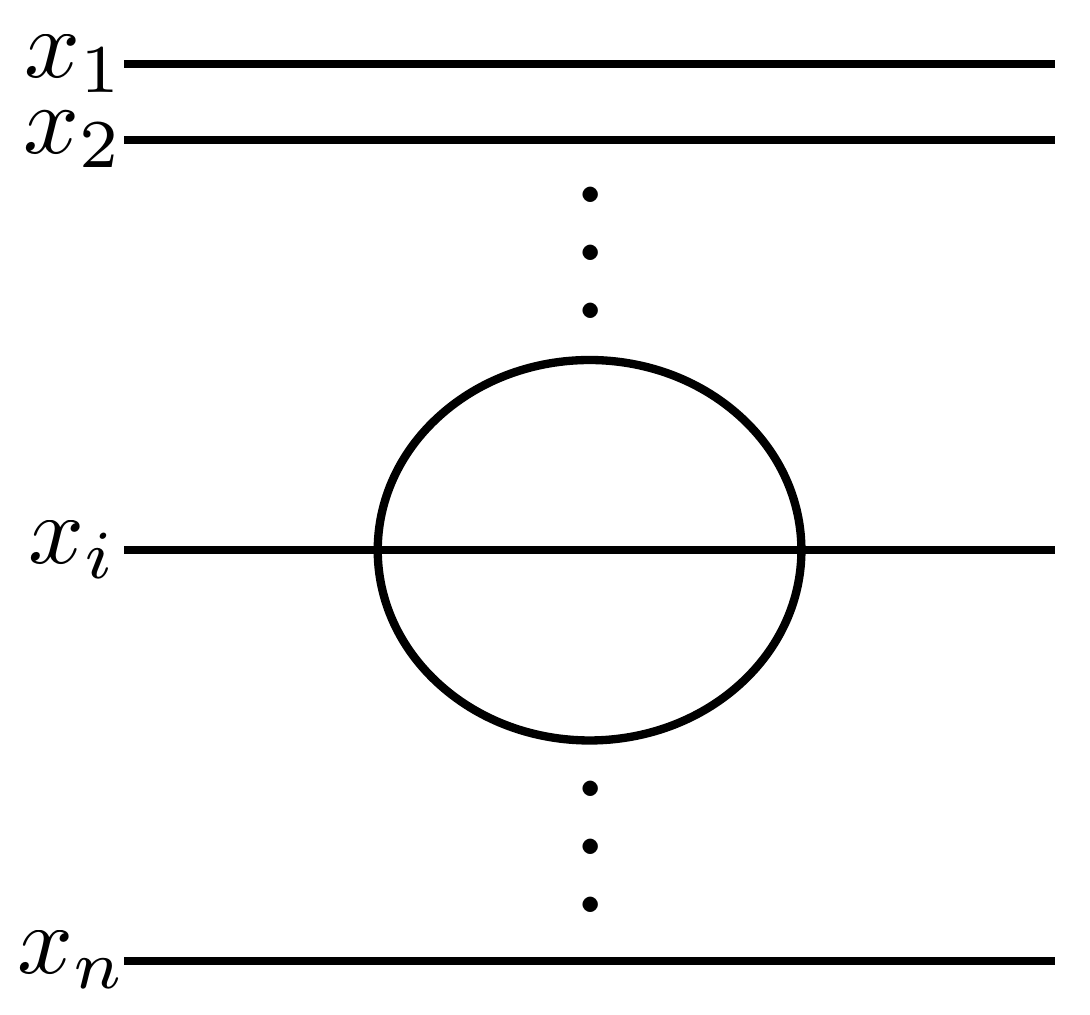} = -\fr{g^2}{96\pi^2}\log\fr{\Lambda}{8m} - \fr{g^2}{192\pi^2}\log x_i + O\Big(\fr{1}{\Lambda}\Big),
\label{eq:StateDepMassShift}
\ee
where we have simply plugged the resulting loop momentum cutoff from~\eqref{eq:MuIntCutoff} into the logarithmic divergence~\eqref{eq:1pMassShift}. As we can see, the state-dependence is a finite shift $\sim\!\log x_i$ set by the momentum fraction carried by the interacting particle.

While we have focused on the specific case of the mass shift due to the sunset diagram, we can draw several important general lessons from the inequalities in (\ref{eq:MuIntCutoff}). Let us enumerate some of the main punchlines:
\begin{itemize}
\item Energy cutoffs (rather than loop momentum cutoffs) lead to state-dependent UV sensitivities. Indeed, the right-hand sides of~(\ref{eq:MuIntCutoff}) obviously depend on the individual momenta $\vec{p}_i$ of the incoming particles, which vary from state to state. This means that the counterterms we introduce to cancel UV sensitivities must also be state-dependent. This is true of both ET and LC quantization. In both quantization schemes, a local operator counterterm like (\ref{eq:MassCT}) will not suffice. 
\item To make matters worse, one expects to have to add state-dependent counterterms at every order in perturbation theory. The diagram we considered in Figure~\ref{fig:Sunset}(b) occurs at $O(g^2)$. However, this process could be a sub-diagram within a higher-order term. With an energy cutoff, one should not expect that counterterms designed to cancel state-dependent UV sensitivities at $O(g^2)$ will continue to cancel UV sensitivities at higher order. This is quite different from the usual case of loop momentum cutoffs.
\item Regarding the nature of the state-dependence, in ET quantization the state dependence in (\ref{eq:MuIntCutoff}) comes as an additive shift in $\Lambda$. In LC quantization, in addition to the additive shift, there is a \emph{multiplicative} rescaling of $\Lambda^2$ by the factor $x_i$, which is the fraction of $p_{\text{tot}-}$ carried by the incoming particle that participates in the interaction. It is worth noting that the factor $x_i$ is insensitive to the spectators (apart from their total momentum). This will be important in the next section. 
\item The formula (\ref{eq:MuIntCutoff}) has nothing to do with truncation (\emph{i.e.}, restricting the Hilbert space to low-dimension operators). It is simply a consequence of having an energy cutoff.\footnote{In finite volume, Hamiltonian truncation typically \emph{is} simply placing an energy cutoff, so this distinction is less meaningful. In infinite volume, however, state-dependent UV sensitivities arise even in continuum QFT when using a total energy cutoff.} Eventually, we will be interested in applying Hamiltonian truncation methods. In that case, there will be additional corrections to the right hand sides of (\ref{eq:MuIntCutoff}), which vanish as the truncation threshold is taken to infinity (\emph{e.g.}, $\frac{1}{\Dmax}$ corrections in the context of LCT). 
\end{itemize}

At first glance, overcoming the state-dependence in (\ref{eq:MuIntCutoff}) seems daunting due to the proliferation of counterterms. However, as we will discuss in the next section, there is a simple, albeit brute force, solution for (2+1)d $\phi^4$-theory in LC quantization. We will have to introduce state-dependent counterterms; however, we will only need to introduce them at $O(g^2)$ in perturbation theory due to certain simplifications of LC quantization. Moreover, evaluating these counterterms in practice will be computationally trivial. 

\subsection{A Simple Counterterm Prescription for Lightcone Quantization}
\label{subsec:Counterterm}

In the previous section, we discussed why regulating a UV-divergent QFT with an overall energy cutoff necessitates the addition of state-dependent counterterms. In particular, one generically expects to have to add state-dependent counterterms order-by-order in perturbation theory, making the situation quite daunting due to the proliferation of counterterms. However, as we will now discuss, there are two crucial simplifications that will allow us to overcome these obstacles in $\phi^4$-theory with a simple prescription:
\begin{itemize}
\item In LC quantization, the vacuum is trivial~\cite{Klauder:1969zz,Maskawa:1975ky,Brodsky:1997de}. In particular, there is no vacuum renormalization and there are no vacuum bubble divergences. In 3d $\phi^4$-theory, the vacuum energy divergence is linear, whereas the mass divergence is logarithmic. Thus, working in LC quantization, one avoids linear divergences and only has to deal with logarithmic ones.
\item The state-dependence in the logarithmic divergence is insensitive to any details of the spectators, and only depends on the momentum fraction $x_i \equiv \fr{p_{i-}}{p_{\textrm{tot}-}}$ carried by the sunset diagram, as we can see in eq.~\eqref{eq:StateDepMassShift}. This is true even if the sunset is a subdiagram within a higher-order contribution or for multiple sunset diagrams in parallel. It is therefore sufficient to only introduce state-dependent counterterms at $O(g^2)$ in order to cancel state-dependencies at \emph{all higher orders} in $g$ (up to corrections suppressed by $\Lambda$ or $\Dmax$).\footnote{Schematically, for every contribution containing a sunset subdiagram with momentum fraction $x_i$, there is a compensating contribution where the sunset is replaced by our $O(g^2)$ counterterm for that same $x_i$.} Thus, although the counterterms we introduce will be state-dependent, we will only have to compute them at second-order in perturbation theory. This is a major simplification.
\end{itemize}

Given these simplifications, we propose the following prescription for removing the $O(g^2)$ state-dependence due to sunset diagrams. First, at a given truncation $\Dmax$ (see section~\ref{sec:Implementation} for the details of our truncation scheme), diagonalize the finite-dimensional Hamiltonian at $g=0$ in order to find the eigenstates of the free massive theory. Then for every $n$-particle mass eigenstate $|\mu_i^2\>$, we numerically compute the $O(g^2)$ perturbative shift in its mass due to the $(n+2)$-particle states in our truncated basis. This shift, $\delta \mu_i^2$, is precisely the contribution of the sunset process. The crucial next step is to add a counterterm at $O(g^2)$ that exactly cancels these perturbative shifts state-by-state. In the free massive basis, the counterterm is simply a diagonal matrix with entries $-\de\mu_i^2$. It is worth emphasizing that although this counterterm is state-dependent, it is trivial to compute. The final step in our scheme is to add a \emph{state-independent} local counterterm $\delta\CL = \frac{1}{2} c_L g^2  \phi^2$, for some constant $c_L$. This is simply a redefinition of the physical mass, and as we explain in section~\ref{sec:Results}, is useful for improving the convergence at finite truncation and ensuring we observe the IR critical point.\footnote{As shown in~\cite{SeroneUpcoming}, one expects that the critical point is visible (for real values of the coupling $g$) only for $c_L$ above some threshold value. We thank Giacomo Sberveglieri, Marco Serone, and Gabriele Spada for discussions on this point.}

Let us summarize our counterterm prescription in step-by-step fashion:
\begin{enumerate}
\item Diagonalize the truncated Hamiltonian at $g=0$ to obtain the free massive eigenstates. 
\item For every mass eigenstate $|\mu_i^2\>$, use second-order perturbation theory to compute $\de\mu_i^2$, which is the $O(g^2)$ shift due to all $(n+2)$-particle states in our truncated basis. 
\item Construct the state-dependent counterterm $g^2 \de P_+^{(\textrm{state-dep.})}$, which in the free massive basis is simply a diagonal matrix with entries $-\de\mu_i^2$.\footnote{One minor subtlety in this prescription is that, due to truncation effects, there are a small number of high-mass states whose $O(g^2)$ shifts have the incorrect sign. The counterterms for these states are set to zero (see appendix~\ref{app:CountertermExample} for more details).}${}^{,}$\footnote{Technically, the $O(g^2)$ $(n+2)$-particle contribution also includes a $t$-channel diagram that does not correspond to the sunset diagram. However, this counterterm prescription only removes the diagonal piece of this diagram, which is a set of measure zero.} This matrix can be re-expressed in the original CFT basis via a unitary transformation.
\item In addition to the state-dependent counterterm above, add a local mass shift $\delta\CL = \frac{1}{2} c_L g^2  \phi^2$ for some constant $c_L$ to be determined. 
\end{enumerate}
All in all, including counterterms, our renormalized LC Hamiltonian thus takes the form
\be
\boxed{P_+ = P_+^{(\textrm{CFT})} + \int d^2\vec{x} \left( \frac{1}{2}(m^2 - c_L g^2) \phi^2 + \frac{1}{4!}g\phi^4 \right) + g^2 \de P_+^{(\textrm{state-dep.})},}
\label{eq:HplusCT}
\ee
where the last term is our state-dependent counterterm.

\section{Lightcone Conformal Truncation Setup}
\label{sec:Implementation}

In this work, we will study nonperturbative $\phi^4$-theory in (2+1)d using Lightcone Conformal Truncation (LCT). To remove UV sensitivities, we will utilize the counterterm prescription presented above. In this section, we briefly review the basics of LCT, including a discussion of the different truncation parameters involved. Our goal is to provide the reader with enough background to understand the results presented in the subsequent sections without going into too many technical details. For the interested reader, the appendices contain all of the details of our LCT implementation, including the construction of the basis and computation of Hamiltonian matrix elements.

\subsection{Brief Review of LCT}

Hamiltonian truncation methods all follow the same basic steps. First, the QFT Hamiltonian is expressed as a matrix in a well-motivated, but infinite-dimensional, basis. Second, the basis and corresponding Hamiltonian matrix are truncated to a finite size according to some prescription. Finally, the truncated Hamiltonian is diagonalized (usually numerically) to obtain an approximation to the physical spectrum and eigenstates of the QFT. 

LCT is a specific version of Hamiltonian truncation that can be applied whenever the QFT of interest can be described as a deformation of a UV CFT by one or more relevant operators. The LCT basis is defined in terms of the primary operators of the CFT, while Hamiltonian matrix elements are related to OPE coefficients. Thus, the input is UV CFT data and the output is IR QFT dynamics. For the particular example of 3d $\phi^4$-theory, the UV CFT is free massless scalar field theory, and the relevant deformations are the mass term $\phi^2$ and quartic interaction $\phi^4$.

LCT is formulated in lightcone quantization~\cite{Dirac:1949cp,Weinberg:1966jm,Bardakci:1969dv,Kogut:1969xa,Chang:1972xt}. Our conventions for lightcone coordinates are $x^\pm = \tfrac{1}{\sqrt{2}} \left( x^0 \pm x^1 \right)$ and $x^\perp = x^2$, with $ds^2 = 2dx^+ dx^- - dx^{\perp 2}.$ In this quantization scheme, one takes $x^+$ to be time and $\vec{x} = (x^-,x^\perp)$ to be spatial. The lightcone momenta are defined by $P_{\pm} \equiv  \tfrac{1}{\sqrt{2}} \left( P_0 \pm P_1 \right) $ and $P_\perp \equiv P_2$. In particular, $P_+$ is the Hamiltonian. Concretely, the lightcone Hamiltonian we will study in this work is given by eq.~\eqref{eq:HplusCT}. The next step is to evaluate the matrix elements of this Hamiltonian in the LCT basis. 

In general, the LCT basis is constructed in momentum space and consists of Fourier transforms of primary operators $\CO$ in the UV CFT. We start by defining the states
\be
\ket{\CO,\mu} \equiv \int d^3x \, e^{-ip\cdot x} \, \CO(x) \ket{0},
\label{eq:LCTState}
\ee
where the label $\mu$ is defined by $\mu^2 \equiv p^2 = 2p_+p_- - p_\perp^2$. This notation requires some explanation. Strictly speaking, the Fourier transform appearing on the RHS should be labeled by the operator $\CO$ and the momentum $p_\mu = (p_+, p_-, p_\perp) = \left( \frac{\mu^2 + p_\perp^2}{2p_-}, \vec{p}   \right)$. However, because the spatial momentum generators $\vec{P}$ commute with the Hamiltonian $P_+$, we can always choose to work in a fixed ``momentum frame" with a fixed value for $\vec{p}$. Equivalently, Hamiltonian matrix elements are always proportional to $\delta^{(2)}(\vec{p}-\vec{p}^{\,\prime})$. Consequently, the spatial momentum label $\vec{p}$ just goes along for the ride, and we drop it for notational simplicity. This leaves the label $\mu$ (or equivalently, $p_+$) on the LHS.

The states $|\Ocal,\mu\>$ provide a complete basis for the Hilbert space of the UV CFT (as well as the IR QFT obtained by deforming this CFT). We then truncate this basis by setting a maximum scaling dimension $\Dmax$ and only keeping the finite set of primary operators below this threshold (\emph{i.e.}, with $\De \leq \Dmax$).

However, the label $\mu$ is still a continuous parameter that needs to be discretized in some way. To discretize it, we follow the prescription proposed in~\cite{Katz:2016hxp}. First we introduce a hard cutoff $\LUV$ on the range of $\mu$, restricting to $\mu^2 \leq \LUV^2$. Then we introduce \emph{smearing functions} $b_{\mathfrak{i}}(\mu)$, where $\mathfrak{i} = 1,\dots,\mathfrak{i}_{\text{max}}$, and define the discrete set of states
\be
\ket{\CO,\mathfrak{i}} \equiv \frac{1}{\sqrt{2\pi}} \int_0^{\Lambda_\UV^2} d\mu^2 \, b_{\mathfrak{i}}(\mu) \, \ket{\CO,\mu} \hspace{10mm} (\mathfrak{i} = 1,\dots,\mathfrak{i}_{\text{max}}).
\label{eq:LCTStateDiscrete}
\ee
Once we specify the precise form of the smearing functions, (\ref{eq:LCTStateDiscrete}) defines the LCT basis.

\begin{figure}[t!]
\begin{center}
\includegraphics[width=0.65\textwidth]{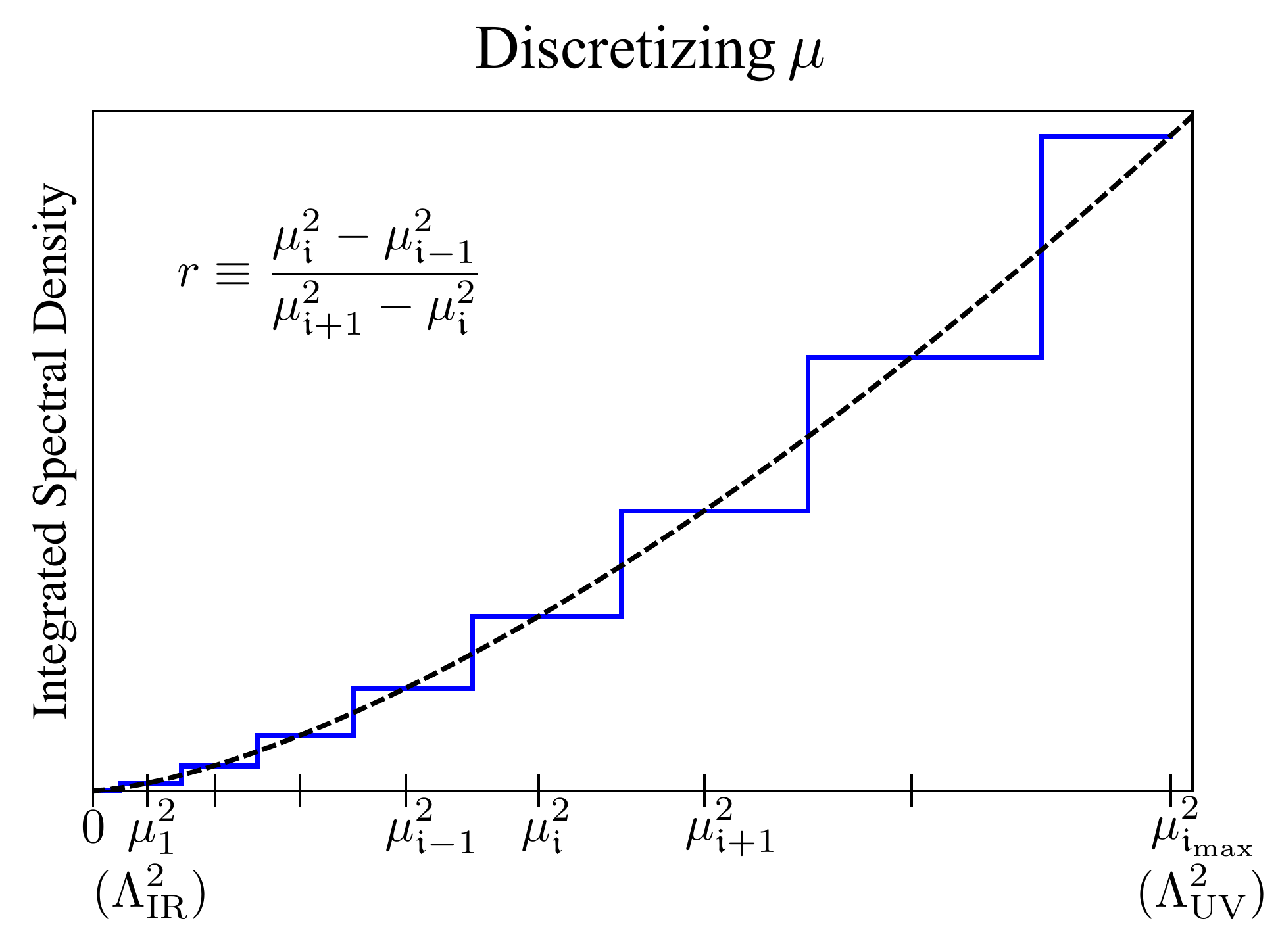}
\caption{Schematic representation of the integrated spectral density (see \eqref{eq:ISDFormula}) of a generic operator $\Ocal$ in the undeformed CFT, with our choice of $\mu$ discretization~\eqref{eq:gDef} (blue line), compared to the exact expression (black dashed line). There are $\imax$ bins, each with constant support on the intervals $[\mu_{\ifrak-1}^2,\mu_\ifrak^2]$. The largest value $\mu_{\imax}^2$ sets the UV cutoff, and the lowest value $\mu_1^2$ sets the IR resolution (see \eqref{eq:LUVIR}). The parameter $r$ controls the relative width of successive bins, such that $r=1$ corresponds to uniform bins, and $r < 1$ has smaller bins in the IR.}
\label{fig:BinExample} 
\end{center}
\end{figure}

There is obviously a large amount of freedom in the choice of functions $b_{\mathfrak{i}}(\mu)$. In this work, we will use non-overlapping bins that span the interval $[0,\LUV^2]$, as shown schematically in figure~\ref{fig:BinExample}. Specifically, we define 
\be
b_{\mathfrak{i}}(\mu) \equiv \frac{  \Theta(\mu^2 - \mu_{\mathfrak{i}-1}^2) - \Theta(\mu^2 - \mu_{\mathfrak{i}}^2) }{\sqrt{\mu_{\mathfrak{i}}^2 - \mu_{\mathfrak{i}-1}^2}}, \quad \mu_{\mathfrak{i}}^2 \equiv \LUV^2 \frac{r^{\mathfrak{i}_{\mathrm{max}}}(1-r^\ifrak)}{r^\ifrak(1-r^{\mathfrak{i}_{\mathrm{max}}})}, \quad(\mathfrak{i} = 1,\dots,\mathfrak{i}_{\text{max}})
\label{eq:gDef}
\ee
where $\Theta(x)$ is the Heaviside step function. In other words, we partition the interval $[0,\LUV^2]$ into $\ifrak_\max$ bins $[\mu_{\mathfrak{i}-1}^2, \mu_{\mathfrak{i}}^2]$, and define $b_{\mathfrak{i}}(\mu)$ to be a constant with support on a single bin. Note that in (\ref{eq:gDef}), we have introduced a parameter $r<1$ that sets the relative widths of successive bins,
\be
r \equiv \fr{\mu_\ifrak^2 - \mu_{\ifrak-1}^2}{\mu_{\ifrak+1}^2 - \mu_\ifrak^2},
\label{eq:rDef}
\ee
so that there are more bins in the IR (small $\mu^2$) than in the UV (large $\mu^2$). We reiterate that this definition of smearing functions is a choice, and it is certainly possible that there are better alternatives.

There is one final subtlety in the construction of our basis, which we discuss in more detail in appendix~\ref{sec:Overview}. In $\phi^4$-theory, there are IR divergences in the Hamiltonian matrix elements associated with the $\phi^2$ deformation, which have the effect of removing a subset of our truncated basis of states from the low-energy Hilbert space. In practice, our basis thus consists only of so-called ``Dirichlet'' operators, which are all linear combinations of primary operators which have at least one $\p_-$ acting on every insertion of $\phi$. When we set a truncation level $\Dmax$, we therefore only keep the subset of operators with $\De \leq \Dmax$ which satisfy this Dirichlet condition. Conceptually, we can still think of the basis as being comprised of primary operators, just with this added restriction on the Hilbert space.

The basis states (\ref{eq:LCTStateDiscrete}) up to $\Dmax$, along with the choice (\ref{eq:gDef}) for the smearing functions $b_{\mathfrak{i}}(\mu)$ up to $\imax$, define our basis. We express the $\phi^4$-theory Hamiltonian (\ref{eq:HplusCT}) in this basis, and then numerically diagonalize the Hamiltonian in order to obtain an approximation to the physical spectrum and eigenstates of the theory. The eigenstates $\ket{\alpha}$ of the full Hamiltonian have corresponding masses $\mu_\alpha$ and, crucially, they can be used to compute other physical observables. 

One of the main deliverables of LCT that we will consider in this work are K\"all\'en-Lehmann spectral densities of local operators $\rho_\Ocal(\mu)$. Recall that spectral densities encode the decomposition of two-point functions in terms of the physical mass eigenstates, 
\be
\<\mathcal{T}\{\Ocal(x) \Ocal(0)\}\> = \int d\mu^2 \rho_\Ocal(\mu) \int \fr{d^dp}{(2\pi)^d} e^{-ip\cdot x} \fr{i}{p^2 - \mu^2 + i\epsilon}.
\label{eq:SD2PF}
\ee
In our Hamiltonian truncation setup, the spectral density is simply computed by
\be
\rho_\Ocal(\mu) \equiv \sum_\alpha |\<\Ocal(0)|\alpha\>|^2 \, \de(\mu^2 - \mu_\alpha^2).
\label{eq:SDFormula}
\ee
Because this observable is formally a sum over delta functions, it is simpler in practice to study its integral,
\be
I_\Ocal(\mu) \equiv \int_0^{\mu^2} d\mu^{\prime 2} \, \rho_\Ocal(\mu^\prime) = \sum_{\mu_\alpha \leq \mu} |\<\Ocal(0)|\alpha\>|^2.
\label{eq:ISDFormula}
\ee
In this work, we will compute spectral densities and corresponding two-point functions of operators like $\phi^n$ and the stress tensor $T_{\mu\nu}$.

\subsection{Parameters of LCT}

Let us quickly summarize the truncation parameters involved in LCT. First, there is $\Dmax$, which is the maximum scaling dimension of the operators $\CO$ appearing in (\ref{eq:LCTStateDiscrete}). Next, there is $\LUV$, which is the hard cutoff on the invariant mass-squared of any basis state, $\mu^2 \leq \LUV^2$. Then there is $\mathfrak{i}_{\mathrm{max}}$, which is the number of smearing functions (or the number of bins in our case) used to probe $\mu^2 \in [0,\LUV^2]$. Finally, there is $r$, which is specific to our use of bins as smearing functions and  sets the relative size of successive bins. 

In addition to $\LUV$, our smearing functions $b_\ifrak(\mu)$ also set an IR cutoff $\LIR$. In our particular setup, this cutoff is simply the width of the first bin, since this sets our resolution for invariant masses $\mu^2$. To be more specific, referring back to (\ref{eq:gDef}), our IR cutoff is $\LIR^2 = \mu_1^2$. We therefore have the relation
\be
\LUV = \LIR \, \sqrt{\frac{r(1-r^{\imax})}{r^{\imax}(1-r)}  }.
\label{eq:LUVIR}
\ee

Now imagine that $\Dmax$ and $r$ are fixed. We then have two choices for how to compare results as we vary $\imax$. The first choice is to hold $\LUV$ fixed, such that $\LIR$ \emph{decreases} as we increase $\imax$. The second choice is to hold $\LIR$ fixed, such that $\LUV$ \emph{increases} as we increase $\imax$. In this work, we will always make the second choice, \emph{i.e.}, we will fix $\LIR$ and then increase the UV cutoff $\LUV$ by increasing $\imax$. This approach is more consistent when working at finite $\Dmax$, where in practice there is an effective lower bound on the allowed value for $\LIR$, as we explain in appendix~\ref{app:LambdaEffective}. So long as $\LIR$ is small compared to the physical scales of the theory, we expect observables to converge as $\LUV\rightarrow \infty$. Thus, in practice, we take our truncation parameters to be $\Dmax$, $\LIR$, $\imax$, and $r$. These parameters are summarized in Table~\ref{table:Parameters}. 

\begin{table}[t!]
\begin{center}
\begin{tabular}{| l | l |}
\hline &  \\[-7pt]
LCT Parameter & Description  \\
\hline &  \\[-7pt]
$\Dmax$  & Maximum scaling dimension of operators in basis, see (\ref{eq:LCTStateDiscrete})   \\
\hline &  \\[-5pt]
$\LIR^2$    &  Smallest bin used to discretize invariant masses $\mu^2$, see (\ref{eq:gDef})    \\
\hline &  \\[-5pt]
$\imax$ &  Total number of bins used to discretize $\mu^2$, see (\ref{eq:gDef})   \\
\hline &  \\[-5pt]
$r$ &  Ratio of successive bin sizes, see (\ref{eq:rDef})    \\
\hline &  \\[-5pt]
$\LUV^2$ (not independent) & UV cutoff on $\mu^2$, fixed in terms of $\LIR^2$, $\imax$, and $r$ via (\ref{eq:LUVIR})   \\
\hline
\end{tabular}
\end{center}
\caption{LCT truncation parameters along with their descriptions, specific to our setup of using bins as smearing functions to discretize the continuous invariant masses $\mu^2$ of basis states.}
\label{table:Parameters}
\end{table}

In this work, the maximum scaling dimension we consider is $\Dmax = 16$ (although for certain perturbative checks we can reach higher $\Dmax$). The Hamiltonian conserves ``transverse-parity'' $x^\perp \rightarrow -x^\perp$, and in this work we will restrict our attention to the even-parity sector, which at $\Dmax=16$ contains 545 primary operators. For each of these operators, the maximum number of bins we use to discretize $\mu^2$ is $\imax = 65$. Thus, at maximum truncation level, our basis contains $545 \times 65 = 35,425$ states. As for $\LIR$ and $r$, we have found experimentally that good values for these parameters are $\LIR/m=0.5$ and $r=0.8$ (at $\imax=65$ the corresponding UV cutoff is $\LUV/m \approx 1411$). We will use these values for $\LIR$ and $r$ unless otherwise noted. In appendix~\ref{app:VaryParameters}, we vary these parameters to ensure that our results are insensitive to their precise value.

\section{Consistency Checks}
\label{sec:Checks}

In this section, we perform several consistency checks of our method in the limit that the quartic coupling $g$ either vanishes or is perturbatively small. First, in section~\ref{subsec:Checks-Free}, we set $g=0$ (with $m\neq 0$), which corresponds to free massive field theory, and verify that spectral densities of the operators $\phi^n$ match their theoretical predictions. Then, in section~\ref{subsec:Checks-PT}, we consider small, perturbative values of $g/m$ and check that our results for the mass gap agree with perturbation theory.

\subsection{Free Massive Theory}
\label{subsec:Checks-Free}

Here, we consider free massive field theory by setting $g=0$ (with $m\neq 0$). In this limit, the lightcone Hamiltonian does not mix different particle-number sectors.\footnote{This is an important simplification compared to equal-time quantization.} As a result, the Hamiltonian for each sector can be diagonalized independently. In particular, computing the spectral density of an $n$-particle operator only requires diagonalizing the Hamiltonian in the $n$-particle sector. 

We set our parameters to be $\Dmax=16$, $\mathfrak{i}_{\max}=65$, $\Lambda_{\text{IR}}/m=0.5$, and $r=0.8$. Recall that the parameters $\Dmax$ and $\mathfrak{i}_{\max}$ set the size of our basis. For this choice of parameters, the number of states in the 2-, 3-, 4-, and 5-particle sectors is $455$, $6045$, $13520$, and $8060$, respectively. Meanwhile, the parameters $\mathfrak{i}_{\max}$, $\Lambda_{\text{IR}}$, and $r$ together set the scale of the UV cutoff $\LUV$, which for the values listed above is $\LUV/m \approx 1411$. For the given parameters, we diagonalize the Hamiltonian in the $n$-particle sector and use the resulting mass eigenstates to compute the spectral density of $\phi^n$. 

\begin{figure}[t!]
\begin{center}
\includegraphics[width=\textwidth]{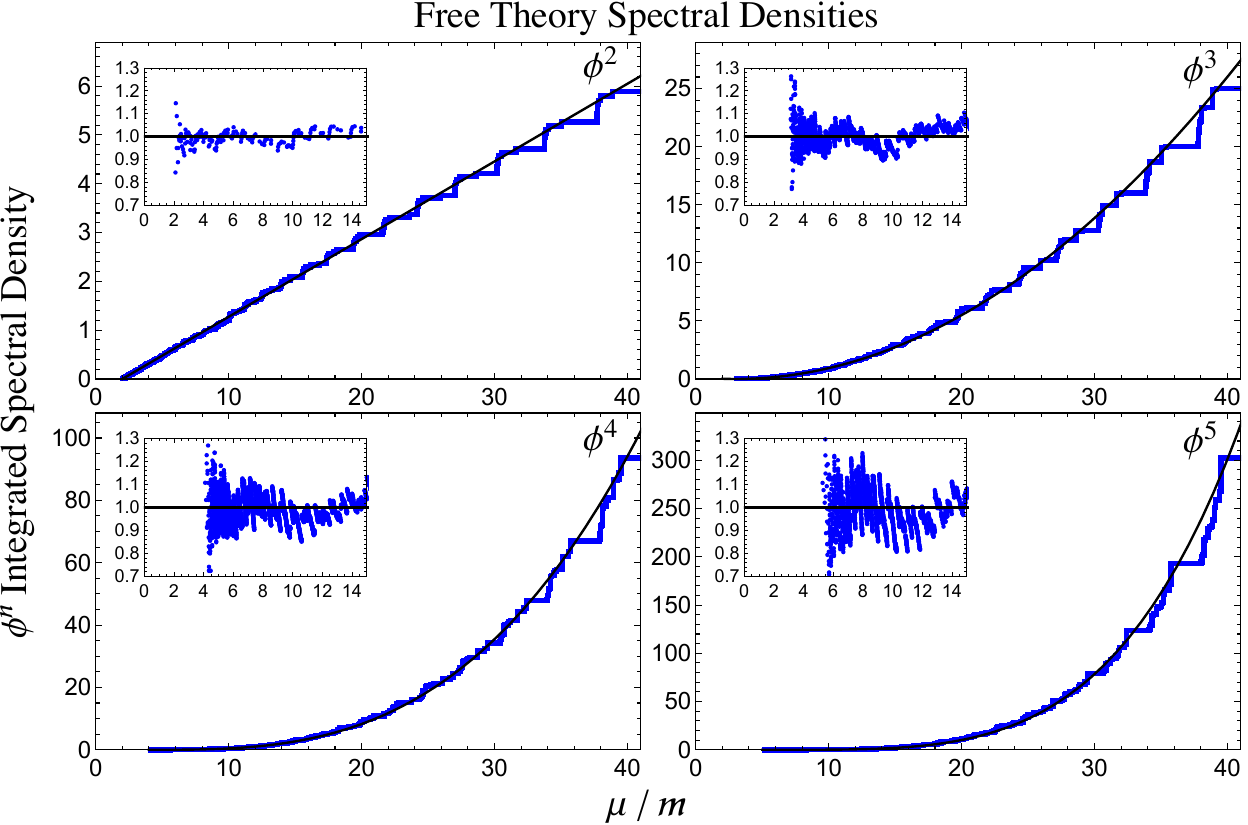}
\caption{Integrated spectral densities of $\phi^2$ (top left), $\phi^3$ (top right), $\phi^4$ (bottom left), and $\phi^5$ (bottom right) in free massive field theory ($g=0$). Each plot shows the LCT data (blue), computed using $\Dmax=16$, $\mathfrak{i}_{\max}=65$, $\frac{\LIR}{m}=0.5$, and $r=0.8$ (corresponding to $\frac{\LUV}{m} = 1411$), along with the known analytical result (black line). The insets show the ratio of the data to the analytical result.}
\label{fig:FreeSpectralDensities} 
\end{center}
\end{figure}

Figure~\ref{fig:FreeSpectralDensities} shows our results for the integrated spectral densities of $\phi^2$, $\phi^3$, $\phi^4$, and $\phi^5$ in free massive field theory. The blue curves correspond to our data, while the black lines show the known analytical result, given by
\begin{equation}
I_{\phi^n}(\mu) \equiv \int_{n^2m^2}^{\mu^2} d\mu^{\prime 2} \rho_{\phi^n}(\mu^{\prime 2})  = \frac{n}{(4\pi)^{n-1}} (\mu-nm)^{n-1}.
\label{eq:FreeSD}
\end{equation}  
In each plot, we have included an inset which shows the ratio of our data to the analytical prediction. Overall, looking at the main plots and the insets, we see excellent agreement between our data and the known theoretical results over a wide range of $\mu/m$. Similar plots can be made for $\phi^n$ with $n>5$.

\begin{figure}[t!]
\begin{center}
\includegraphics[width=\textwidth]{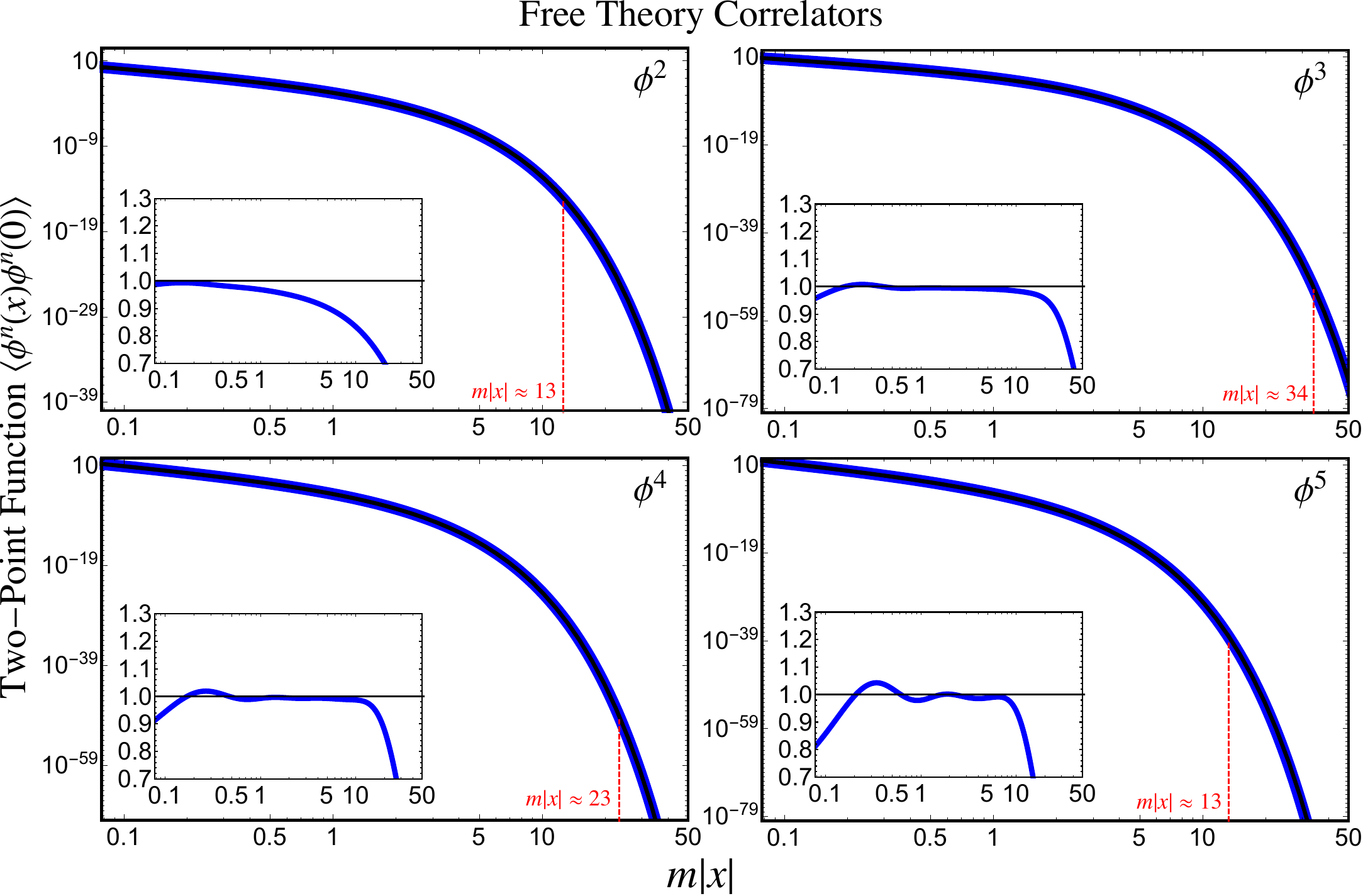}
\caption{The free massive theory ($\bar{g}=0$) two-point functions $\corr{\phi^n(x)\phi^n(0)}$ for $\phi^2$ (top left), $\phi^3$ (top right), $\phi^4$ (bottom left), and $\phi^5$ (bottom right). The LCT data (blue) was obtained with the same parameters as figure~\ref{fig:FreeSpectralDensities}, and is compared to the theoretical prediction (black line). The insets show the ratio of the data to the analytical result. To guide the reader, red dashed lines and text show the scale at which the correlator deviates from the theory prediction by 20 percent.}
\label{fig:PositionSpaceCorrelators} 
\end{center}
\end{figure}

It is also useful to look at the position space correlators $\corr{\phi^n(x)\phi^n(0)}$, which can be obtained from the $\phi^n$ spectral densities via
\be
\<\phi^n(x) \phi^n(0)\> = \int d\mu^2 \rho_{\phi^n}(\mu) \fr{e^{-\mu |x|}}{4\pi |x|} = \sum_\alpha |\<\phi^n(0)|\alpha\>|^2 \, \fr{e^{-\mu_\alpha |x|}}{4\pi |x|}, \label{eq:correlatorsfromspec}
\ee
where we have specifically considered the case where the operators are spacelike separated ($x^2 < 0$), such that the correlation function is real, though one can also obtain the correlator for timelike separation. The resulting correlation functions are shown in figure~\ref{fig:PositionSpaceCorrelators} in blue, compared to the exact analytical expressions in black. It is quite striking that the truncation results follow the theoretical prediction so closely over a wide range of length scales; for example the data for $\corr{\phi^3(x)\phi^3(0)}$ agrees with the theoretical value to within 20 percent up to $m |x| \approx 30$! Note that the $\phi^2$ correlator departs from the theory prediction at a smaller value of $m|x|$; this is a consequence of the fact that there are fewer two-particle states at $\Dmax=16$. Although we are only working in free field theory at the moment, these plots provide an important and encouraging quantitative consistency check of our method and moreover demonstrate our capacity to compute both spectral densities and correlation functions.

\subsection{Perturbation Theory}
\label{subsec:Checks-PT}

In this section, we now turn on a small quartic coupling $g$ and confirm that our result for the mass gap agrees with perturbation theory at $O(g^2)$ and $O(g^3)$, shown diagramatically in figure~\ref{fig:PertDiagrams}.

\begin{figure}[t!]
\begin{center}
\includegraphics[width=0.8\textwidth]{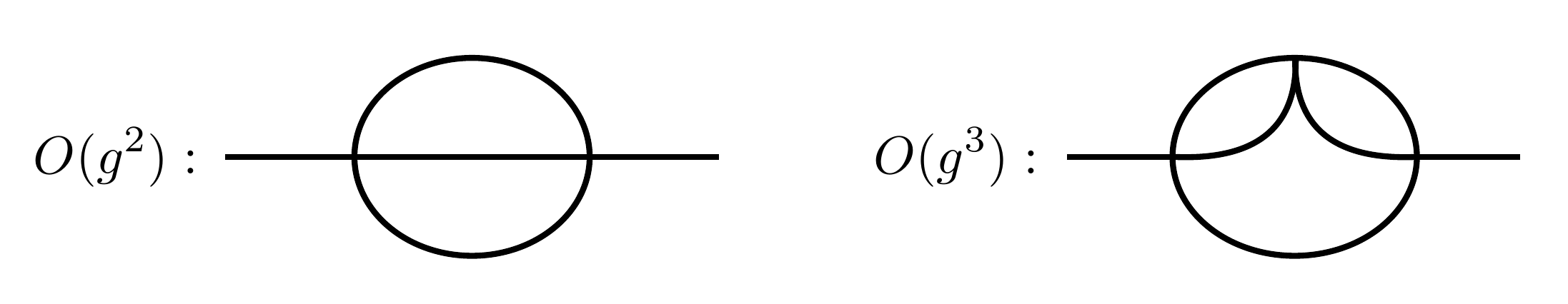}
\caption{Perturbative corrections to the 1-particle mass at $O(g^2)$, given by eq.~\eqref{eq:1pMassShift}, and $O(g^3)$, eq.~\eqref{eq:1pShiftg3}, in LC quantization. Both corrections only involve three-particle intermediate states.}
\label{fig:PertDiagrams} 
\end{center}
\end{figure}

We start at $O(g^2)$. Recall from section~\ref{subsec:Counterterm} that our counterterm prescription involves completely canceling the divergent, state-dependent, $O(g^2)$ sunset contribution to every free massive eigenstate. The addition of this counterterm renders the theory finite. Nevertheless, it is useful to check that \emph{before} adding any counterterms, the 1-particle mass shift is indeed UV divergent and matches the theoretical prediction given by (\ref{eq:1pMassShift}). This will be our first perturbative check. 

A useful observation here is that, up to $O(g^3)$, the 1-particle state only interacts with 3-particle states, and in fact, it only interacts with a subset of all 3-particle states.\footnote{Specifically, the only contribution comes from 3-particle states built from operators with no $\p_\perp$ derivatives.} Consequently, for our perturbative checks at $O(g^2)$ and $O(g^3)$, we are able to push our computations all the way up to $\Dmax=50$. For the nonperturbative computations in the next section, we will no longer have this luxury.

\begin{figure}[t!]
\begin{center}
\includegraphics[width=1.02\textwidth]{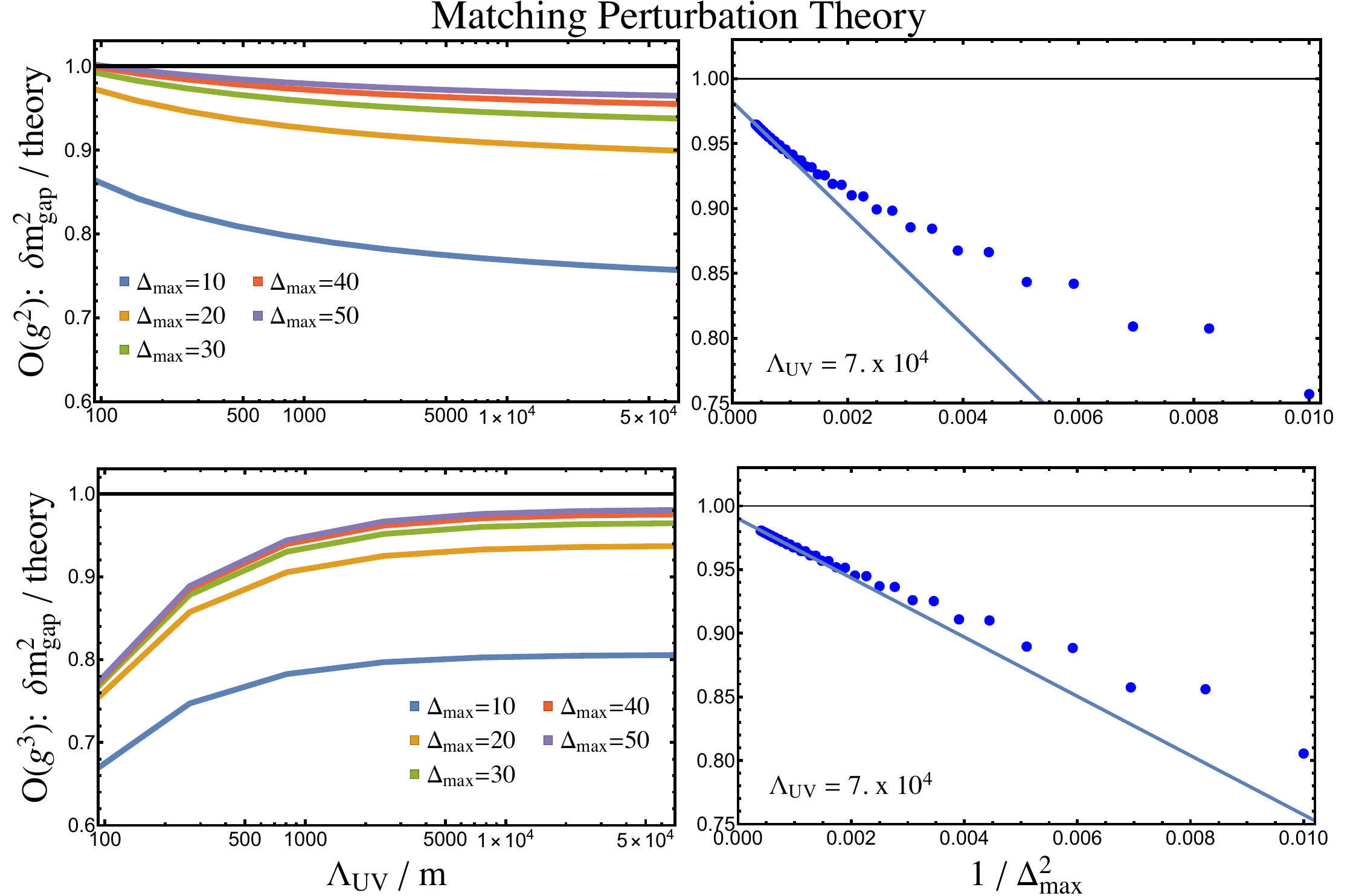}
\caption{\textbf{Top Row:} Perturbative, UV-divergent 1-particle mass shift at $O(g^2)$, computed \emph{before} adding any counterterms. \textbf{Bottom Row:} Perturbative, UV-finite 1-particle mass shift at $O(g^3)$. In all four plots, $\LIR/m=0.5$ and $r=0.8$.}
\label{fig:PerturbationTheory} 
\end{center}
\end{figure}

The top row of Figure~\ref{fig:PerturbationTheory} shows our results at $O(g^2)$. In particular, these are \emph{pre-counterterm} results for the 1-particle mass shift.  All of the plots in this figure are constructed for $m=1.0$, $\LIR/m=0.5$ and $r=0.8$. The parameter $\imax$ then sets $\LUV$ via (\ref{eq:LUVIR}), and we can increase $\LUV$ by dialing up $\imax$. The top left plot shows the $O(g^2)$ mass shift as a function of $\LUV / m$ for $\Dmax=10,20,30,40$, and $50$, all divided by the theoretical prediction $\delta m^2_{\mathrm{thy}}$ given by (\ref{eq:1pMassShift}). Note that the horizontal scale is logarithmic, with $\LUV / m \sim 7\times 10^4$ for $\imax = 100$. We see that as $\Dmax$ increases, the truncation results converge to $\delta m^2_{\mathrm{thy}}$ for asymptotically large $\LUV$. 

In the top right plot of Figure~\ref{fig:PerturbationTheory}, we set $\LUV/m=7\times 10^4$ ($\imax=100$) and track the behavior of the $O(g^2)$ mass shift as a function of $\Dmax$. For sufficiently large $\Dmax$, the trend appears to be captured by a $1/ \Dmax^2$ dependence (note the horizontal axis of the plot). Extrapolating the best-fit line set by $40 \leq \Dmax \leq 50$, we find agreement with $\delta m^2_{\mathrm{thy}}$ to approximately two percent. 

We reiterate that the top row of Figure~\ref{fig:PerturbationTheory} corresponds to $O(g^2)$ results obtained \emph{before} adding any counterterms. It is reassuring to see that we reproduce the correct UV-divergence of the 1-particle mass shift. However, our counterterm prescription completely cancels this $O(g^2)$ shift by construction, so it is useful to also consider the next order in perturbation theory. 

Let us turn to the 1-particle mass shift at $O(g^3)$. This particular quantity is UV-finite, and hence independent of our counterterm prescription. The theoretical result in LC quantization is 
\be
\left. \delta m^2_{\mathrm{thy}} \right|_{O(g^3)} = \frac{7\zeta(3)}{4096 \, \pi^3}\frac{g^3}{m}.
\label{eq:1pShiftg3}
\ee

The bottom row in Figure~\ref{fig:PerturbationTheory} shows our results at $O(g^3)$. These plots are analogous to the top row, except that now the theory prediction is independent of $\LUV$. In the bottom left plot, we see that as $\Dmax$ increases, the results again converge to the theoretical prediction for sufficiently large $\LUV$. In the bottom right plot, we again set $\LUV/m=7\times 10^4$ ($\imax=100$) and extrapolate in $\Dmax$,  We find that the $O(g^3)$ mass shift is also fit by a $1/ \Dmax^2$ dependence for large $\Dmax$. Extrapolating, we find agreement with $\delta m^2_{\mathrm{thy}}$ at $O(g^3)$ to approximately one percent. 

These consistency checks give us confidence that our truncated basis reproduces free theory and perturbation theory correctly. Now, we will dial up the coupling $g$ and compute the spectrum and correlation functions of the fully nonperturbative, interacting theory.


\section{Strong-Coupling Results}
\label{sec:Results}

In this section, we use the full machinery of LCT, along with the counterterm prescription described in section~\ref{subsec:Counterterm}, to compute the spectrum and correlation functions of (2+1)d $\phi^4$-theory. To recap, the Hamiltonian including counterterms is given by~(\ref{eq:HplusCT}), which we reproduce here for the reader's convenience
\be
P_+ = P_+^{(\textrm{CFT})} + \int d^2\vec{x} \left( \frac{1}{2}(m^2 - c_L g^2) \phi^2 + \frac{1}{4!}g\phi^4 \right) + g^2 \de P_+^{(\textrm{state-dep.})}.
\label{eq:HplusCT2}
\ee
LCT allows us to compute physical observables, such as the spectrum and two-point functions of local operators, at arbitrary values of the (scheme-dependent) dimensionless coupling
\be
\gbar \equiv \frac{g}{4\pi m}.
\ee 

Our main results are as follows. First, we demonstrate the closing of the mass gap as $\gbar$ is dialed up from zero. The smooth closing of the gap signals a second-order critical point that should correspond to the 3d Ising CFT. We show that our state-dependent counterterm is crucial for 
ensuring that certain eigenvalue ratios match theoretical predictions at strong coupling, and in particular that higher mass eigenstates approach zero self-consistently with the gap as we approach the critical point. Then, we compute the spectral densities of some example operators at generic, nonperturbative values of $\gbar$ in order to illustrate the types of observables that are computable using LCT. Finally, we study the vicinity of the critical point and demonstrate the onset of universal behavior in correlation functions as well as the vanishing of the trace of the stress tensor, both of which strongly suggest that we are beginning to probe 3d Ising physics.

\subsection{Spectrum and Closing of the Mass Gap}
\label{subsec:Results-Spectrum}

We begin by computing the spectrum of $\phi^4$-theory as a function of the dimensionless coupling $\gbar$ and show the closing of the mass gap. In particular, we show explicitly that the inclusion of our state-dependent counterterm is crucial to ensure that higher eigenvalues approach zero self-consistently, whereas the usual state-independent local operator counterterm, $\delta \CL_{\textrm{c.t.}} \propto \phi^2$ (see eq.~(\ref{eq:MassCT})), does not lead to such self-consistency.

The $\phi^4$-theory Hamiltonian factorizes into odd and even particle-number sectors due to the $\mathbb{Z}_2$ symmetry $\phi\rightarrow -\phi$, and each sector can be diagonalized independently. Figure~\ref{fig:Spectrum} shows the lowest eigenvalue in the odd-particle sector ($\mu_{1\text{-part.}}^2$, green), the lowest eigenvalue in the even-particle sector ($\mu_{2\text{-part.}}^2$, blue), and the second-lowest eigenvalue in the odd-particle sector ($\mu_{3\text{-part.}}^2$, red) as functions of $\gbar$. These states correspond to the 1-, 2-, and 3-particle thresholds, respectively, hence the notation. In particular $\mu_{1\text{-part.}}^2 \equiv \mgap^2$ is the mass gap squared. We see that the eigenvalues decrease smoothly to zero as we increase $\gbar$, signaling a second-order phase transition. This critical point should be in the same universality class as the 3d Ising CFT, which we will study in more detail in section~\ref{subsec:Results-Ising}. Note that the actual numerical value of the critical coupling where the gap closes is scheme-dependent and hence not physical. 

\begin{figure}[t!]
\begin{center}
\includegraphics[width=.9\textwidth]{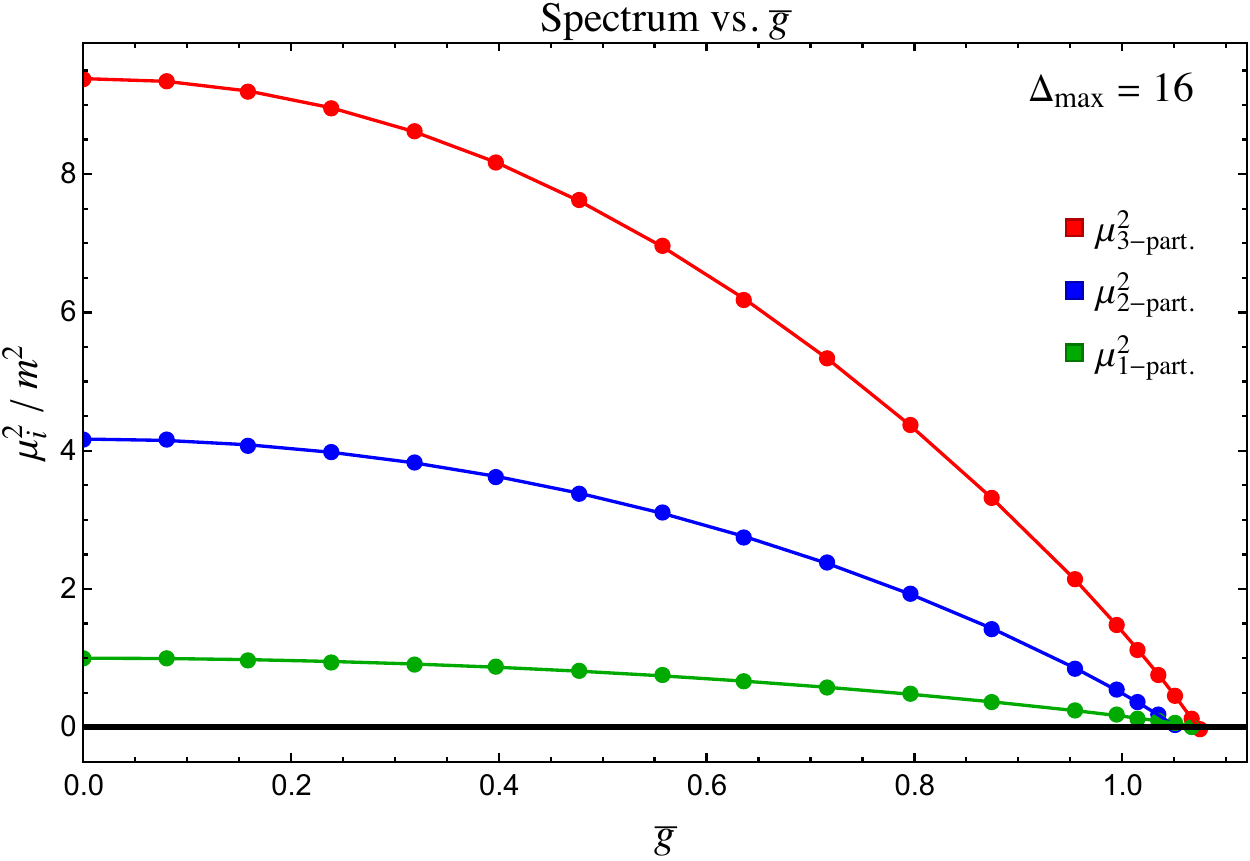}
\caption{Mass spectrum of $\phi^4$-theory as a function of $\gbar = \fr{g}{4\pi m}$ at $\Dmax=16$, with $\hat{c}_L = 1$ (see text). $\mu_{1\text{-part.}}^2$ (green) and $\mu_{3\text{-part.}}^2$ (red) are the lowest and second-lowest eigenvalues in the odd-particle sector, while $\mu_{2\text{-part.}}^2$ (blue) is the lowest eigenvalue in the even-particle sector.} 
\label{fig:Spectrum} 
\end{center}
\end{figure}

Figure~\ref{fig:Spectrum} was constructed at $\Dmax=16$, with $\imax$ extrapolated to infinity. Specifically, we have set $m=1$, $\LIR/m=0.5$, and $r=0.8$, and $\imax$ sets $\LUV$ via (\ref{eq:LUVIR}). The largest value of $\imax$ we have used is $65$, which corresponds to 35,425 total states and a UV cutoff of $\LUV^{(\imax=65)}/m \approx 1411$. At large $\LUV$, the spectrum has corrections that decay as $\frac{m}{\LUV}$, which allows us to easily extrapolate to $\LUV \rightarrow \infty$. Finally, in this figure we have also fixed the coefficient $c_L$ of the local counterterm in \eqref{eq:HplusCT2}. It is natural to parametrize this coefficient in terms of the $O(g^2)$ correction to the mass in eq.~\eqref{eq:1pMassShift}, leading to the rescaled coefficient $\hat{c}_L$,
\be
c_L \equiv \hat{c}_L \cdot \frac{1}{96\pi^2} \log\left( \frac{\left(\frac{\Lambda}{m}+1\right)^2}{8\left(\frac{\Lambda}{m}-1\right)} \right), \hspace{10mm} \text{where } \Lambda = \LUV^{(\imax=65)}.
\label{eq:cLhat}
\ee In Figure~\ref{fig:Spectrum}, we have set $\hat{c}_L=1$, which just corresponds to choosing a particular definition for the bare mass. As we will discuss below, $\hat{c}_L$ should be greater than some threshold in order for the mass gap to close, and at finite truncation, we find in practice that the IR results converge most quickly for values of $\hat{c}_L$ within a particular range. 

With the spectrum in hand, we can consider the following ratios between the eigenvalues plotted in Figure~\ref{fig:Spectrum}:
\be
R_{2:1} \equiv  \frac{1}{4}\frac{\mu_{2,\text{-part.}}^2}{\mu_{1\text{-part.}}^2}, \hspace{10mm} R_{3:1} \equiv  \frac{1}{9}\frac{\mu_{3\text{-part.}}^2}{\mu_{1\text{-part.}}^2}. 
\ee
$R_{2:1}$ is the ratio of the two-particle threshold to four times the one-particle mass-squared, and $R_{3:1}$ is the ratio of the three-particle threshold to nine times the one-particle mass-squared. An important physical requirement is that $R_{2:1}$ and $R_{3:1}$ should both be equal to 1 all the way to the critical point. This is a consequence of the fact that the $\phi^4$ interaction is repulsive, and hence there are no bound states in the spectrum. Of course, in a truncation computation, these ratios will only be approximately equal to 1. The deviation of these ratios from unity provides an important rubric for deciding how far into the strongly-coupled regime we should trust any truncation result. As the coupling $\gbar$ is increased and the mass gap approaches zero, one expects that beyond some coupling $R_{2:1}$ and $R_{3:1}$ will eventually deviate significantly from 1, due to the finite resolution of any Hamiltonian truncation scheme. The question is: how far into strong coupling can one reach, and in particular, how close to the critical point can one go? 

\begin{figure}[t!]
\begin{center}
\includegraphics[width=1\textwidth]{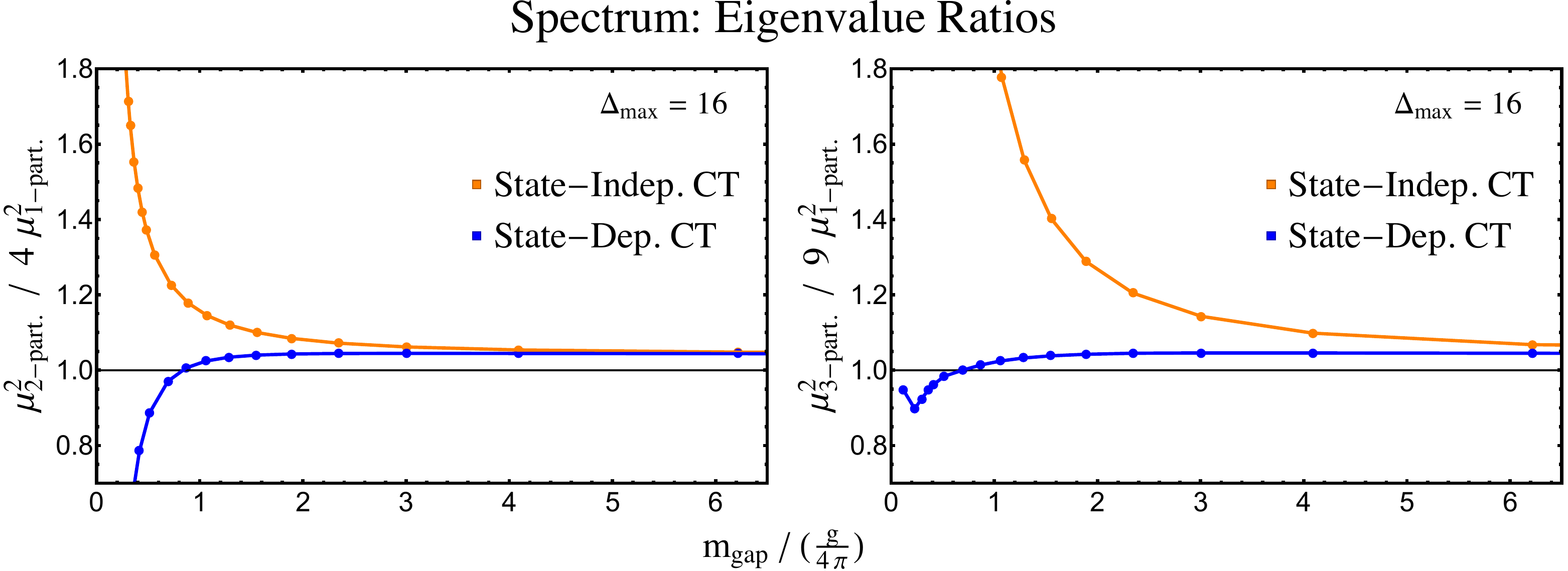}
\caption{ Eigenvalue ratios $R_{2:1} \equiv  \frac{1}{4}\frac{\mu_{2,\text{-part.}}^2}{\mu_{1\text{-part.}}^2}$ (left) and $R_{3:1} \equiv  \frac{1}{9}\frac{\mu_{3\text{-part.}}^2}{\mu_{1\text{-part.}}^2}$ (right) versus $\fr{\mgap}{g/4\pi}$ at $\Dmax=16$, compared to the theoretical prediction of $1$ (black line). Our data, obtained using our state-\emph{dependent} counterterm, is shown in blue. For comparison, in orange we show the result obtained by replacing our counterterm with a state-\emph{independent} one. Only the state-dependent counterterm yields reliable ratios in the nonperturbative regime $\fr{\mgap}{g/4\pi} \lesssim O(1)$.}
\label{fig:SpectrumRatios} 
\end{center}
\end{figure}

To address this question, it is useful to plot $R_{2:1}$ and $R_{3:1}$ as functions of $\fr{\mgap}{g/4\pi}$. Both $\mgap$ and $g$ are physical scales, and their ratio provides a well-defined parametrization of the theory. In particular, $\fr{\mgap}{g/4\pi} \rightarrow \infty$ is free field theory, $\fr{\mgap}{g/4\pi} \rightarrow 0 $ is the critical point, and very roughly $\fr{\mgap}{g/4\pi} \lesssim O(1) $ indicates nonperturbative physics. 

Figure~\ref{fig:SpectrumRatios} shows our truncation results for $R_{2:1}$ (left) and $R_{3:1}$ (right) plotted versus $\fr{\mgap}{g/4\pi}$ at $\Dmax=16$. In both plots the data shown in blue was obtained by using the state-\emph{dependent} counterterm prescription described in section~\ref{subsec:Counterterm}. We see that both $R_{2:1}$ and $R_{3:1}$ agree with the expected value of $1$ to within five percent for a wide range of $\fr{\mgap}{g/4\pi}$. Below $\fr{\mgap}{g/4\pi} \approx 0.5$, the data begins to trend away from 1, signaling that our finite truncation approximation is beginning to break down. 

For comparison, in both plots we have also included the result (shown in orange) obtained if one does \emph{not} use our state-dependent counterterm and insists on using only the usual state-\emph{independent} local-operator counterterm $\delta \CL = \frac{1}{2}c \phi^2$ (where $c$ is chosen to cancel the leading one-particle mass shift, see eq.~\eqref{eq:MassCT}). We see that the state-independent counterterm is not at all trustworthy at strong coupling, as these results rapidly deviate away from $1$ as we lower $\fr{\mgap}{g/4\pi}$, with significant deviations by the time we reach $\fr{\mgap}{g/4\pi} \sim  O(1)$, especially in $R_{3:1}$.

Figure~\ref{fig:SpectrumRatios} is one of our main results, which illustrates that our state-dependent counterterm is crucial for reliably probing strongly-coupled physics and that the usual state-independent counterterm does not work.

Finally, as promised, let us comment on the role of the local shift in the bare mass, parametrized by $\hat{c}_L$. In principle, changing the value of this coefficient just amounts to a redefinition of the bare mass. However, based on the analysis of~\cite{SeroneUpcoming}, we expect that $\hat{c}_L$ should be greater than some threshold value $\hat{c}_{L,\min}$ in order for the IR critical point to be visible with real coupling $g$. So long as $\hat{c}_L > \hat{c}_{L,\min}$, its precise value should have no effect on physical observables such as the eigenvalue ratios in figure~\ref{fig:SpectrumRatios}. The value of $\hat{c}_{L,\min}$ was recently computed for ET quantization in~\cite{SeroneUpcoming} using Borel resummation techniques developed in~\cite{Serone:2018gjo,Serone:2019szm,Sberveglieri:2019ccj}, but the map of this ET value to LC quantization is not currently known.

In principle, we can use LCT to determine $\hat{c}_{L,\min}$ by varying $\hat{c}_L$, computing the resulting mass spectrum, and seeing whether the mass gap closes. At finite truncation, however, we find experimentally that if $\hat{c}_L$ is too large or too small the ratios $R_{2:1}$ and $R_{3:1}$ begin to deviate from $1$ more quickly as we decrease $\fr{\mgap}{g/4\pi}$. At $\Dmax=16$, there is a finite range of values $0.5 \lesssim \hat{c}_L \lesssim 2$ that lead to reliable results, and for all values in this range we find that the mass gap closes, indicating that in LC quantization $\hat{c}_{L,\min} \lesssim 0.5$. Within this range, we also see no significant sensitivity to the precise value of $\hat{c}_L$ until close to the critical point. The range of reliable $\hat{c}_L$ grows as $\Dmax$ increases, and our expectation is that as $\Dmax \ra \infty$ it should be possible to study the theory for any value of $\hat{c}_L$. It would be useful to confirm this behavior by pushing to higher values of $\Dmax$, in particular to determine if there is a finite value $\hat{c}_{L,\min}$ below which the mass gap does not close.

For Figure~\ref{fig:SpectrumRatios}, we have chosen $\hat{c}_L=1$ as a generic value from the preferred range for $\Dmax=16$. The behavior of figure~\ref{fig:SpectrumRatios} as we vary $\hat{c}_L$, as well as the truncation level $\Dmax$ and the IR cutoff $\LIR$, is shown explicitly in appendix~\ref{app:VaryParameters}. The main punchline is that observables such as $R_{2:1}$ and $R_{3:1}$ do not change significantly as we vary these parameters, indicating that our numerical results are robust.

\subsection{Nonperturbative Spectral Densities}
\label{subsec:Results-SD}

In addition to the spectrum, one of the main deliverables of LCT is K\"all\'en-Lehmann spectral densities, $\rho_{\CO}(\mu)$ (see eqs.~(\ref{eq:SD2PF})-(\ref{eq:SDFormula})). In principle, spectral densities can be computed for any local operator $\CO$ at any value of the coupling $\gbar$, with the primary barrier being whether or not one can reach high enough truncation levels $\Dmax$ to see convergence.

In this section, as examples, we compute the spectral densities of the operators $\phi^2$ and $\phi^4$ at a nonperturbative value of the coupling $\gbar \approx 0.85$, where $\mu_{2\text{-part.}}/m = 1.25$. Recall that $\mu_{2\text{-part.}}/m$ is the lowest eigenvalue in the even-particle sector, corresponding to the two-particle threshold, and therefore takes the value 2 in free field theory ($\gbar=0$). As usual, we specifically plot integrated spectral densities $I_{\CO}(\mu)$ (see eq.~(\ref{eq:SDFormula})).

Figure~\ref{fig:SDStrong} shows the integrated spectral densities of $\phi^2$ (left) and $\phi^4$ (right) at $\gbar\approx 0.85$\footnote{Strictly speaking, as in previous work~\cite{Anand:2017yij,Anand:2020gnn}, we adjust $\gbar$ with $\Dmax$ in order to keep the even-particle gap fixed (in this case, at $\mu_{2\text{-part.}}/m = 1.25$) for every $\Dmax$. This allows us to better study the convergence of the functional form of the spectral density as we increase $\Dmax$. For the values of $\Dmax$ shown in the figure, $\gbar$ varies between 0.85-0.87.} for different values of $\Dmax$ and at our maximum binning level of $\imax = 65$. For comparison with free field theory, we have included a black dashed line that shows the free $\phi^2$ (left) and $\phi^4$ (right) integrated spectral densities, which start at $\mu/m = 2$ and $\mu/m=4$, respectively. Recall that the free $\phi^2$ integrated spectral density is linear in $\mu$ whereas the $\phi^4$ integrated spectral density is cubic (see eq.~(\ref{eq:FreeSD})). At this value of $\gbar$, we see a clear deviation in the spectral densities from free field behavior, as expected at strong coupling.

For $\phi^2$, the spectral density does not change significantly as we vary $\Dmax$, indicating that the results have largely converged throughout the entire range of $\mu/m$ shown. For $\phi^4$, we find that the functional behavior appears to have converged, particularly in the IR, but the spectral density slowly changes by approximately an overall constant as we vary $\Dmax$.

This behavior is due to an important subtlety in the spectral densities of local operators. The operator $\phi^2$ is well-defined and unambiguous, but in the interacting theory higher-dimensional operators such as $\phi^4$ are actually sensitive to our choice of UV cutoff. Concretely, at leading order in perturbation theory the operator $\phi^4$ has a logarithmically divergent contribution which mixes it with $\phi^2$, coming from the familiar sunset diagram. The two-point function of $\phi^4$ computed in figure~\ref{fig:SDStrong} therefore depends on the effective UV cutoff set by $\Dmax$.\footnote{See appendix~\ref{app:LambdaEffective} for more details of the effective cutoff set by truncation.}

\begin{figure}[t!]
\begin{center}
\includegraphics[width=1\textwidth]{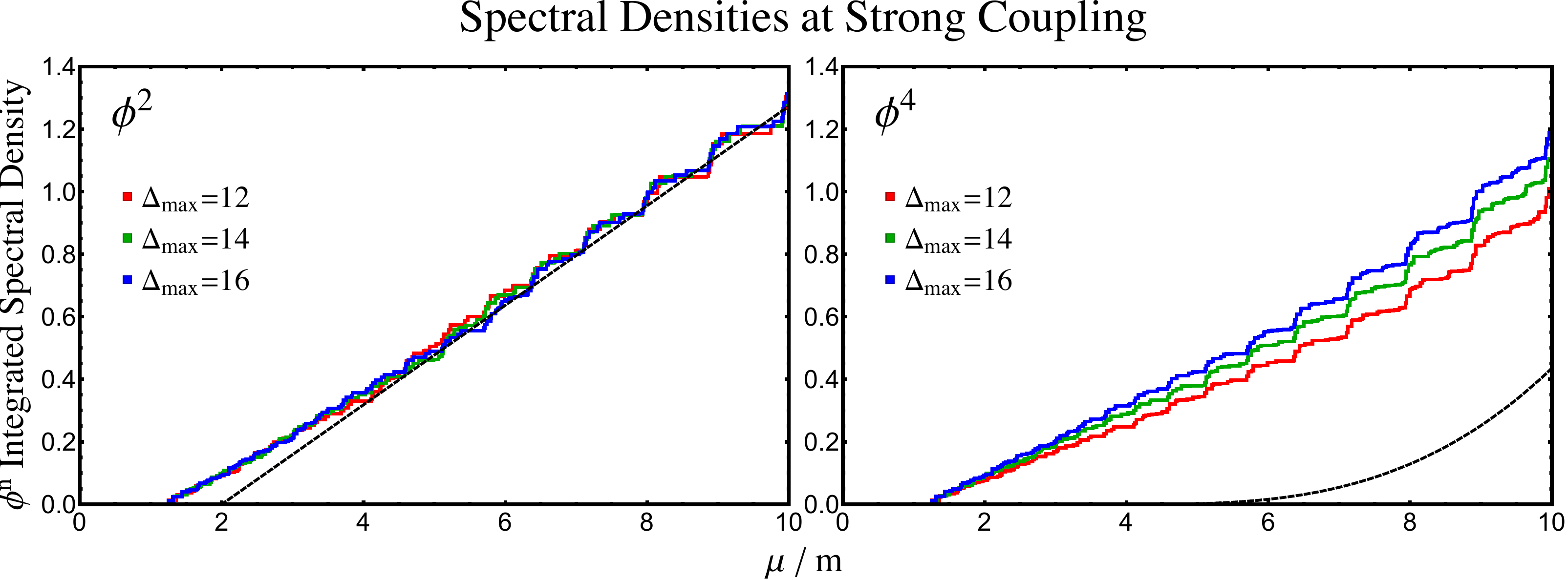}
\caption{Spectral density of $\phi^2$ (left) and $\phi^4$ (right) at nonperturbative coupling $\gbar\approx 0.85$ (where $\mu_{2\text{-part.}}/m = 1.25$). The dashed lines show the free theory result for comparison.}
\label{fig:SDStrong} 
\end{center}
\end{figure}

This does \emph{not} mean that the $\phi^4$ spectral density in figure~\ref{fig:SDStrong} is unphysical or does not contain meaningful new data about the interacting theory. It simply means that to obtain a cutoff-independent correlation function one should construct a ``normal-ordered'' $\phi^4$ operator, with the cutoff-dependent $\phi^2$ contribution removed. For example, in section~\ref{subsubsec:trace} we construct the cutoff-independent operator $T^\mu_{\,\,\mu}$ from $\phi^2$ and $\phi^4$, which can be thought of as exactly such a ``normal-ordering'' procedure. More generally, when comparing these truncation results to those obtained with other computational methods, one must be careful to ensure a consistent normal-ordering definition of local operators.

Overall, we see that our truncated basis can be used to compute nonperturbative correlation functions of local operators at generic values of the coupling $\bar{g}$, converging most rapidly in the IR. This is a generic feature encountered in previous studies: LCT spectral densities tend to converge from the IR up, which is a sign that low-dimension basis states have the most overlap with the physical IR degrees of freedom.

\subsection{Critical Point and the 3d Ising Model}
\label{subsec:Results-Ising}

In this section, we turn our attention to the critical point, where the mass gap closes and the low-energy physics is described by an IR CFT. We compute correlation functions near the critical point and verify that they exhibit behavior consistent with criticality. First, we compute the spectral densities and position space correlators of the operators $\phi^n$ and demonstrate that they have universal IR behavior. Then, we compute the spectral density for the trace of the stress tensor and demonstrate that it vanishes in the IR, as would be expected for a critical point described by a CFT. 

The specific CFT describing the critical point is the 3d Ising model. In this work, our IR precision is currently insufficient to reliably extract precise 3d Ising critical exponents. Nevertheless, our results provide strong evidence that we are beginning to probe Ising physics, and we hope this will open the door to new approaches to studying the 3d Ising model and its deformations.

\subsubsection{Universal Behavior}

In this section, we compute the two-point functions of the operators $\phi^n$ near the critical point. The expectation is that $\phi^n$ will flow in the IR to the lowest-dimension $\mathbb{Z}_2$ odd or even operator in the Ising model, \emph{i.e.}, to either $\sigma$ or $\epsilon$, depending on parity. In other words, we expect that in the IR
\be
\phi^{2n-1} \Rightarrow a_{2n-1}\, \sigma + \dots, \hspace{10mm} \phi^{2n} \Rightarrow a_{2n}\, \epsilon + \dots,
\label{eq:UniversalFlow}
\ee
where the coefficients $a_i$ are proportionality constants and the dots denote higher-dimension Ising operators. The expected flow in (\ref{eq:UniversalFlow}) implies that near the critical point, the operators $\phi^n$ should exhibit \emph{universal} behavior in the IR. Concretely, up to overall proportionality constants, the spectral densities of $\phi^{2n}$ should be identical in the IR, and similarly, the spectral densities of $\phi^{2n-1}$ should be identical in the IR.

\begin{figure}[t!]
\begin{center}
\includegraphics[width=1\textwidth]{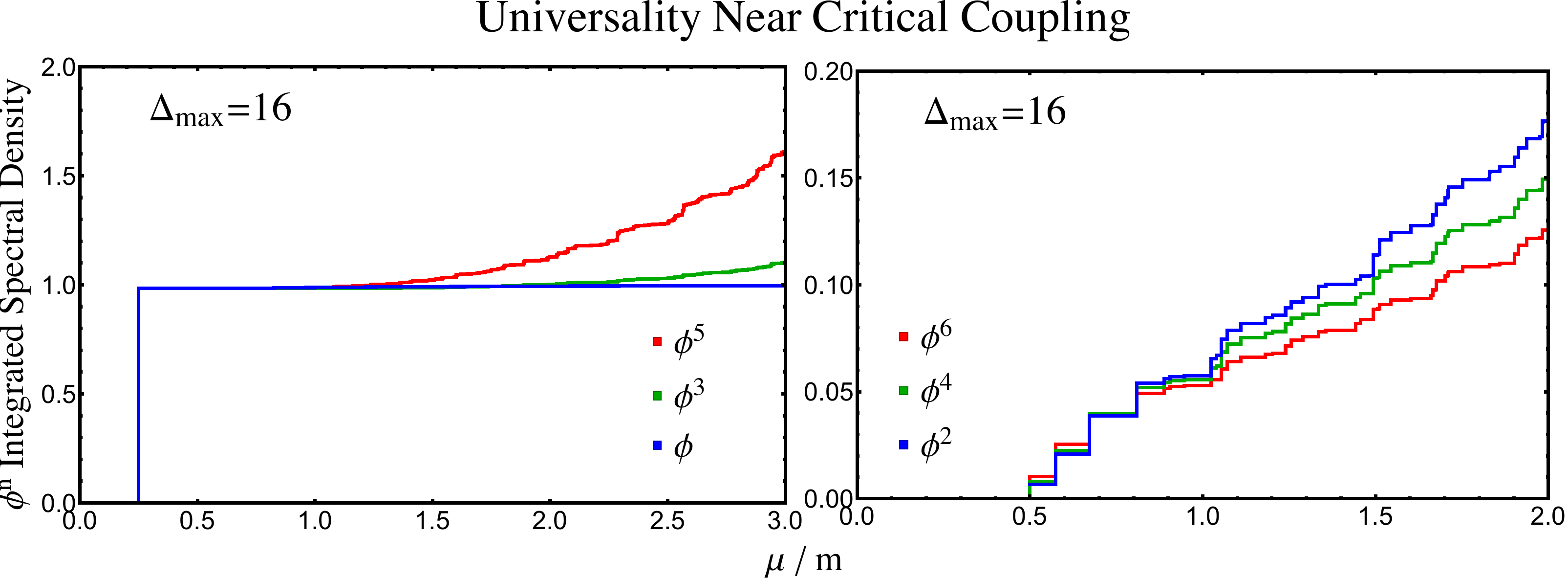}
\caption{Integrated spectral densities of $\phi^{2n-1}$ (left) and $\phi^{2n}$ (right) near the critical point at $\gbar \approx 1.03$ (where $2 \mu_{1\text{-part.}}/m = \mu_{2\text{-part.}}/m = 0.5$; see footnote \ref{footnote:uni}). The agreement in the IR signals universal behavior.}
\label{fig:Universality} 
\end{center}
\end{figure}

Figure~\ref{fig:Universality} shows the integrated spectral densities of $\phi$, $\phi^3$, and $\phi^5$ (left) and the integrated spectral densities of $\phi^2$, $\phi^4$, and $\phi^6$ (right) near the critical point. Specifically, we have set $\gbar \approx 1.03$\footnote{Strictly, $\gbar=1.05$ in the odd-particle sector and $\gbar=1.03$ in the even-particle sector.\label{footnote:uni}} such that $2 \mu_{1\text{-part.}} /m = \mu_{2\text{-part.}}/m = 0.5$. We clearly see universality in the IR behavior of these spectral densities. In the IR, the spectral densities of the operators $\phi^n$ with the same $\mathbb{Z}_2$-parity all match. These plots were constructed at our maximum truncation level of $\Dmax=16$ and $\imax=65$. We have rescaled each $\phi^{n}$ spectral density (for $n\geq 3$) by an overall multiplicative constant to allow for the proportionality constants $a_i$ in (\ref{eq:UniversalFlow}). The universal behavior we see in these plots is a clear indication that the IR physics is being controlled by the critical point.

\begin{figure}[t!]
\begin{center}
\includegraphics[width=.95\textwidth]{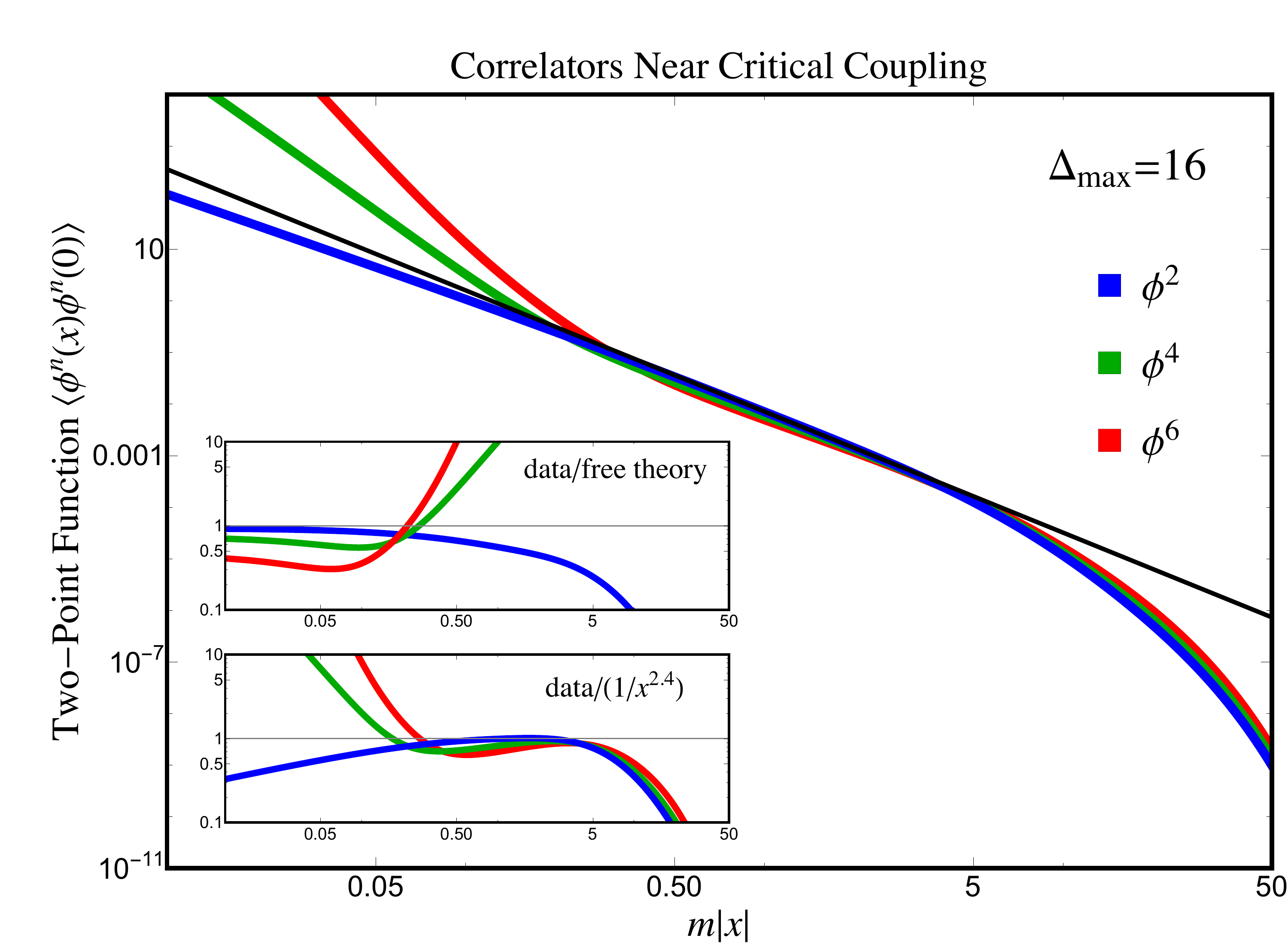}
\caption{Position space two-point functions $\corr{\phi^n(x)\phi^n(0)}$ for $n=2$, $4$, and $6$ near the critical coupling ($\bar{g} =1.05$).  Solid black line shows the fit in the region $1 \leq m|x| \leq 4$, given by $\frac{0.007}{|x|^{2.4}}$. Top inset shows the ratio of the correlators to the free theory correlators, given by $\frac{n!}{(4\pi)^n |x|^n}$. Bottom inset shows the ratio of the correlators to the power law fit. The extracted exponent $2.4$ is within $20\%$ of the 3d Ising prediction $2\Delta_\epsilon \approx 2.83$. In comparing to the fit in the bottom inset, we have rescaled the data by an overall constant for visual clarity.}
\label{fig:CriticalCouplingCorrelators} 
\end{center}
\end{figure}

As we did for the free theory in section~\ref{sec:Checks}, we can also compute the position space correlators $\corr{\phi^n(x)\phi^n(0)}$ at spacelike separation, by means of eq.~\eqref{eq:correlatorsfromspec}. Figure~\ref{fig:CriticalCouplingCorrelators} shows the $\phi^2$, $\phi^4$, and $\phi^6$ correlators closer to the critical point (with $\gbar = 1.05$).\footnote{Note that now $\gbar = 1.05$ in the \textit{even} sector (compared to the value of $\gbar \approx 1.03$ used in figure~\ref{fig:Universality}), so that the value of the gap $\mu_{2\text{-part.}}/m \approx 0.02$ is smaller than in figure~\ref{fig:Universality}.} Again, we have rescaled $\phi^4$ and $\phi^6$ to account for the proportionality constants in (\ref{eq:UniversalFlow}). We can see that there are three schematic regions. In the left region for $m|x| \lesssim 0.5$, the correlators approach free scalar correlators, as is expected from the UV CFT. In the middle IR region, for $0.5 \lesssim m |x| \lesssim 5$, the correlators exhibit clear universality and are all well-fit by the power law $\sim 1/|x|^{2.4}$.\footnote{We also note that this fit is insensitive to the precise start and end points of the middle interval.} Despite the relatively low value of $\Delta_{\textrm{max}}$ used in this work, it is encouraging that this approximate exponent is within $20\%$ of the 3d Ising prediction~\cite{Kos:2016ysd}
\be
\corr{\epsilon(x) \epsilon(0)} \approx \frac{1}{|x|^{2.83}}.
\ee
Finally, in the right region $m |x| \gtrsim 5$, the correlator starts to decay rapidly due to the nonzero mass gap. Overall, these plots provide strong evidence that we are accurately capturing the vicinity of the critical point with our truncated basis. With larger values of $\Dmax$, we expect to be able to push the mass gap even lower and reliably extract critical exponents from such correlation functions.

\subsubsection{Stress Tensor Trace}
\label{subsubsec:trace}

In this section, we compute the spectral density of the trace of the stress tensor, $T^\mu_{\,\,\mu}$. As we approach the critical coupling, we show that this spectral density vanishes in the IR, indicating that the critical point is described by a CFT, as expected.

Due to the presence of the state-dependent counterterm in the Hamiltonian~\eqref{eq:HplusCT2}, computing $T^\mu_{\,\,\mu}$ is nontrivial in our setup. This counterterm is a complicated object that cannot be written as the integral of a simple local operator. However, it does contain a \emph{state-independent piece}, which is clearly proportional to $\phi^2$ by construction (the counterterm is designed to cancel a UV-sensitive correction to the mass). Thus, generally speaking, our counterterm can be thought of schematically as
\be
g^2 \de P_+^{(\textrm{state-dep.})} (\Dmax) = \frac{1}{2} g^2 \delta_T(\Dmax) \int d^2 \vec{x} \, \phi^2(x)  + \dotsb, 
\label{eq:CTSchematic}
\ee
where $\delta_T(\Dmax)$ is an unknown $\Dmax$-dependent coefficient and the ellipses denote the remaining strictly state-dependent contributions. This state-independent part of the counterterm is designed to cancel the $O(g^2)$ contribution to the mass~\eqref{eq:1pMassShift}, so as we increase $\Dmax$ we expect
\be
\delta_T(\Dmax) \rightarrow \frac{1}{96\pi^2} \log \frac{\left(\frac{\Lambda}{m}+1 \right)^2}{8\left(\frac{\Lambda}{m}-1\right)} \qquad (\Dmax \ra \infty).
\ee
In analogy with~\eqref{eq:cLhat}, it will therefore be useful to parametrize this coefficient as
\begin{equation}
\delta_T(\Dmax) \equiv \hat{\de}_T(\Dmax) \cdot \frac{1}{96\pi^2} \log \frac{\left(\frac{\Lambda}{m}+1 \right)^2}{8\left(\frac{\Lambda}{m}-1\right)} \qquad \left( \Lambda \equiv \LUV^{(\imax=65)} \right), \label{eq:deltaThatdef}
\end{equation}
such that $\hat{\de}_T(\Dmax) \to 1$ in the $\Dmax \to \infty$ limit (at $\imax = 65$).

Including the state-independent contribution from our counterterm, we thus expect that at finite $\Dmax$ the trace of the stress tensor takes the form
\be
T^\mu_{\,\,\mu} (\Dmax) = \Big( m^2 - c_L g^2 + \delta_T(\Dmax)g^2 \Big) \phi^2 + \frac{g}{4!}\phi^4 .
\label{eq:Tmumu}
\ee
This is the operator we expect to vanish at the critical point. However, at any finite truncation, we do not know \textit{a priori} the value of the constant $\delta_T(\Dmax)$. Fortunately, we can measure $\delta_T(\Dmax)$ from the equation of motion for the operator $\phi$, which also receives a correction from the state-independent piece of our counterterm,
\begin{equation}
\p^2 \phi = -\Big( m^2 - c_L g^2 + \delta_T(\Dmax)g^2 \Big) \phi - \frac{g}{3!}\phi^3, \label{eq:eomconstraint}
\end{equation} 
giving us a separate, independent determination of this parameter.

\begin{figure}[t!]
\begin{center}
\includegraphics[width=.7\textwidth]{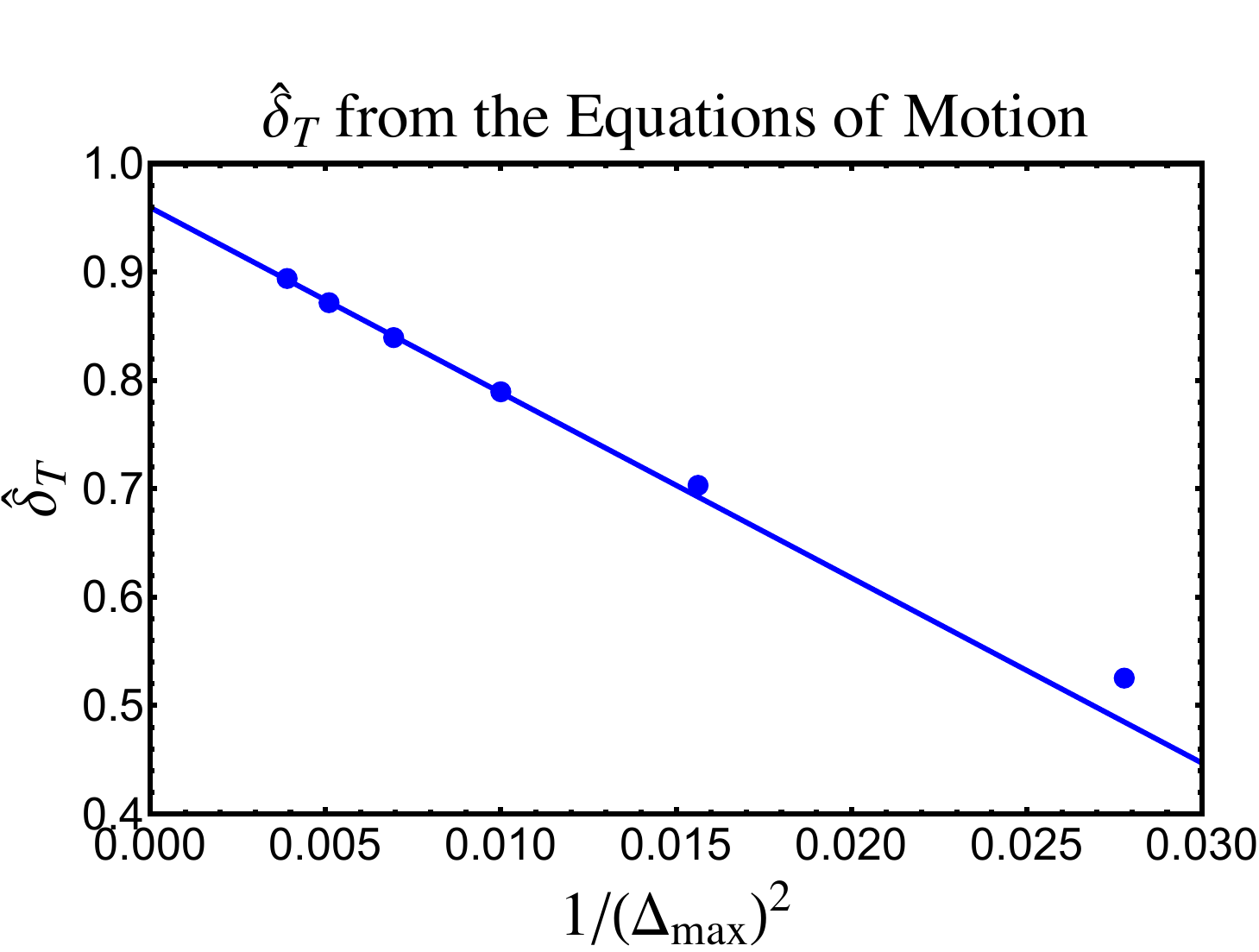}
\caption{Measured values of $\hat{\delta}_T$ (as defined in~\eqref{eq:deltaThatdef} and~\eqref{eq:deltaTstatedep}) for different values of $\Dmax$. The blue line shows the best fit to the four highest $\Dmax$ values, $\hat{\delta}_T \approx 0.96 - \frac{17}{\Dmax^2}$.}
\label{fig:deltaTvsdMax} 
\end{center}
\end{figure}

Concretely, at any fixed $\Dmax$, we can determine $\delta_T$ by requiring that the equation of motion~\eqref{eq:eomconstraint} is satisfied when acting on the vacuum. By acting from the left with an arbitrary state $|\psi\>$, we then obtain a relation between Hamiltonian matrix elements and the overlaps with $\phi$ and $\phi^3$,
\begin{equation}
\<\psi|2P_+P_-|\phi(0)\> = \Big( m^2 - c_L g^2 + \delta_T(\Dmax)g^2 \Big) \<\psi|\phi(0)\> + \frac{g}{3!}\<\psi|\phi^3(0)\>.
\end{equation}
If we choose the state to be the free one-particle Fock space state $|p\>$ and use the Hamiltonian~\eqref{eq:HplusCT2}, we find that every contribution from the Hamiltonian except for the counterterm explicitly cancels with a term on the RHS, leaving only the constraint
\begin{equation}
2p_-\<p|\de P_+^{(\textrm{state-dep.})}|\phi(0)\> = \delta_T(\Dmax).
\label{eq:deltaTstatedep}
\end{equation}
In other words, $\delta_T$ is simply the one-particle matrix element of $\delta P_+^{(\textrm{state-dep.})}$. Once we perform this measurement and obtain $\de_T$ for a given $\Dmax$, we can check that the spectral density $\rho_{T_{\,\,\mu}^\mu}(\mu)$ vanishes at the critical point for the \emph{same value} of $\delta_T$.

\begin{figure}[t!]
\begin{center}
\includegraphics[width=\textwidth]{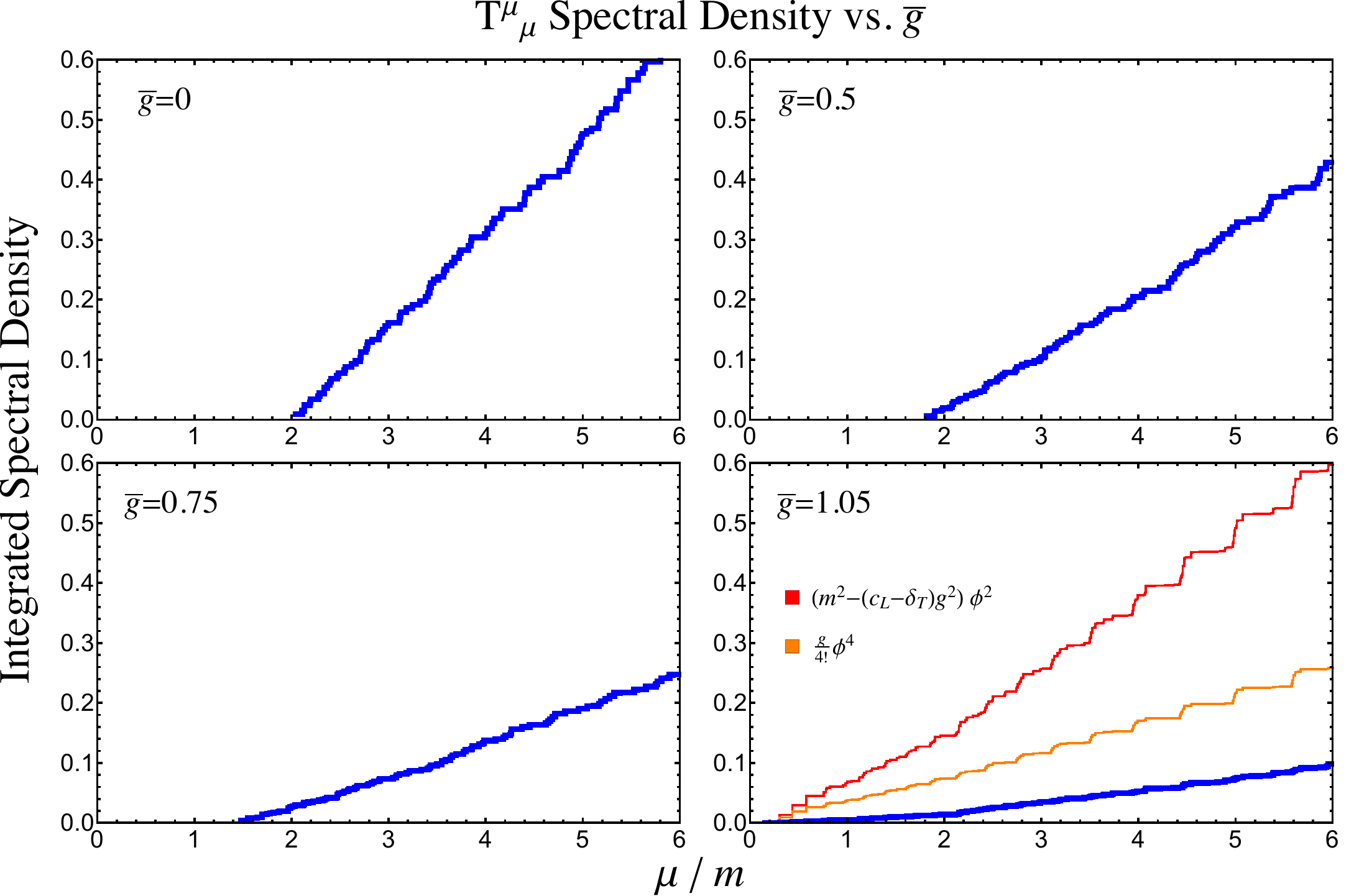}
\caption{Integrated spectral density of $T^\mu_{\,\,\mu}$ (blue), defined in (\ref{eq:Tmumu}), for different values of $\gbar$ approaching the critical point. Here $\Dmax=16$ and $\imax=65$, with $\hat{\delta}_T \approx 0.89$ fixed by the equations of motion. Note that the spectral density vanishes in the IR as one approaches the critical point.}
\label{fig:StressTensorTrace}
\end{center}
\end{figure}

Figure~\ref{fig:deltaTvsdMax} shows the values of $\hat{\delta}_T$ obtained from the equation of motion as we vary $\Dmax$. As we can see, $\hat{\de}_T$ approaches $1$ as we increase $\Dmax$, with the corrections scaling as approximately $1/\Dmax^2$. For $\Dmax=16$, we specificially obtain the value $\hat{\de}_T \approx 0.89$.

Given this value of $\hat{\de}_T$, we can now compute the spectral density of $T^\mu_{\,\,\mu}$. Figure~\ref{fig:StressTensorTrace} shows the resulting integrated spectral density (blue) for four different values of the coupling starting from free field theory (top left) and increasing to near the critical point (bottom right). Note that as we approach criticality, the spectral density steadily goes to zero in the IR. For comparison, in the final plot we have also included the spectral densities for the $\phi^2$ (red) and $\phi^4$ (orange) contributions to the trace. Note that the individual spectral densities of these operators are nonvanishing, and only the linear combination defining $T^\mu_{\,\,\mu} $ in (\ref{eq:Tmumu}), with our chosen value of $\delta_T$, vanishes near the critical point.\footnote{We note that the spectral density of $T_{\,\,\mu}^\mu$ contains a cross term between $\phi^2$ and $\phi^4$, so it is not given by just the sum of these two spectral densities.}

\begin{figure}[t!]
\begin{center}
\includegraphics[width=\textwidth]{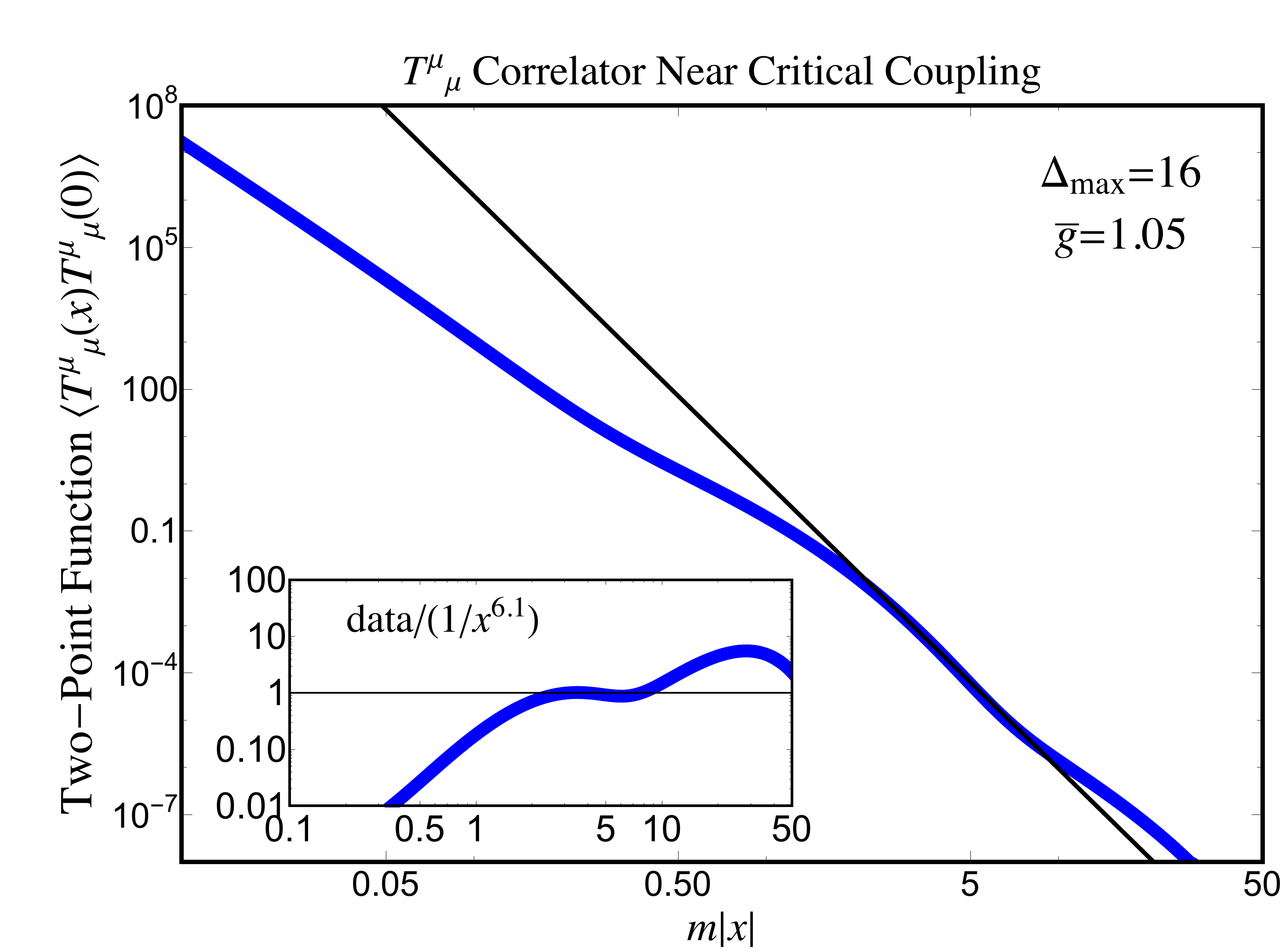}
\caption{The two-point function $\corr{T^\mu_{\,\,\mu}(x)T^\mu_{\,\,\mu}(0)}$ at $\gbar = 1.05$. Black line shows the best fit in the region $3 \leq m |x| \leq 7$, given by a power law $\approx \frac{1.1}{|x|^{6.1}}$. Inset shows the data normalized by the power law fit. The extracted exponent $6.1$ is within $15\%$ of the Ising prediction $2\Delta_\epsilon+4 \approx 6.82$.}
\label{fig:TraceCorrelator}
\end{center}
\end{figure}

From this spectral density, we can also compute the two-point function $\corr{T^\mu_{\,\,\mu}(x)T^\mu_{\,\,\mu}(0)}$ at spacelike separation near the critical point. In the vicinity of the critical point, where the dynamics should be controlled by the Ising model, we expect that the trace behaves schematically like\footnote{Note that $T^\mu_{\,\,\mu}$ on the LHS of~\eqref{eq:UVStressTensor} is the trace of the improved stress tensor of the \emph{UV theory}, which does not generically correspond to the improved stress tensor of the \emph{IR theory}, hence we expect a contribution from the descendant $\p^2\epsilon$ which does not vanish as $\mgap \to 0$. We thank Slava Rychkov and Riccardo Rattazzi for helpful discussions on this point.} 
\begin{equation}
T^\mu_{\,\, \mu}(x) \sim m_{\textrm{gap}}^{3-\Delta_\epsilon} \, \epsilon(x) + \frac{1}{\Lambda_{\textrm{Ising}}^{\Delta_\epsilon-1}}\partial^2 \epsilon(x) + \dotsb,
\label{eq:UVStressTensor}
\end{equation} 
where $\Lambda_{\textrm{Ising}}$ is the cutoff of the Ising effective theory (set by the UV parameter $g$) and $\cdots$ indicates other higher-dimensional Ising operators. As $m_{\textrm{gap}} \to 0$, we therefore expect that the correlator behaves like a power law $\corr{T^\mu_{\,\,\mu}(x)T^\mu_{\,\,\mu}(0)} \sim \frac{1}{|x|^{2\Delta_\epsilon+4}} \approx \frac{1}{|x|^{6.82}}$. This expectation is qualitatively confirmed in figure~\ref{fig:TraceCorrelator}, which shows $\corr{T^\mu_{\,\,\mu}(x)T^\mu_{\,\,\mu}(0)}$ near criticality at $\gbar = 1.05$. In the IR, there is a regime $3 \lesssim m |x| \lesssim 7$ that is well-fit by a power law $\sim 1/|x|^{6.1}$, which is within 15\% of the expected exponent. Although we are quite far from precision physics due to our low truncation cutoff and our error bars should be more rigorously understood, it is nevertheless surprising and encouraging that we appear to be able to observe subleading behavior of the trace correlator. It is our hope that this proof-of-concept will pave the way for more precise Hamiltonian truncation studies of this theory, both near the critical point and at generic coupling.


\section{Discussion and Outlook}
\label{sec:Discussion}

Hamiltonian truncation is a potentially very powerful tool for computing the nonperturbative dynamics of general quantum field theories. However, there has long been a significant obstacle to its implementation in most QFTs, especially those in $d \geq 3$: the necessity of state-dependent counterterms for UV divergences. In this work, we have focused on addressing this issue in the specific case of (2+1)d $\phi^4$-theory in lightcone quantization.

In this particular setting, we have presented a simple prescription for constructing the necessary state-dependent counterterms for the logarithmic correction to the mass due to the ``sunset'' diagram. Our prescription is admittedly rather brute force, where we explicitly cancel the $O(g^2)$ correction to each $n$-particle mass eigenstate due to $(n+2)$-particle intermediate states and replace them with a local, state-independent shift in the bare mass. However, this prescription allows us to reproduce the closing of the mass gap as the $\phi^4$ coupling increases, with consistent ratios between the one-, two-, and three-particle thresholds until near the critical point. In addition, we have computed the two-point functions of $\phi^n$ and the stess tensor trace $T^\mu_{\,\,\mu}$ at various couplings, providing the first nonperturbative calculation of these observables in the symmetric phase of (2+1)d $\phi^4$-theory.

This theory was also recently studied in~\cite{EliasMiro:2020uvk}, in order to address the same issue of state-dependent counterterms in the context of finite volume and equal-time quantization. As discussed in appendix~\ref{app:StateDependence}, the details of this state-dependence are different in the two quantization schemes, requiring distinct prescriptions for constructing the necessary counterterms. Nevertheless, the approaches presented here and in~\cite{EliasMiro:2020uvk} should be readily generalizable to other theories, and we hope that these two works will motivate much future work on the use of Hamiltonian truncation methods for strongly-coupled QFT.

An alternative prescription for constructing state-dependent counterterms was presented in~\cite{Anand:2019lkt} for the case of (1+1)d Yukawa theory, which also contains logarithmic divergences. This approach involves first computing the supercharge $Q_+$ of a supersymmetric theory containing the desired QFT, and constructing the state-dependent counterterm from the square of the truncated supercharge, $\de P_+^{(\textrm{state-dep.})} \sim Q_+^2$. The advantage of this prescription is that the resulting counterterm is automatically built from a truncated sum over intermediate states, giving it the same state-dependence as the UV divergence. It would be exciting if this approach could be applied to (2+1)d $\phi^4$-theory, using the $Q_+$ from $\mathcal{N}=1$ SUSY Yukawa theory. It would also be interesting to study 3d SUSY Yukawa theory itself and obtain the RG flow to the minimal 3d SCFT in the IR~\cite{Atanasov:2018kqw}. Note that no new technology needs to be developed to apply LCT to this SUSY theory, making it a perfect target for future Hamiltonian truncation studies in $d=3$.

In LCT, the main computational difficulty is the construction of the basis of primary operators and the calculation of Hamiltonian matrix elements. As a result, in this work we have needed to develop many tools to improve the efficiency of the evaluation of inner products and matrix elements, detailed in appendices~\ref{sec:Basis} and \ref{sec:MatrixElements}. This is in contrast with methods which use a Fock space basis in finite volume, such as ET Hamiltonian truncation~\cite{Rychkov:2014eea,Rychkov:2015vap,Bajnok:2015bgw,Elias-Miro:2015bqk,Elias-Miro:2017xxf,Elias-Miro:2017tup,EliasMiro:2020uvk} and discrete lightcone quantization (DLCQ)~\cite{Pauli:1985pv,Pauli:1985ps}, for which the computation of the basis and matrix elements is largely trivial. However, LCT appears to require fewer states than such Fock space methods to obtain a given IR resolution and UV cutoff. For example, in this work the maximum basis we consider ($\Dmax=16$, $\imax=65$) has 35,425 states. We can estimate the associated UV and IR cutoffs from the free massive theory results in figures~\ref{fig:FreeSpectralDensities} and~\ref{fig:PositionSpaceCorrelators}. The effective UV cutoff corresponds to the value of $\mu$ at which the truncation results for spectral densities significantly deviate from the theoretical predictions, which we can conservatively estimate from figure~\ref{fig:FreeSpectralDensities} as $\LUV/m \approx 30$. The effective IR cutoff corresponds to the length scale at which the correlation functions in figure~\ref{fig:PositionSpaceCorrelators} deviate from the theory prediction, which is approximately $mL \approx 10$. These cutoff values would naively require a much larger basis in any Fock space method (see, \emph{e.g.}, figure 1 in~\cite{EliasMiro:2020uvk}).

It is intriguing that there is this apparent tradeoff in complexity between different methods: simplicity of matrix elements versus size of the basis. It would be interesting to compare the relative computational complexity of various truncation methods more quantitatively, in order to better understand if this ``conservation of difficulty'' is fundamental to obtaining nonperturbative physics or if there are certain choices of basis or approaches which are most efficient for particular observables. 

It would also be useful to push the LCT basis to higher values of $\Dmax$, in order to extract critical exponents and confirm that the critical point is described by the 3d Ising CFT. One useful tool for reaching higher values of $\Dmax$ would be to compute the general expression for CFT three-point functions of spinning operators in momentum space, building on recent results~\cite{Gillioz:2018mto,Gillioz:2019lgs,Bautista:2019qxj,Anand:2019lkt,Jain:2020rmw,Bzowski:2020kfw,Jain:2020puw}, which would greatly improve the efficiency for computing Hamiltonian matrix elements. Though in practice our Dirichlet basis states do not individually correspond to primary operators, one can construct a map between the two bases, in order to more efficiently construct matrix elements from CFT data. Relatedly, it would also be useful to implement the ``OPE method'' of~\cite{Hogervorst:2014rta} to evaluate matrix elements more efficiently than via Wick contraction. It is also worth noting that the data presented in this work was generated on personal laptops with an approximate runtime of a few days. With more computational resources and improved efficiency, future work should be able to significantly increase $\Dmax$.

Further increasing the basis size would also allow us to better understand the effects of truncation in the context of UV divergences. At finite $\Dmax$, we have seen that our counterterm prescription requires the addition of a local shift in the bare mass, parametrized by $c_L g^2$, to ensure that we reach the correct universality class in the IR. While the precise value of this counterterm does not appear to affect scheme-independent physical observables, there is only a finite range of values for $c_L$ which do not lead to significant truncation errors. Our expectation is that as $\Dmax$ increases, this allowed range for the counterterm will continue to grow. However, from~\cite{SeroneUpcoming} we expect that only values of $c_L$ above some lower bound reach the same universality class, and below this threshold the gap no longer closes. With higher values of $\Dmax$, we could potentially observe both the existence of this threshold, as well as the Chang-Magruder duality~\cite{Chang:1976ek,Magruder:1976px} relating low and high values of the coupling $g$ in the $\mathbb{Z}_2$-symmetric phase, studied recently in~\cite{EliasMiro:2020uvk,SeroneUpcoming}.

As discussed in appendix~\ref{app:LambdaEffective}, there is also an interesting connection between the IR resolution set by the discretization of $\mu$ for CFT basis states and the effective UV cutoff set by truncation. It would be useful to study this interplay between $\Dmax$ and $\Lambda_\IR$ in more detail, in order to better understand the convergence of LCT in the presence of UV divergences and improve the extrapolation of our numerical results in $\Dmax$.

\section*{Acknowledgments}

Firstly, we would like to thank Charles Hussong for collaboration during the initial stages of this project. We are grateful to Joan Elias Mir\'{o}, Liam Fitzpatrick, Slava Rychkov, and Marco Serone for comments on the draft. We would also like to thank Simon Caron-Huot, Joan Elias Mir\'{o}, Liam Fitzpatrick, Marc Gillioz, Edward Hardy, Brian Henning, Matthijs Hogervorst, Denis Karateev, Jo\~{a}o Penedones, Riccardo Rattazzi, Slava Rychkov, Giacomo Sberveglieri, Marco Serone, Gabriele Spada, Balt van Rees, and Yuan Xin for valuable discussions. We are grateful to the Abdus Salam International Centre for Theoretical Physics for hospitality while parts of this work were completed. This research was supported in part by the Perimeter Institute for Theoretical Physics. Research at Perimeter Institute is supported by the Government of Canada through the Department of Innovation, Science and Economic Development and by the Province of Ontario through the Ministry of Research and Innovation. EK is supported in part by the US Department of Energy Office of Science under Award Number DE-SC0015845. NA, ZK, and MW are supported by the Simons Collaboration on the Nonperturbative Bootstrap. ZK is also supported by the DARPA, YFA Grant D15AP00108. This research received funding from the Simons Foundation grant \#488649  (Simons Collaboration on the Nonperturbative Bootstrap). MW is partly supported by the National Centre of Competence in Research SwissMAP funded by the Swiss National Science Foundation.

\appendix
\section{Overview: Lightcone Conformal Truncation in 3d}
\label{sec:Overview}

Lightcone conformal truncation (LCT) is an example of a \emph{Hamiltonian truncation} method. Hamiltonian truncation broadly refers to an array of computational methods used to study QFTs nonperturbatively. While the implementation details and physical deliverables vary significantly from one method to another, all truncation methods involve restricting the Hilbert space to some finite-dimensional subspace and diagonalizing the resulting truncated Hamiltonian to obtain an approximation to the low-energy eigenstates of the QFT.

LCT is a specific version of Hamiltonian truncation that can be applied whenever the QFT of interest has a UV CFT fixed point. The LCT basis is defined in terms of the primary operators of the CFT, and the Hamiltonian matrix elements can be obtained from the OPE coefficients of these primary operators. Thus, one uses UV CFT input to compute IR QFT dynamics. We now describe more concretely the details of the LCT setup in three spacetime dimensions. 

\subsection{Lightcone Hamiltonian}
\label{subsec:Overview-Hamiltonian}

First, to even define the QFT Hamiltonian, we need to choose a quantization scheme, \emph{i.e.}, a definition of `time' versus `space'. As its name suggests, LCT is formulated in lightcone quantization. In our conventions, lightcone coordinates $(x^+,x^-,x^\perp)$ are related to $(x^0,x^1,x^2)$ by
\be
\begin{aligned}
& x^\pm \equiv \tfrac{1}{\sqrt{2}} \left( x^0 \pm x^1 \right), \hspace{10mm} x^\perp \equiv x^2, \\
& \hspace{10mm} ds^2 = 2dx^+ dx^- - dx^{\perp 2}.
\end{aligned}
\label{eq:LCcoords}
\ee
In this quantization scheme, one takes $x^+$ to be time and $\vec{x} = (x^-,x^\perp)$ to be spatial. The lightcone momenta are defined by $P_{\pm} \equiv \left( P_0 \pm P_1 \right) / \sqrt{2}$ and $P_\perp \equiv P_2$. In particular, $P_+$ is the Hamiltonian: 
\be
P_+ = \text{Hamiltonian}.
\ee

The Hamiltonian has two contributions: the original Hamiltonian of the UV CFT and the relevant deformations which lead to the desired IR QFT. For the specific case of $\phi^4$-theory, with the Lagrangian given by eq.~\eqref{eq:L}, we have
\be
\boxed{
\begin{gathered}
\text{Hamiltonian:} \hspace{114mm} \\
P_+ = P_+^{\textrm{(CFT)}} + \delta P_+^{(m)} + \delta P_+^{(g)}, \\[10pt]
P_+^{\textrm{(CFT)}} = \frac{1}{2} \int d^2\vec{x} \,  (\p_\perp \phi)^2, \hspace{5mm} \delta P_+^{(m)} =  \frac{1}{2}m^2 \int d^2\vec{x} \,  \phi^2, \hspace{5mm} \delta P_+^{(g)} =  \frac{1}{4!}g \int d^2\vec{x} \,  \phi^4.
\label{eq:P+Appendix}
\end{gathered}
}
\ee
This is the Hamiltonian we will study (plus the state-dependent counterterm, which we discuss in more detail in appendix~\ref{app:CountertermExample}). The next step is to express the Hamiltonian in a particular basis, so now let us turn to the LCT basis. 

\subsection{LCT Basis} 
\label{subsec:Overview-Basis}

Our basis is constructed in momentum space and consists of Fourier transforms of operators $\CO$ in the UV CFT. We start by defining the states
\be
\ket{\CO,\mu} \equiv \int d^3x \, e^{-ip\cdot x} \, \CO(x) \ket{0},
\label{eq:LCTStateAppendix}
\ee
where the label $\mu$ is the invariant mass $\mu^2 \equiv 2p_+p_- - p_\perp^2$. Because we are interested in relevant deformations that preserve Poincar\'{e} invariance, we can choose to work in a fixed spatial momentum frame $\vec{p}$. For simplicity, we will only label states in this momentum frame by the associated local operator $\Ocal$ and invariant mass $\mu$.

However, $\mu$ is still a continuous parameter that needs to be discretized in some way in order to implement Hamiltonian truncation. Following the prescription in~\cite{Katz:2016hxp}, we choose to impose a hard cutoff $\LUV$, restricting the invariant mass to $\mu^2 \leq \LUV^2$, and introduce \emph{smearing functions} $b_{\mathfrak{i}}(\mu)$ on this interval, resulting in the discrete set of states
\be
\ket{\CO,\mathfrak{i}} \equiv \frac{1}{\sqrt{2\pi}} \int_0^{\LUV^2} d\mu^2 \, b_{\mathfrak{i}}(\mu) \, \ket{\CO,\mu},
\label{eq:LCTStateDiscreteAppendix}
\ee
where $\ifrak = 1,\ldots,\imax$, and the prefactor $1/\sqrt{2\pi}$ has been added for future convenience. There is clearly freedom in the choice of the smearing functions $b_{\mathfrak{i}}(\mu)$, and we discuss our specific choice in appendix~\ref{sec:Basis}.

The states $\ket{\CO,\mathfrak{i}}$ defined in (\ref{eq:LCTStateDiscreteAppendix}) come with \emph{two} truncation parameters: $\Dmax$ and $\mathfrak{i}_{\text{max}}$. The first parameter, $\Dmax$, is the cutoff on the maximum scaling dimension of the operators $\CO$. That is, for a given $\Dmax$, we only include operators in our basis with scaling dimension $\Delta \leq \Dmax$. The second parameter, $\mathfrak{i}_{\text{max}}$, is the cutoff on the number of smearing functions $b_{\mathfrak{i}}(\mu)$ that we will use to probe the interval $\mu^2 \in [0,\LUV^2]$. The parameter $\mathfrak{i}_{\text{max}}$ sets our resolution in $\mu^2$ and is computationally inexpensive, whereas increasing $\Dmax$ is more complicated, since it involves constructing additional higher-dimension operators and computing their Hamiltonian matrix elements.

The precise definition of the LCT basis depends on which operators $\CO$ from the UV CFT are used to construct the momentum space states in (\ref{eq:LCTStateAppendix}) (and their discretized versions in (\ref{eq:LCTStateDiscreteAppendix})). In a CFT, the Fourier transform is an operation that packages together a full conformal multiplet (primary plus descendants). Thus, the standard strategy for constructing a complete LCT basis would be to select one representative operator, say the primary operator, from each conformal multiplet. For 3d $\phi^4$-theory, however, we will actually choose operators that are linear combinations of representatives from different multiplets, due to a subtlety introduced by the mass deformation $\delta P_+^{(m)}$ in (\ref{eq:P+Appendix}). More concretely, matrix elements of $\delta P_+^{(m)}$ exhibit \emph{IR divergences} in 3d, and a practical and efficient way to handle these divergences is to choose ``Dirichlet" operators $\CO_D$ from the UV CFT, which are linear combinations of operators from different conformal multiplets defined to satisfy a particular boundary condition. 

The fact that matrix elements of $\delta P_+^{(m)}$ can exhibit divergences when evaluated between states of the form $\ket{\CO,\mu}$ in~(\ref{eq:LCTStateAppendix}) was discussed in detail in~\cite{Katz:2016hxp}. These divergences arise when the Fock space wavefunction of an operator $\CO$ fails to satisfy a particular boundary condition. Specifically, let $\ket{p_1,\dots,p_n}$ denote an $n$-particle lightcone Fock space state. Given an $n$-particle operator $\CO(x)$ (\emph{i.e.}, an operator built out of $n$ $\phi$'s), its Fock space wavefunction is $F_{\CO}(p) \equiv \langle p_1,\dots,p_n|\CO(0) \rangle$. We say that $\CO$ is ``Dirichlet" (in that it satisfies a Dirichlet boundary condition) if $F_{\CO}(p) \rightarrow 0$ whenever any $p_{i-} \rightarrow 0$. It is straightforward to check that Dirichlet operators have finite $\delta P_+^{(m)}$ matrix elements, while non-Dirichlet operators have divergent matrix elements. Thus, unsurprisingly, the effect of the divergences in $\delta P_+^{(m)}$ is to lift out all non-Dirichlet states from the spectrum, \emph{i.e.}, these states become infinitely massive in the presence of a $\phi^2$ deformation, leaving only Dirichlet states in the low-energy Hilbert space. We refer the reader to~\cite{Katz:2016hxp} for more details.  

For our purposes, all we need is the punchline, which is simple. Since all non-Dirichlet states are lifted out anyway, it is more efficient to define our basis using Dirichlet operators from the outset. Thus, instead of choosing primary operators when defining the states in~(\ref{eq:LCTStateAppendix})-(\ref{eq:LCTStateDiscreteAppendix}), we use only Dirichlet operators $\CO_{D}$. In particular, the definitions of the states $\ket{\CO,\mu}$ in~(\ref{eq:LCTStateAppendix}) and the discrete version $\ket{\CO,\mathfrak{i}}$ in~(\ref{eq:LCTStateDiscreteAppendix}) remain unchanged except that $\CO$ is restricted to Dirichlet operators. 

Identifying Dirichlet operators is easy. In position space language, they are simply operators where every $\phi$ has at least one $\p_-$ acting on it~\cite{Katz:2016hxp}. To say it another way, Dirichlet operators are built from $\p_- \phi$ instead of $\phi$. For example, $(\p_-\phi)^2$ and $(\p_-\phi)(\p_\perp \p_- \phi)$ are Dirichlet, whereas $\phi^2$ and $\phi \p_-\phi$ are not. Note that in momentum space, each insertion of $\p_-\phi$ yields a factor of $p_{i-}$, which of course vanishes as $p_{i-} \rightarrow 0$, thus satisfying the Dirichlet condition. 

To summarize, our final basis consists of the discretized momentum space states in~(\ref{eq:LCTStateDiscreteAppendix}), where $\CO\in\{\CO_D\}$. This ensures that all Hamiltonian matrix elements are finite and avoids the circuitous route of keeping non-Dirichlet states around only to see them be subsequently lifted out. 

There is one significant drawback to working with Dirichlet operators, however. As we have just discussed, Dirichlet operators are themselves not primary. Rather, a Dirichlet operator is a linear combination of a primary operator and descendants of lower-dimension primaries and thus cannot be described as living in the kernel of the special conformal generator. As a consequence, we do not know of a recursive algorithm for generating higher-particle Dirichlet operators from lower-particle ones. In 2d scalar field theory, where Dirichlet operators \emph{are} primary, such a recursive algorithm provided an efficient way of generating basis states~\cite{Anand:2020gnn}. We will not have this tool available to us in 3d. Nevertheless, we will show that conformal symmetry still plays an important role and can be harnessed to efficiently generate Dirichlet basis states. 

Let us conclude this subsection by summarizing the main formulas. Our basis consists of the states $\{ \ket{\CO,\mathfrak{i}} \}$ defined by:
\be
\boxed{
\begin{aligned}
\text{Basis:} \hspace{11mm} & \\
& \ket{\CO,\mathfrak{i}} \equiv \frac{1}{\sqrt{2\pi}} \int_0^{\LUV^2} d\mu^2 \, b_{\mathfrak{i}}(\mu) \, \ket{\CO,\mu}, \hspace{10mm} (\Delta \leq \Dmax, \hspace{2mm} \mathfrak{i} \leq \mathfrak{i}_{\text{max}})\\[10pt]
\text{where}  \hspace{10mm} \\
& \CO = \text{Dirichlet operator $\CO_D$, built from $(\p_-\phi)$'s plus derivatives}, \hspace{10mm} \\
& \ket{\CO,\mu} = \int d^3x \, e^{-ip\cdot x} \, \CO(x) \ket{0},  \hspace{5mm} (\mu^2 \equiv 2p_+p_- - p_\perp^2) \\
& \LUV^2 = \text{hard UV cutoff on $\mu^2$}, \\
& b_{\mathfrak{i}}(\mu) = \text{smearing functions (to be specified)}.
\label{eq:Basis}
\end{aligned}
}
\ee
With our basis defined, let us now turn to the structure of inner products and Hamiltonian matrix elements.

\subsection{Inner Products and Matrix Elements}
\label{subsec:Overview-MatrixElements}

With our basis defined in (\ref{eq:Basis}), inner products and matrix elements between states take the following forms, respectively,
\be
\begin{aligned}
\langle \CO,\mathfrak{i} | \CO^\prime, \mathfrak{j} \rangle &= \frac{1}{2\pi} \int_0^{\LUV^2} d\mu^2 \, d\mu^{\prime 2} \, b_{\mathfrak{i}}(\mu) \, b_{\mathfrak{j}}(\mu^\prime) \, \langle \CO,\mu | \CO^\prime, \mu^\prime \rangle, \\[10pt]
\langle \CO,\mathfrak{i} | P_+ | \CO^\prime, \mathfrak{j} \rangle &= \frac{1}{2\pi} \int_0^{\LUV^2} d\mu^2 \, d\mu^{\prime 2} \, b_{\mathfrak{i}}(\mu) \, b_{\mathfrak{j}}(\mu^\prime) \, \langle \CO,\mu | P_+ | \CO^\prime, \mu^\prime \rangle.
\label{eq:InnerAndME}
\end{aligned}
\ee
Note that the integrals on the RHS are smearings of the ``continuous $\mu$" inner products $\langle \CO,\mu | \CO^\prime, \mu^\prime \rangle$ and matrix elements $\langle \CO,\mu | P_+ | \CO^\prime, \mu^\prime \rangle$.

The undiscretized inner products and matrix elements always have the following general form
\be
\begin{aligned}
\langle \CO,\mu | \CO^\prime, \mu^\prime \rangle &\equiv (2\pi)^3 \delta^{(3)}(p-p^\prime) \, G_{\CO\CO^\prime}, \\[10pt]
\langle \CO,\mu | P_+ | \CO^\prime, \mu^\prime \rangle &\equiv (2\pi)^2 \delta^{(2)}(\vec{p}-\vec{p}^{\,\prime}) \, \CM_{\CO\CO^\prime}(\mu,\mu^\prime).
\label{eq:ContinuousInnerAndME}
\end{aligned}
\ee
Recall that when we write $\ket{\CO,\mu}$, we are suppressing the explicit spatial momentum label $\vec{p}$, as discussed below (\ref{eq:LCTStateAppendix}). Additionally, we will always choose to normalize $\ket{\CO,\mu}$ with appropriate powers of $p_-$ and $\mu$ such that the inner products $G_{\CO\CO^\prime}$ in the first line of (\ref{eq:ContinuousInnerAndME}) are all \emph{constant}. We will be explicit in later appendices about precisely what factors should be included.

For now, let us note the delta functions appearing on the RHS. Their presence is a manifestation of the fact that we can always choose to work in a particular spatial momentum frame with a fixed value for $\vec{p}=(p_-,p_\perp)$. In this paper, we will always work in the following frame,
\be
\text{ Frame for $\vec{p}$ :} ~~(p_- = \text{constant}, \, p_\perp = 0).
\ee
This is a convenient choice for two reasons. First, in this frame the Lorentz invariant mass-squared operator $M^2 \equiv 2P_+P_- - P_\perp^2$ simplifies to $M^2 = 2p_- P_+$. Thus, in this frame \emph{diagonalizing $M^2$ is equivalent to diagonalizing $P_+$}. Consequently, we will often refer to these two operators interchangeably. The second reason this frame is convenient is that the expressions for Hamiltonian matrix elements become simpler.  

The objects $G_{\CO\CO^\prime}$ and $\CM_{\CO\CO^\prime}(\mu,\mu^\prime)$ were studied in great detail in~\cite{Anand:2019lkt}. In particular, substituting the definition of $\ket{\CO,\mu}$ from (\ref{eq:LCTStateAppendix}), it is evident that the continuous-$\mu$ inner products and matrix elements are Fourier transforms of UV CFT 2- and 3-point functions, respectively. Using this fact (and working without loss of generality in the same $p_\perp = 0$ frame as above), the authors of~\cite{Anand:2019lkt} were able to derive a complete set of reference formulas for evaluating $G_{\CO\CO^\prime}$ and $\CM_{\CO\CO^\prime}(\mu,\mu^\prime)$ in 3d $\phi^4$-theory. We will be able to directly use these formulas, as we explain in detail in appendix~\ref{sec:MatrixElements}.

For now, let us simply insert (\ref{eq:ContinuousInnerAndME}) into (\ref{eq:InnerAndME}). For the inner product, a useful intermediate relation is that $\delta^{(3)}(p-p^\prime) = 2p_- \delta(\mu^2 - \mu^{\prime 2}) \delta^{(2)}(\vec{p}-\vec{p}^{\,\prime})$. The resulting formulas can be summarized as follows
\be
\boxed{
\begin{aligned}
& \hspace{-30mm} \text{Inner Product \& Matrix Elements:}  \\
\langle \CO,\mathfrak{i} | \CO^\prime, \mathfrak{j} \rangle &\equiv 2p_- (2\pi)^2 \delta^{(2)}(\vec{p}-\vec{p}^{\,\prime}) \cdot G_{\CO\CO^\prime} \cdot \int_0^{\LUV^2} d\mu^2 \,  b_{\mathfrak{i}}(\mu) \, b_{\mathfrak{j}}(\mu), \\
\langle \CO,\mathfrak{i} | 2p_-P_+ |\CO^\prime, \mathfrak{j} \rangle &= 2p_- (2\pi)^2 \delta^{(2)}(\vec{p}-\vec{p}^{\,\prime}) \cdot \frac{1}{2\pi} \int_0^{\LUV^2} d\mu^2 \, d\mu^{\prime 2} \, b_{\mathfrak{i}}(\mu) \, b_{\mathfrak{j}}(\mu^\prime ) \, \CM_{\CO\CO^\prime}(\mu,\mu^\prime), \\[10pt]
\text{where} \hspace{20mm} & \\
& \hspace{-15mm} G_{\CO\CO^\prime} = \frac{\langle \CO,\mu | \CO^\prime, \mu^\prime \rangle}{(2\pi)^3 \delta^{(3)}(p-p^\prime)}, \hspace{10mm} \CM_{\CO\CO^\prime}(\mu,\mu^\prime) = \frac{\langle \CO,\mu | P_+ | \CO^\prime, \mu^\prime \rangle}{ (2\pi)^2 \delta^{(2)}(\vec{p}-\vec{p}^{\,\prime})}.
\label{eq:InnerAndMEBox}
\end{aligned}
}
\ee

\section{Details: Constructing the Basis}
\label{sec:Basis}

\subsection{Monomial Operators}
\label{subsec:Basis-Monomials}

The LCT basis, defined in (\ref{eq:Basis}), is constructed using operators in the UV CFT, which in our case is 3d free massless scalar field theory. The CFT operators are thus composed of the free field $\phi$ along with the derivatives $\p_+$, $\p_-$, and $\p_\perp$. The equation of motion $\p^2 \phi=0$ allows us to eliminate $\p_+$ in favor of only working with $\p_-$ and $\p_\perp$. Making this choice, we now introduce the concept of \emph{monomial} operators, which will be our building blocks for more general operators. We also introduce some useful shorthand notation. 

A monomial operator is simply a string of $\phi$'s with accompanying derivatives. Let us start with single-particle monomials, \emph{i.e.}, those with one insertion of $\phi$. We let $k=(k_-,k_\perp)$ denote a two-component variable having a minus and transverse component and introduce the notation $\p^k\phi \equiv \p_-^{k_-} \p_\perp^{k_\perp} \phi$. More generally, for $n$-particle monomials, we use a vector $\bk = (k_1,\dots, k_n)$ and define $\p^{\bk}\phi \equiv \p^{k_1}\phi\cdots \p^{k_n} \phi$, where each $k_i = (k_{i-}, k_{i\perp})$. To summarize our notation, 
\be
\boxed{
\begin{aligned}
& \text{Monomial Operators:} \\[10pt]
& \text{1-particle:} \hspace{10mm} k = (k_{-}, k_{\perp}), \hspace{16mm} \p^k\phi \equiv \p_-^{k_-} \p_\perp^{k_\perp} \phi,  \\
& \text{$n$-particle:} \hspace{10mm} \bk = (k_1,\dots, k_n), \hspace{10mm} \p^{\bk}\phi \equiv \p^{k_1}\phi\cdots \p^{k_n} \phi, \hspace{10mm} k_i = (k_{i-}, k_{i\perp}). \\[10pt]
& \hspace{35mm} \text{The Dirichlet condition is } k_{i-} \geq 1 \, \forall \, i. 
\end{aligned}
}
\ee

Monomial operators will be our building blocks, because general operators can be written as linear combinations of monomials,
\be
\CO(x) = \sum_{\bk} C_{\bk}^{\CO} \, \p^{\bk}\phi(x). \label{eq:genopexpansion}
\ee
In particular, our goal is to construct the Dirichlet basis in terms of monomials. Recall that the definition of a Dirichlet operator is that every $\phi$ has a $\p_-$ attached to it. Consequently, we can restrict our attention to monomials that satisfy this condition, which is that $k_{i-} \geq 1$ for all $i$.

It will also be useful to introduce shorthand notation for the following summations, 
\be
|\bk_-| = \sum_{i=1}^n k_{i-}, \hspace{10mm} |\bk_\perp| = \sum_{i=1}^n k_{i\perp}, \hspace{10mm} |\bk| = |\bk_-| + |\bk_\perp|.
\ee
We will refer to $|\bk|$, which is simply the total number of derivatives appearing in $\p^{\bk}\phi$, as the \emph{degree} of the monomial. Note that the scaling dimension of $\p^{\bk}\phi$ is equal to $|\bk| + \frac{n}{2}$, since $\phi$ has scaling dimension $\frac{1}{2}$.

\subsection{`Minimal' Monomials}
\label{subsec:Basis-Minimal}

For a given particle number $n$, the naive strategy for computing the Dirichlet basis up to a maximum scaling dimension $\Dmax$ is to simply list all Dirichlet monomials up to $\Dmax$ and then Gram-Schmidt them according to the LCT inner product in (\ref{eq:InnerAndME}). However, this strategy is inefficient, because the number of monomials is much larger than the number of final basis states. This is a consequence of the fact that descendant operators, \emph{i.e.}, those that are total derivatives of lower-dimension operators, do not yield independent LCT basis states.\footnote{From (\ref{eq:LCTStateAppendix}), it is clear that $\CO$ and $\p_\mu \CO$ yield states that simply differ by a factor of $p_\mu$.} Thus, we can do better by systematically eliminating descendants from the full list of Dirichlet monomials before applying Gram-Schmidt. 

We will now describe an algorithm for eliminating descendants from a list of monomials. Given the full list of Dirichlet monomials for $n$ particles up to some maximum degree, the output of the algorithm is a \emph{minimal} list of monomials that spans the final Dirichlet basis without being overcomplete. In other words, the number of monomials in the minimal list is equal to the number of basis states, and to obtain the final basis, one simply needs to Gram-Schmidt the minimal monomials. 

The algorithm works by identifying linear combinations of monomials that can be rewritten as a descendant. To be more precise, given a set of $n$-particle monomials $\p^{\bk}\phi$ with degree $\mathsf{d}$, we say that they are \emph{linearly dependent} if there exist coefficients $c_{\bk}$ (not all zero) such that 
\be
\hspace{15mm} \sum_{\bk: \, |\bk| = \mathsf{d}} c_{\bk} \, \p^{\bk} \phi ~ = ~ \p_\mu \Big(  \p^{\bkp} \phi \Big), \hspace{15mm} |\bkp| = \mathsf{d}-1
\label{eq:MonomialConstraint}
\ee
where $\p_\mu \in \{\p_-,\, \p_\perp,\,  \p_+ = \tfrac{\p_\perp^2}{2\p_-}\}$. In words, this relation says that the linear combination of degree-$\mathsf{d}$ monomials appearing on the left is really just a descendant of the degree-$(\mathsf{d}-1)$ monomial appearing on the right. Crucially, for every independent relation of the form (\ref{eq:MonomialConstraint}), we can eliminate a monomial on the left-hand-side from our list, because its corresponding LCT state will not be independent. 

Thus, our goal is to generate all possible relations of the form (\ref{eq:MonomialConstraint}). A straightforward way to do this is to proceed degree-by-degree and act with $\p_\mu$ on all of the monomials at degree $\mathsf{d}-1$ in order to generate all of the possible relations at degree $\mathsf{d}$. Then, once we have the full set of relations in hand, we can successively eliminate linearly dependent monomials in favor of monomials with equal or lesser degree. 

Let us illustrate all of these ideas with a simple example: constructing the minimal set of 2-particle Dirichlet monomials up to degree 3. The minimum degree of a 2-particle Dirichlet monomial is 2, so initially our full list of monomials is 
\be
\begin{aligned}
& \text{degree 2}:  \hspace{10mm} \bk_1 = ((1,0),(1,0)) ~~\leftrightarrow~~ \p^{\bk_1}\phi = (\p_-\phi)^2, \\
& \text{degree 3}:  \hspace{10mm} \bk_2 = ((2,0),(1,0)) ~~\leftrightarrow~~ \p^{\bk_2}\phi = (\p_-^2 \phi)(\p_- \phi), \\
& \hspace{27mm} \bk_3 = ((1,1),(1,0)) ~~\leftrightarrow~~ \p^{\bk_3}\phi = (\p_-\p_\perp \phi)(\p_- \phi). 
\end{aligned}
\ee
Acting with $\p_\mu$ on $\p^{\bk_1}\phi$ we obtain two relations of the form of (\ref{eq:MonomialConstraint}),\footnote{In this particular example, acting with $\p_+$ takes us outside the space of Dirichlet monomials (since $\p_+ (\p^{\bk_1}\phi) = 2\p_+\p_-\phi\p_-\phi = \p_\perp^2\phi\p_-\phi$) and hence is not a relation we can use. More generally, though, $\p_+$ will also yield constraints.}
\be
\p^{\bk_2}\phi = \p_- \left( \tfrac{1}{2} \p^{\bk_1}\phi \right), \hspace{10mm} \p^{\bk_3}\phi = \p_\perp \left( \tfrac{1}{2} \p^{\bk_1}\phi \right).
\ee
Consequently, both $\p^{\bk_2}\phi$ and $\p^{\bk_3}\phi$ can be eliminated in favor of keeping $\p^{\bk_1}\phi$. Thus, our minimal monomial list for 2 particles up to degree 3 is just $\p^{\bk_1}\phi$. 

In the next section, we turn to the details of orthonormalizing the basis of minimal monomials via Gram-Schmidt.

\subsection{Gram-Schmidt}
\label{subsec:GramSchmidt}

Once we have chosen a minimal set of Dirichlet monomials for a given particle number up to some maximum degree, the next step is to construct the Gram matrix for these monomials and then apply Gram-Schmidt in order to construct the Dirichlet basis. Following (\ref{eq:LCTStateAppendix}), a monomial operator $\p^{\bk}\phi(x)$ has the corresponding momentum space state (before $\mu$-discretization) given by 
\be
\ket{\p^{\bk}\phi, \mu} \equiv \int d^3x \, e^{-ip\cdot x} \, \p^{\bk}\phi(x)\, \ket{0}.
\label{eq:LCTStateMon}
\ee
The inner product and $\phi^4$-theory Hamiltonian matrix elements of these specific states have been computed previously in~\cite{Anand:2019lkt}.\footnote{In that reference, $|\p^{\bk}\phi, \mu\>$ is written as $|\p^{\bk}\phi(P) \>$.} 

Referring to~\cite{Anand:2019lkt}, we note that the mass-dimension of the inner product is given by 
\be
\langle \p^{\bk}\phi, \mu \,|\, \p^{\bkp}\phi, \mu^\prime \rangle =  (2\pi)^3\delta^{(3)}(p-p^\prime) \, p_-^{|\bk_-|+|\bkp_-|} \mu^{|\bk_\perp|+|\bkp_\perp|+n-3}\cdot G_{\bk\bk'}.
\label{eq:InnerSchematic}
\ee 
Since the factors of $p_-$ and $\mu$ appearing on the right will eventually be canceled anyway when we properly normalize the basis states, in practice it is convenient to work directly with rescaled monomial states, defined as 
\be
 \ket{\p^{\bk}\phi,\mu}_R \equiv \frac{1}{ p_-^{|\bk_-|} \mu^{\frac{n-3}{2}+|\bk_\perp|} } \, \ket{\p^{\bk}\phi, \mu},
 \label{eq:rescale}
\ee
where the subscript `R' on the LHS stands for `rescaled'. 
Then, up to an overall factor of $(2\pi)^3\delta^{(3)}(p-p^\prime)$, the Gram matrix of these rescaled monomial states just consists of momentum-independent numbers. Referring back to (\ref{eq:InnerAndMEBox}),  
\be
G_{\bk\bkp} = \frac{{}_R\langle \p^{\bk}\phi, \mu \,|\, \p^{\bkp}\phi, \mu^\prime \rangle_R}{(2\pi)^3\delta^{(3)}(p-p^\prime) }.
\label{eq:RescaledInnerSchematic}
\ee

For a given $\Dmax$ (or equivalently maximum degree), we can then compute the Gram matrix for the set of minimal monomials using the expressions for $G_{\bk\bk'}$ from~\cite{Anand:2019lkt}, and use Gram-Schmidt to construct a complete, orthogonal basis of Dirichlet states,
\be
|\Ocal,\mu\> = \sum_{\bk: \, |\bk| \leq \Dmax} C_{\bk}^{\CO} |\p^{\bk}\phi,\mu\>_R, \qquad \<\Ocal,\mu|\Ocal',\mu'\> = (2\pi)^3 \delta^{(3)}(p-p') \cdot \de_{\Ocal\Ocal'}.
\ee

\subsection{$\mu$-Discretization}
\label{subsec:muDiscretization}

The momentum space states defined in (\ref{eq:LCTStateMon}), and the rescaled versions in (\ref{eq:rescale}), are not fully discretized yet, since $\mu$ is still a continuous parameter. Recall from section~\ref{subsec:Overview-Basis} that our prescription for discretizing $\mu$ is as follows. We first introduce a UV cutoff $\LUV$ such that $\mu^2 \leq \LUV^2$ and then introduce a set of smearing functions $b_{\mathfrak{i}}(\mu)$, where  $\mathfrak{i}=1,\dots,\mathfrak{i}_{\text{max}}$, to define the discrete set of states
\be
\ket{\p^{\bk}\phi, \mathfrak{i}}_R \equiv \frac{1}{ \sqrt{2\pi}} \int_0^{\Lambda^2} d\mu^2 \, b_{\mathfrak{i}}(\mu) \,  \ket{\p^{\bk}\phi,\mu}_R, \hspace{10mm} \mathfrak{i}=1,\dots,\mathfrak{i}_{\text{max}}. 
\ee
There is freedom in the precise choice of smearing functions, but they are normalized such that 
\be
\int_0^{\LUV^2} d\mu^2 \, b_{\mathfrak{i}}(\mu) \, b_{\mathfrak{j}}(\mu) = \delta_{\mathfrak{i}\mathfrak{j}}.
\ee
With this convention, it follows from (\ref{eq:InnerAndMEBox}) that
\be
{}_R\langle \p^{\bk}\phi, \mathfrak{i} \, | \, \p^{\bkp}\phi, \mathfrak{j} \rangle_R =  2p_- (2\pi)^2 \delta^{(2)}(\vec{p}-\vec{p}^{\,\prime}) \, \delta_{\mathfrak{i}\mathfrak{j}} \cdot G_{\bk\bkp}.
\label{eq:DiscretizedInnerSchematic}
\ee 
The resulting orthogonalized Dirichlet states are then normalized such that
\be
|\Ocal,\ifrak\> = \sum_{\bk: \, |\bk| \leq \Dmax} C_{\bk}^{\CO} |\p^{\bk}\phi,\ifrak\>_R, \qquad \<\Ocal,\ifrak|\Ocal',\mathfrak{j}\> = 2p_-(2\pi)^2 \delta^{(2)}(\vec{p}-\vec{p}^{\,\prime}) \de_{\mathfrak{i}\mathfrak{j}} \cdot \de_{\Ocal\Ocal'}.
\ee

We now specify our choice for the smearing functions $b_{\mathfrak{i}}(\mu)$. First, we partition the interval $\mu^2\in [0,\LUV^2]$ into $\mathfrak{i}_{\text{max}}$ bins $[\mu_{\mathfrak{i}-1}^2, \mu_{\mathfrak{i}}^2]$ for some choice of bin endpoints $\{\mu_{\mathfrak{i}}^2\}$. Then we define
\be
b_{\mathfrak{i}}(\mu) \equiv \frac{1}{\sqrt{\mu_{\mathfrak{i}}^2 - \mu_{\mathfrak{i}-1}^2}} \Big(   \Theta(\mu^2 - \mu_{\mathfrak{i}-1}^2) - \Theta(\mu^2 - \mu_{\mathfrak{i}}^2) \Big), \label{eq:smearingfuncdef}
\ee
for $\mathfrak{i}=1,\dots,\mathfrak{i}_{\text{max}}$. Thus, $b_{\mathfrak{i}}(\mu)$ is a normalized ``bar function" with constant support on the bin $[\mu_{\mathfrak{i}-1}^2, \mu_{\mathfrak{i}}^2]$.

There is again a large amount of freedom in the spacing of the endpoints $\{\mu_{\mathfrak{i}}^2\}$. In this work, we choose to keep a fixed ratio $r$ between the widths of successive bins,
\be
r \equiv \fr{\mu_\ifrak^2 - \mu_{\ifrak-1}^2}{\mu_{\ifrak+1}^2 - \mu_\ifrak^2},
\ee
such that the endpoints are given by
\be
\mu_{\mathfrak{i}}^2 = \LUV^2 \frac{r^{\mathfrak{i}_{\mathrm{max}}}(1-r^\ifrak)}{r^\ifrak(1-r^{\mathfrak{i}_{\mathrm{max}}})}.
\ee
Because we are specifically interested in IR physics, we keep $r < 1$, such that there is a higher density of bins at small $\mu^2$ than at large $\mu^2$.

\section{Details: Computing Matrix Elements}
\label{sec:MatrixElements} Now that we have outlined the method for constructing the basis in Appendix \ref{sec:Basis}, we can turn to the evaluation of their Hamiltonian matrix elements.

Recall from~\eqref{eq:InnerAndME}, reproduced here for convenience, that our final discretized matrix elements are given by \begin{equation}
	\begin{aligned}
		\langle \CO,\mathfrak{i} | 2p_-P_+ |\CO^\prime, \mathfrak{j} \rangle &= 2p_- (2\pi)^2 \delta^{(2)}(\vec{p}-\vec{p}^{\,\prime}) \cdot \frac{1}{2\pi} \int_0^{\LUV^2} d\mu^2 \, d\mu^{\prime 2} \, b_{\mathfrak{i}}(\mu) \, b_{\mathfrak{j}}(\mu^\prime ) \, \CM_{\CO\CO^\prime}(\mu,\mu^\prime), \\
		\CM_{\CO\CO'}(\mu, \mu') &\equiv \frac{\bra{\CO, \mu} P_+ \ket{\CO', \mu'}}{(2\pi)^2\delta^{(2)}(\vec{p}-\vec{p}^{\,\prime})},\quad\quad\quad \ket{\cO, \mu} = \int d^3 x e^{-i p \cdot x} \cO(x)\ket{\textrm{vac}}.
	 \label{eq:discretizedme1}
	\end{aligned}
\end{equation} To avoid potential ambiguity, we refer to the objects on the left-hand side $\langle \CO,\mathfrak{i} | 2p_-P_+ |\CO^\prime, \mathfrak{j} \rangle$ as `discretized matrix elements' and the functions $\CM_{\CO\CO^\prime}(\mu,\mu^\prime)$ as just `matrix elements'. Determining discretized matrix elements thus amounts to two steps: \begin{enumerate}
	\item Computing the  matrix elements, given by sandwiching the Hamiltonian with the basis states defined in~\eqref{eq:discretizedme1}. The exact expressions for these objects, which boil down to momentum space CFT three-point functions, were first presented in \cite{Anand:2019lkt}. We will schematically review those results in the following section. 
	\item Second, discretizing $\mu$ and $\mu'$. This amounts to performing the integrals over $\mu$, $\mu'$ weighted by the smearing functions $b_{\mathfrak{i}}(\mu)$, $b_{\mathfrak{j}}(\mu)$. Since our smearing functions are chosen to be step functions with support over an interval, as in~\eqref{eq:smearingfuncdef}, discretizing the matrix elements amounts to integrating over a window in $\mu$ and $\mu'$. As we will soon describe, this procedure is technically somewhat nontrivial. The reason is that there are naive divergences within the matrix elements $\CM_{\CO\CO'}(\mu, \mu')$ near the edges of the integration regions. These divergences completely cancel amongst each other at the end of the day to give rise to finite answers. However, performing the discretization integrals in a manifestly finite and numerically stable way requires an efficient way to track the singular terms.
\end{enumerate}

The outline of this section is thus as follows: in section \ref{sec:momspacereview} we will do a lightning review of the calculations presented in \cite{Anand:2019lkt}, and describe how these formulas are implemented in practice in our truncation setup. We also present a few examples of matrix elements to guide the reader. Then, in section \ref{sec:discretizingme} we explain the discretization procedure and address technical subtleties in obtaining finite expressions for the final discretized matrix elements. 

\subsection{Implementation of Momentum Space Formulas \label{sec:momspacereview}} In LCT, matrix elements are obtained from Fourier transforming CFT three-point functions where the middle operator is the relevant deformation and the external operators are LCT basis states\footnote{For a more pedagogical introduction, see \cite{Anand:2020gnn}.}.

 Recall that from~\eqref{eq:P+Appendix} that the Hamiltonian can be broken up into three distinct pieces: $P_+^{(\textrm{CFT})}$, $\delta P_+^{(m)}$, $\delta P_+^{(g)}$, corresponding to the kinetic term, mass term, and quartic interaction, respectively. The first of these, $P_+^{(\textrm{CFT})}$, has a trivial structure due to the CFT equations of motion. In particular, because our basis states are automatically eigenstates of $P_+^{(\textrm{CFT})}$, with the eigenvalue set by $\mu$,
 \begin{equation}
 	P_+^{(\textrm{CFT})} \ket{\cO, \mu} = \frac{\mu^2}{2p_-} \ket{\cO, \mu},
 \end{equation} the corresponding matrix elements are trivial,
 \begin{equation}
 	\Mcal^{(\textrm{CFT})}_{\Ocal\Ocal'}(\mu,\mu') = (2\pi)\delta(\mu^2-{\mu'}^2) \frac{\mu^2}{2p_-} \cdot \de_{\Ocal\Ocal'}.
 \end{equation} We will therefore focus on the more nontrivial matrix elements of $\delta P_+^{(m)}$, $\delta P_+^{(g)}$. In order to match notation with \cite{Anand:2019lkt}, we shall refer to these pieces of the Hamiltonian as $m^2 V_{\phi^2}$ and $g V_{\phi^4}$. It will also be useful to further divide the interacting part of the Hamiltonian, $V_{\phi^4}$ into two pieces: a term that preserves particle number, $V_{\phi^4}^{(n\to n)}$, and a piece that changes particle number by two, $V_{\phi^4}^{(n \to n+2)}$. Naively, one might also expect an $n$-to-$(n+4)$ contribution, but this is absent due to lightcone momentum conservation; particles all carrying positive $p_-$ cannot be spontaneously created from the lightcone vacuum.

We can therefore define the associated matrix elements \begin{equation}
	\begin{aligned}
		\CM_{\cO \cO'}^{(m)}(\mu,\mu') &\equiv m^2 \frac{\bra{\CO, \mu} V_{\phi^2} \ket{\CO', \mu'}}{(2\pi)^2\delta^{(2)}(\vec{p}-\vec{p}^{\,\prime})}, \\
		\CM_{\cO \cO'}^{(n\to n)}(\mu,\mu') &\equiv g \frac{\bra{\CO, \mu} V_{\phi^4}^{(n\to n)} \ket{\CO', \mu'}}{(2\pi)^2\delta^{(2)}(\vec{p}-\vec{p}^{\,\prime})}, \\
		\CM_{\cO \cO'}^{(n\to n+2)}(\mu,\mu') &\equiv g \frac{\bra{\CO, \mu} V_{\phi^4}^{(n\to n+2)} \ket{\CO', \mu'}}{(2\pi)^2\delta^{(2)}(\vec{p}-\vec{p}^{\,\prime})}. \label{eq:matrixelemOOpdef}
	\end{aligned}
\end{equation} Recall that the operators $\CO$, $\CO'$ in our basis correspond to operators built from the free scalar $\phi$. As stated in appendix \ref{subsec:Basis-Monomials}, their building blocks are monomials\footnote{The fact that we are working  with these rescaled states is the only modification to directly using the formulas in \cite{Anand:2019lkt}.} \begin{equation}
	\ket{\cO, \mu} = \sum_{\bk}C_{\bk}^{\cO} \, \ket{ \p^{\bk} \phi, \mu}_R, \quad\quad\quad \ket{\p^{\bk} \phi, \mu}_R \equiv \frac{1}{p_-^{|\bk_-|}\mu^{\frac{n-3}{2}+|\bk_\bot|}} \ket{\p^{\bk}\phi, \mu}. \label{eq:rescalereminder}
\end{equation} We can therefore focus on the matrix elements of these monomials, as the full matrix elements can be obtained by summing over the individual monomial matrix elements times the coefficients $C^\Ocal_\bk$ appearing in the above expansion. The building block matrix elements are then \begin{equation}
	\boxed{\begin{aligned}
		\CM_{\bk \bk'}^{(m)}(\mu,\mu') &\equiv m^2 \frac{{_R}\bra{\p^{\bk} \phi, \mu} V_{\phi^2} \ket{\p^{\bk'} \phi, \mu'}_R}{(2\pi)^2\delta^{(2)}(\vec{p}-\vec{p}^{\,\prime})}, \\
		\CM_{\bk \bk'}^{(n\to n)}(\mu,\mu') &\equiv g \frac{{_R}\bra{\p^{\bk} \phi, \mu} V_{\phi^4}^{(n \to n)} \ket{\p^{\bk'} \phi, \mu'}_R}{(2\pi)^2\delta^{(2)}(\vec{p}-\vec{p}^{\,\prime})}, \\
		\CM_{\bk \bk'}^{(n\to n+2)}(\mu,\mu') &\equiv g\frac{{_R}\bra{\p^{\bk} \phi, \mu} V_{\phi^4}^{(n \to n+2)} \ket{\p^{\bk'} \phi, \mu'}_R}{(2\pi)^2\delta^{(2)}(\vec{p}-\vec{p}^{\,\prime})} . \label{eq:genericME}
	\end{aligned}}
\end{equation}

The expressions for these objects can be found in \cite{Anand:2019lkt}. The main idea is that computing these matrix elements amounts to computing momentum space CFT three-point functions, where the middle operator carries zero momentum. That is to say, they are Fourier transforms of CFT three-point functions. Although conceptually straightforward, performing the Fourier transforms is a technically involved computation for two reasons. First, the external operators can have spin, so that the right-hand side of~\eqref{eq:genericME} is typically a linear combination of multiple tensor structures. Second, the correlators appearing on the right-hand side are Wightman functions, with a fixed time ordering. Computing the Fourier transform thus requires careful accounting of the $i\epsilon$ prescription to obtain well defined in- and out-states. The authors of \cite{Anand:2019lkt} presented expressions for the matrix elements~\eqref{eq:genericME} in $d=3$, for both the general case (for any relevant deformation to the CFT Hamiltonian), as well as the specific implementation of $\phi^2$, $\phi^4$ interactions. In the latter setting, the OPE coefficients that reside inside the three-point functions can be explicitly calculated using Wick contractions. We refer readers to \cite{Anand:2019lkt} for specific formulas for the matrix elements, and in the following paragraphs only comment on their general structure.

From the results of \cite{Anand:2019lkt}, the mass term matrix elements take a very simple form in terms of the inner product between states created by $\cO$ and $\cO'$ (similar to the free theory matrix elements): \begin{equation}
	\CM_{\bk \bk'}^{(m)}(\mu,\mu') = (2\pi)\delta(\mu^2-{\mu'}^2) \cdot \textrm{number}.
\end{equation} Because of their resemblance to the inner product, the implementation of these matrix elements is trivial once we compute the Gram matrix. The interaction monomial matrix elements on the other hand take the very rough form (focusing on $n\to n$ for simplicity, but a similar equation holds for $n \to n+2$) \begin{equation}
	\CM_{\bk \bk'}^{(n\to n)}(\mu,\mu') \sim  \sum \dotsb \sum \textrm{coefficient} \times \, \mu^a {\mu'}^b \, {_2 F_1}\Big(\cdots;\tfrac{\mu^2}{{\mu'}^2}\Big). \label{eq:ntonschematic}
\end{equation} We have omitted a great deal of structure (see \cite{Anand:2019lkt} for the precise equations), but the key point is that these monomial matrix elements are effectively given by multiple sums over powers of $\mu$, $\mu'$ times $_2 F_1$ hypergeometric functions whose first three arguments depend on the indices summed over, and whose last argument is the ratio \begin{equation}
	\boxed{\alpha \equiv \frac{\mu^2}{{\mu'}^2}.}
\end{equation} Despite the multiple powers and arguments appearing in~\eqref{eq:ntonschematic}, there are only three free parameters to keep track of (the arguments of the hypergeometric function). Once those are known, the scaling in $\mu$ and $\mu'$ can be fixed. It is worth noting that the formula in~\eqref{eq:ntonschematic} only holds when $\mu < \mu'$. This means that the interaction matrix elements are asymmetric in $\bk$, $\bk'$. Furthermore, for $n\to n$, these hypergeometric functions can be simplified to a much smaller basis consisting of a few special functions, such as elliptic integrals and inverse hyperbolic functions. This fact will be useful when we consider the discretization procedure. On the other hand, for the $n\to n+2$ matrix elements, it turns out that the hypergeometrics simplify even more dramatically, such that the monomial matrix elements are simply sums of powers in $\mu$, $\mu'$. It would be interesting to understand these simplifications in more detail, as it is not manifest in the formulas of \cite{Anand:2019lkt}; here, we will use it as an empirical fact. The general strategy to computing the interacting matrix elements is therefore straightforward: compute the monomial matrix elements using the formulas in \cite{Anand:2019lkt}, simplify the hypergeometrics as much as possible, and then sum over monomials to obtain the final basis matrix element.

Before we move on to the discretization procedure, let us look at a few examples of the monomial matrix elements.

\textbf{\underline{Example 1}:} Consider the following three-particle Dirichlet monomial operators \begin{equation}
	\begin{aligned}
		\bk &= ((1,1),(1,1),(1,0))\quad \rightarrow \quad \p^{\bk}\phi = (\p_- \p_\bot \phi)^2 \p_- \phi, \\
		\bk' &= ((2,0),(1,1),(1,1))\quad \rightarrow \quad \p^{\bk'} \phi = (\p_- \p_\bot \phi)^2 \p_-^2 \phi.
	\end{aligned}
\end{equation} Using the formulas given in \cite{Anand:2019lkt}, the $n \to n$ matrix element between these monomials is given by (supressing the coupling $g$ for brevity) \begin{equation}
	\begin{aligned}
		\CM_{\bk \bk'}^{(n\to n)}(\mu,\mu') &=  -\frac{273 \alpha^2 \,
   _2F_1\left(\frac{3}{2},\frac{5}{2};4;\alpha\right)}{8796093022208}+\frac{63 \alpha^2 \,
   _2F_1\left(\frac{3}{2},\frac{7}{2};4;\alpha\right)}{549755813888}+\frac{4329 \alpha \,
   _2F_1\left(\frac{1}{2},\frac{3}{2};3;\alpha\right)}{8796093022208} \\
   &+\frac{24633 \alpha
   \, _2F_1\left(\frac{1}{2},\frac{5}{2};3;\alpha\right)}{8796093022208}-\frac{189 \alpha
   \, _2F_1\left(\frac{1}{2},\frac{7}{2};3;\alpha\right)}{274877906944}+\frac{1755 \,
   _2F_1\left(-\frac{1}{2},\frac{1}{2};2;\alpha\right)}{2199023255552} \\
   &+\frac{5187 \,
   _2F_1\left(-\frac{1}{2},\frac{3}{2};2;\alpha\right)}{2199023255552}+\frac{189 \,
   _2F_1\left(-\frac{1}{2},\frac{5}{2};2;\alpha\right)}{137438953472}.
	\end{aligned}
    \label{eq:NtoNUnsimplified}
\end{equation} We can simplify the hypergeometric functions using the identities \begin{equation}
	\begin{aligned}
		 K(z) = \frac{\pi}{2} {_2F_1}\left( \half, \half; 1; z \right), \quad\quad\quad  E(z) = \frac{\pi}{2} {_2F_1}\left( -\half, \half; 1; z \right), \\
	\end{aligned}
\end{equation} where $K(z)$ and $E(z)$ are the complete elliptic integrals of the first and second kind, respectively. The other hypergeometric functions with different arguments can be related to these by taking derivatives. We can therefore reduce this $n \to n$ matrix element to the simpler form
\begin{equation}
		\CM_{\bk \bk'}^{(n\to n)}(\mu,\mu') = \frac{1}{\mu'} \frac{(3373 \alpha+1268) K(\alpha)+4 (244 \alpha-317) E(\alpha)}{549755813888 \pi  \alpha} \label{ntonexample}.
\ee
This formula (as well as the original unsimplified expression~\eqref{eq:NtoNUnsimplified}) specifically holds for $\mu < \mu'$ (\emph{i.e.}, $\alpha < 1$). We can obtain the other regime by simply swapping $\bk$ and $\bk'$ in the formulas from~\cite{Anand:2019lkt}, which after simplifying results in
\begin{equation}
		\CM_{\bk' \bk}^{(n\to n)}(\mu,\mu') = \frac{1}{\mu'}  \frac{(5617 \alpha-976) K(\alpha)+(976-1268 \alpha) E(\alpha)}{549755813888 \pi  \alpha}.
\end{equation}
Note that in both of these expressions, the monomial matrix element which started its life as a multiple sum over several ${}_2F_1$ hypergeometric functions has reduced to just a few special functions. Second, these are merely monomial matrix elements; to obtain a full matrix element between basis states, we would sum over all monomials that comprise the Dirichlet states.

\textbf{\underline{Example 2}:} We can also look at $n \to n$ matrix elements in the even particle number sector. Consider the four-particle monomials  \begin{equation}
	\begin{aligned}
		\bk &= ((1,1),(1,1),(1,0),(1,0)) \quad \rightarrow \quad \p^{\bk} \phi = (\p_- \p_\bot \phi)^2 (\p_- \phi)^2, \\
		\bk' &= ((2,0),(1,1),(1,1),(1,0))\quad \rightarrow \quad \p^{\bk'} \phi = (\p_-^2 \phi) (\p_- \p_\bot \phi)^2 (\p_- \phi).
	\end{aligned}
\end{equation} The matrix element is (after simplification) \begin{equation}
	\begin{aligned}
		\CM_{\bk \bk'}^{(n\to n)}(\mu,\mu') &= \frac{1}{\mu'} \frac{\left(1725 \alpha^2+36880 \alpha+6171\right) \tanh ^{-1}\left(\sqrt{\alpha}\right)-3
   \sqrt{\alpha} (575 \alpha+2057)}{4678596585062400 \pi ^2 \alpha^{5/4}}, \\
   		\CM_{\bk' \bk}^{(n\to n)}(\mu, \mu') &= \frac{1}{\mu'}\frac{\left(6171 \alpha^2+36880 \alpha+1725\right) \tanh ^{-1}\left(\sqrt{\alpha}\right)-3
   \sqrt{\alpha} (2057 \alpha+575)}{4678596585062400 \pi ^2 \alpha^{5/4}}.
	\end{aligned}
\end{equation} For even sector $n \to n$ matrix elements, we see the appearance of $\tanh ^{-1}$ functions, rather than the elliptic functions for odd $n$.

\textbf{\underline{Example 3}:} Consider the four- and six-particle monomials given by  \begin{equation}
	\begin{aligned}
		\bk &= ((1,1),(1,1),(1,0),(1,0)), \,\quad\quad\quad\quad\quad\quad \rightarrow \quad \p^{\bk} \phi = (\p_- \p_\bot \phi)^2 (\p_- \phi)^2, \\
		\bk' &= ((2,0),(1,1),(1,1),(1,0),(1,0),(1,0))\quad \rightarrow \quad \p^{\bk'} \phi = (\p_-^2 \phi)(\p_- \p_\bot \phi)^2 (\p_- \phi)^3.
	\end{aligned}
\end{equation} The corresponding $n\to(n+2)$ matrix element, after simplifying the hypergeometric functions, is \begin{equation}
\CM_{\bk \bk'}^{(n\to n+2)}(\mu,\mu') = \frac{1}{\mu'} \frac{\alpha^{1/4} (489103 \alpha+1150200)}{697176967140605952000 \pi ^3 }.
\end{equation}
Note that this matrix element (and all other $n \to n+2$ matrix elements) is only nonzero for $\mu < \mu'$.
As we can see, this expression reduces to a sum of powers in $\mu$, $\mu'$, a pattern that continues to hold for all of the $n \to n+2$ sectors. We also note that the matrix elements can always be written as $\frac{1}{\mu'}$ times a function in $\alpha$. This can be seen by dimensional analysis with the rescaling of monomial states we have chosen in~\eqref{eq:rescalereminder}.

It is worth emphasizing that so far, we have not used any numerics. That is, for this work we compute all monomial matrix elements \textit{exactly}; decimal approximations (where in practice we keep 60 digits of precision) are made only when we take linear combinations of these expressions to obtain basis state matrix elements.

\subsection{Discretizing $\mu$, $\mu'$\label{sec:discretizingme}} With the matrix elements at hand, we now turn to the discretization procedure. We wish to compute the integrals over $\mu$ and $\mu'$ in~\eqref{eq:discretizedme1}. We will again focus on monomials, rather than the full discretized matrix elements between basis states. We define these building block  discretized \textit{monomial} matrix elements as \begin{equation}
	\begin{aligned}
		{}_R\langle \bk ,\mathfrak{i} | 2p_-P_+ |\bk^\prime, \mathfrak{j} \rangle_R &= 2p_- (2\pi)^2 \delta^{(2)}(\vec{p}-\vec{p}^{\,\prime}) \cdot \frac{1}{2\pi} \int_0^{\LUV^2} d\mu^2 \, d\mu^{\prime 2} \, b_{\mathfrak{i}}(\mu) \, b_{\mathfrak{j}}(\mu^\prime ) \, \CM_{\bk \bk'}(\mu, \mu'). \label{eq:discretizedmonome}
	\end{aligned}
\end{equation}  As seen in the examples in the previous section, our matrix elements always take the general form \begin{equation}
	\CM_{\bk \bk'}^{(n\to n) \textrm{ or } (n \to n+2)}(\mu,\mu') = g \cdot \frac{1}{\mu'} f_{\bk \bk'} (\alpha),
\end{equation} for some function $f_{\bk \bk'}(\alpha)$, which can be read off from the hypergeometric functions appearing in the matrix elements. Plugging this into~\eqref{eq:discretizedmonome} and our expressions for the smearing functions given in~\eqref{eq:smearingfuncdef}, we therefore compute \begin{equation}
	\boxed{\begin{aligned}
		\frac{{}_R\langle \bk ,\mathfrak{i} | 2p_-P_+ |\bk^\prime, \mathfrak{j} \rangle_R}{2p_- (2\pi)^2 \delta^{(2)}(\vec{p}-\vec{p}^{\,\prime})} &=  \begin{cases}   N_{\fraki \frakj} \cdot \cI_{\, \fraki \, < \, \frakj} \left[f_{\bk \bk'}(\alpha) \right] \quad\quad\,\,\, \fraki < \frakj \\
		N_{\fraki \frakj} \cdot \cIeq \left[ f_{\bk\bk'}(\alpha) \right] \quad\quad\,\, \,\fraki = \frakj \end{cases} , \label{eq:discretizedmonome2}
	\end{aligned}}
\end{equation} where for later convenience we defined an overall normalization factor \begin{equation}
	N_{\fraki \frakj} \equiv \frac{g}{2\pi} \frac{1}{\sqrt{(\mu^2_{\frakj} - \mu^2_{\frakj-1})(\mu^2_{\fraki}-\mu^2_{\fraki-1})}}
\end{equation} and the discretization integrals \begin{equation}\boxed{
	\begin{aligned}
		\cI_{\, \fraki \, < \, \frakj} \left[ f(\alpha) \right] &\equiv \int_{\mu_{\frakj-1}^2}^{\mu_{\frakj}^2} \, d\mu^{\prime 2} \, \mu' \int_{\mu_{\fraki-1}^2/\mu^{\prime 2}}^{\mu_{\fraki}^2/\mu^{\prime 2}} d\alpha  \,f(\alpha), \\
		\cIeq \left[ f(\alpha) \right] &\equiv \int_{\mu_{\frakj-1}^2}^{\mu_{\frakj}^2} \, d\mu^{\prime 2} \, \mu'  \int_{\mu_{\fraki-1}^2/\mu^{\prime 2}}^{1} d\alpha  \, f(\alpha). \label{eq:discreteints}
	\end{aligned}}
\end{equation} In the above integrals, we have changed variables from $\mu$ to $\alpha \equiv \mu^2/{\mu'}^2$. The reason~\eqref{eq:discretizedmonome2} breaks up into two cases is due to the fact that our expressions for the monomial matrix elements $f_{\bk\bk'}(\alpha)$ only hold for $\mu < \mu'$. Obtaining the expressions for $\mu > \mu'$ is equivalent to just swapping the monomials $\bk \leftrightarrow \bk'$. When $\fraki < \frakj$ we are integrating over two distinct windows in $\mu$ and $\mu'$, such that $\mu < \mu'$ over the full integration range. However, when $\fraki = \frakj$, we only integrate $\mu$ up to $\mu'$.\footnote{Although our matrix elements are formally defined only for $\mu < \mu'$, the integral is evaluated all the way to $\mu = \mu'$, as this edge of the integration range is a set of measure zero.} It is therefore useful to split the matrix element up into these two cases. We do not need to consider the case $\fraki > \frakj$, as this can be obtained by exchanging the bra and ket states to obtain an integral of the form $\fraki < \frakj$.

Let us now consider the $n \to n+2$ matrix elements first, as they are technically simpler. In this case, after simplifying the hypergeometric functions, the building block functions that span $f_{\bk \bk'}(\alpha)$ are simply power laws in $\alpha$: \begin{equation}
	f_{\bk\bk'} (\alpha): \, \{ \alpha^L \}, \quad\quad (n \to n+2), \label{eq:nton2reducedfunc}
\end{equation} where $L \ne -1$ is some rational number (not necessarily positive).  We can therefore compute the discretization matrix elements on this reduced basis of functions (in this case, powers of $\alpha$), and then obtain the $n \to n+2$ monomial matrix element by summing over all the powers that appear. Plugging~\eqref{eq:nton2reducedfunc} into~\eqref{eq:discretizedmonome2}, we find that \underline{for $n \to n+2$}:  \begin{equation}
	\begin{aligned}
		\cIless[\alpha^L] = \begin{cases}
			\frac{1}{(L+1)(\frac{1}{2}-L)} \left[ \frac{(\mu^2_{\fraki})^{L+1}}{(\mu^2_{\frakj})^{L+\half}} - \frac{(\mu^2_{\fraki})^{L+1}}{(\mu^2_{\frakj-1})^{L-\half}} - \frac{(\mu^2_{\fraki-1})^{L+1}}{(\mu^2_{\frakj})^{L-\half}} + \frac{(\mu^2_{\fraki-1})^{L+1}}{(\mu^2_{\frakj-1})^{L-\half}} \right] \quad L \ne \half, -1, \\
			\\
			\frac{2}{3} \left[ (\mu_\fraki^2)^{\frac{3}{2}} \log \frac{\mu_{\frakj}^2}{\mu_{\frakj-1}^2} - (\mu_{\fraki-1}^2)^{\frac{3}{2}} \log \frac{\mu_{\frakj}^2}{\mu_{\frakj-1}^2} \right] \quad\quad\quad\quad\quad\quad\quad\quad\quad\quad L = \half.
		\end{cases}
	\end{aligned}
\end{equation}  Similarly, \begin{equation}
	\begin{aligned}
		\cIeq[\alpha^L] = \begin{cases}
			\frac{1}{(L+1)} \left[ \frac{2}{3}(\mu_\fraki^2)^{\frac{3}{2}} - \frac{2(L+1)}{3(L-\half)}(\mu_{\fraki-1}^2)^{\frac{3}{2}} - \frac{(\mu_{\fraki-1}^2)^{L+1}}{(\half - L)(\mu_\fraki^2)^{L-\half}} \right] \quad\quad\quad\quad\quad L \ne \half, -1, \\
			\\
			\frac{2}{3} \left[ \frac{2}{3}(\mu_\fraki^2)^{\frac{3}{2}} - \frac{2}{3}(\mu_{\fraki-1}^2)^{\frac{3}{2}} - (\mu_{\fraki-1}^2)^{\frac{3}{2}} \log \frac{\mu_\fraki^2}{\mu_{\fraki-1}^2} \right] \quad\quad\quad\quad\quad\quad\quad\quad L = \half.
		\end{cases}
	\end{aligned}
\end{equation}

Let us now turn to the case of $n \to n$. As mentioned in the previous examples, the set of functions that span $f_{\bk\bk'}(\alpha)$ are \begin{equation}
		f_{\bk\bk'}(\alpha) : \bigg\{ \alpha^L \cdot \{ K(\alpha), E(\alpha) \}, \hspace{2mm} \frac{\{K(\alpha), E(\alpha) \}}{\alpha^L}, \hspace{2mm} \alpha^L \tanh^{-1} \sqrt{\alpha}, \hspace{2mm} \frac{\tanh^{-1}\sqrt{\alpha}}{\alpha^L} \bigg\}, \,\,\, (n \to n) \label{eq:ntonbuildingblockfcn} 
\end{equation} where $K(\alpha)$ and $E(\alpha)$ are complete elliptic integral functions. We have separated the powers of $\alpha$ that multiply these functions into positive and negative powers for later convenience, \emph{i.e}, $L\ge 0$.

We can now try to compute the discretization integrals~\eqref{eq:discreteints} on these functions. However, here we encounter a technical subtlety: sometimes, the discretization integrals of the individual functions above are divergent. This is best illustrated by an example. Consider our previously computed example matrix element~\eqref{ntonexample}, which we can write as \begin{equation}
	\CM_{\bk\bk'}^{(n \to n)}(\mu, \mu') = \frac{1}{\mu'}\left[ c_1 \cdot K(\alpha) + c_2  \cdot E(\alpha) + c_3 \left(\frac{K(\alpha)}{\alpha} - \frac{E(\alpha)}{\alpha} \right)  \right], \label{eq:divergenceexample}
\end{equation} where the coefficients $c_i$ can  be read off from~\eqref{ntonexample}. Now  consider performing one of the discretization integrals on the third term \begin{equation}
	\cIless\left[ \frac{K(\alpha)}{\alpha} \right] = \int_{\mu_{\frakj-1}^2}^{\mu_{\frakj}^2} \, d\mu^{\prime 2} \, \mu' \int_{\mu_{\fraki-1}^2/\mu^{\prime 2}}^{\mu_{\fraki}^2/\mu^{\prime 2}} d\alpha \, \frac{K(\alpha)}{\alpha}.
\end{equation} As we can see, the integrand  diverges near $ \alpha= 0$ and $\alpha=1$ (since $\fraki < \frakj$, there is only one singularity at $\alpha=0$, which corresponds to $\mu_{\fraki-1} = 0$). However, the same discretization integral on the full matrix element in~\eqref{eq:divergenceexample} is finite. The reason is that the  divergence appearing above is precisely canceled by the \textit{same} divergence appearing in the last term, $\frac{E(\alpha)}{\alpha}$. That is, the combination \begin{equation}
	\frac{K(\alpha)}{\alpha} - \frac{E(\alpha)}{\alpha}  = \frac{\pi}{4} + \frac{3\pi \alpha}{32} + \cO(\alpha^2),
\end{equation} is clearly finite as $\alpha\to 0$. For this to be the case, it was crucial that the same multiplicative constant multiply the singular terms (in this case, $c_3$). On general grounds, we expect this to be true as the Fourier transform formulas given in \cite{Anand:2019lkt} are finite for $\mu < \mu'$. We also see this at empirical level in the structure of our matrix elements; any divergence appearing in functions like $K(\alpha)/\alpha^L$ is accompanied by the same divergence appearing in $E(\alpha)/\alpha^L$ but with opposite sign. Thus, \textit{all divergences in the $n \to n$ matrix elements cancel.} We shall now outline the method to quickly compute the discretization integrals, and extract the finite contribution without having to laboriously keep track of all of the divergences.

The key point is that because all matrix elements are finite, we can get away with keeping track of \textit{only} the finite part of each ``building block" function. Returning to our example, since \begin{equation}
	\frac{K(\alpha)}{\alpha} = \frac{\pi}{2\alpha} + \frac{\pi}{8} + \cO(\alpha),
\end{equation} we can ignore the $\frac{\pi}{2\alpha}$ piece since it will eventually cancel with the same term in $\fr{E(\alpha)}{\alpha}$. Using the series expansions for $K(\alpha)$, $E(\alpha)$, and $\tanh^{-1}$, \begin{equation}
	\begin{aligned}
		K(\alpha) &= \sum_{n=0}^\infty \frac{\pi}{2} \left(\frac{(2n)!}{2^{2n}(n!)^2}\right)^2 \alpha^n, \\
		E(\alpha) &= \sum_{n=0}^\infty \frac{\pi}{2} \left(\frac{(2n)!}{2^{2n}(n!)^2}\right)^2 \frac{1}{1-2n} \alpha^n, \\
		\tanh^{-1} \sqrt{\alpha} &= \sum_{n=0}^\infty \frac{1}{2n+1} \alpha^{n+\half},
	\end{aligned}
\end{equation} we can easily extract their finite contributions to the matrix elements and drop the divergences, with the understanding that they all cancel at the end. For example, the finite contribution from $\fr{K(\alpha)}{\alpha}$ is \begin{equation}
	\frac{K(\alpha)}{\alpha} \bigg|_{\textrm{finite}} = \sum_{n=1}^\infty \frac{\pi}{2} \left(\frac{(2n)!}{2^{2n}(n!)^2}\right)^2 \alpha^{n-1},
\end{equation} and so \begin{equation}
	\begin{aligned}
		\cIless^{(\textrm{finite})}\left[ \frac{K(\alpha)}{\alpha} \right] &= \int_{\mu_{\frakj-1}^2}^{\mu_{\frakj}^2} \, d\mu^{\prime 2} \, \mu' \int_{\mu_{\fraki-1}^2/\mu^{\prime 2}}^{\mu_{\fraki}^2/\mu^{\prime 2}} d\alpha  \sum_{n=1}^\infty \frac{\pi}{2} \left(\frac{(2n)!}{2^{2n}(n!)^2}\right)^2 \alpha^{n-1}.
	\end{aligned}
\end{equation} The sum and integration can be interchanged since the integral converges absolutely, and the result can be resummed to give an answer in terms of ${}_3 F_2$ and ${}_4 F_3$ hypergeometric functions. Using this algorithm, we can thus extract the finite pieces of all the functions in~\eqref{eq:ntonbuildingblockfcn}. The final discretized matrix element is obtained by summing over these pieces, multiplied by their accompanying coefficients in the matrix elements. We summarize the results below. Unfortunately, there are many special cases that depend on the power $L$, but once they have been tabulated they are computationally inexpensive to implement.

\underline{Summary of $n \to n$ discretization integrals for \emph{odd} $n$}: First, we begin with odd $n$ where we only encounter elliptic functions (and powers of $\alpha$), but not the $\atanh$ functions in~\eqref{eq:ntonbuildingblockfcn}. We define the useful functions \begin{equation}
	\begin{aligned}
		G_{L\pm}(x) &\equiv \frac{\pi x^L}{3(L+1)} \left[ \threeFtwo \left(\pm\half,\half,L+1;1,L+2;x \right) - \frac{L+1}{L-\half} \threeFtwo\left( \pm \half, \half, L-\half; 1, L+\half; x\right) \right], \\
		H_{L\pm}(x) &\equiv \frac{\Gamma^2(L\pm \half)}{3\Gamma^2(L+1)} \bigg[2 \, \fourFthree \left( -\half, 1, L\pm \half, L+\half;\half,L+1,L+1;x \right) \\
		& \qquad + \fourFthree\left(1,1,L\pm \half,L+\half;2,L+1,L+1;x \right) \bigg], \\
		J_{L\pm}(x) &\equiv -\frac{\Gamma(L+2)\Gamma(L+\frac{3}{2} \pm \half)}{15\Gamma^2(L+\frac{5}{2})} x \bigg[5 \, \fourFthree\pth{1,1,L+\frac{3}{2}\pm\half,L+2;2,L+\frac{5}{2},L+\frac{5}{2};x} \\
		& \qquad -2 \, \fourFthree\pth{1,\frac{5}{2},L+\frac{3}{2}\pm \half, L+2;\frac{7}{2},L+\frac{5}{2}, L+\frac{5}{2};x}
		\bigg].
	\end{aligned}
\end{equation}
The functions with a ``$+$'' subscript will correspond to integrals of expressions containing $K(\alpha)$, while those with a ``$-$'' subscript will correspond to $E(\alpha)$. It will also be convenient to define the shorthand ``$\pm$ perms'' as \begin{equation}
	g(\mu_{\fraki}^2, \mu_{\frakj}^2) \pm \textrm{perms} \equiv g(\mu_\fraki^2, \mu_\frakj^2)  - g(\mu_{\fraki-1}^2, \mu_\frakj^2) - g(\mu_\fraki^2, \mu_{\frakj-1}^2) + g(\mu_{\fraki-1}^2, \mu_{\frakj-1}^2).
\end{equation} Then, \underline{for $\fraki < \frakj$}, we have
\begin{equation}
\begin{aligned}
\mathcal{I}_{\fraki<\frakj}\left[\alpha^{L} K\left(\alpha\right) \right]_{L\geq0,L\neq\frac{1}{2}} & = \left(\mu_\fraki^2\right)\left(\mu_\frakj^2\right)^{\frac{1}{2}}G_{L+}\left(\frac{\mu_\fraki^2}{\mu_\frakj^2}\right)\pm\text{perms}, \\ \\
\mathcal{I}_{\fraki<\frakj}\left[\alpha^{L} E\left(\alpha\right) \right]_{L\geq0,L\neq\frac{1}{2}} & = \left(\mu_\fraki^2\right)\left(\mu_\frakj^2\right)^{\frac{1}{2}}G_{L-}\left(\frac{\mu_\fraki^2}{\mu_\frakj^2}\right)\pm\text{perms}, \\ \\
\mathcal{I}_{\fraki<\frakj }^{(\text{finite})}\left[\frac{K\left(\alpha\right)}{\alpha^{L}}\right]_{L\in\left\{ 1,2,3,\dots\right\} } & = \left(\mu_\fraki^2\right)\left(\mu_\frakj^2\right)^{\frac{1}{2}}H_{L+}\left(\frac{\mu_\fraki^2}{\mu_\frakj^2}\right)\pm\text{perms},\\ \\
\mathcal{I}_{\fraki<\frakj}^{(\text{finite})}\left[\frac{E\left(\alpha\right)}{\alpha^{L}}\right]_{L\in\left\{ 1,2,3,\dots\right\} } & = \frac{\left(1-2L\right)}{4}\left(\mu_\fraki^2\right)\left(\mu_\frakj^2\right)^{\frac{1}{2}}H_{L-}\left(\frac{\mu_\fraki^2}{\mu_\frakj^2}\right)\pm\text{perms},\\ \\
\mathcal{I}_{\fraki<\frakj}^{(\text{finite})}\left[\frac{K\left(\alpha\right)}{\alpha^{L}}\right]_{L\in\left\{ -\frac{1}{2},\frac{1}{2},\frac{3}{2},\dots\right\} } & = \frac{\Gamma^{2}(L+1)}{3\Gamma^{2}(L+\frac{3}{2})} \left(\mu_\fraki^2\right)^{\frac{3}{2}}\log\left(\mu_\frakj^2\right) + \left(\mu_\fraki^2\right)^{\frac{3}{2}}J_{L+}\left(\frac{\mu_\fraki^2}{\mu_\frakj^2}\right)\pm\text{perms}, \\ \\
\mathcal{I}_{\fraki<\frakj}^{(\text{finite})}\left[\frac{E\left(\alpha\right)}{\alpha^{L}}\right]_{L\in\left\{ -\frac{1}{2},\frac{1}{2},\frac{3}{2},\dots\right\} } & = \frac{-L\,\Gamma^{2}(L)}{6\Gamma^{2}(L+\frac{3}{2})}\left(\mu_\fraki^2\right)^{\frac{3}{2}}\log\left(\mu_\frakj^2\right) - \frac{1}{2}\left(\mu_\fraki^2\right)^{\frac{3}{2}}J_{L-}\left(\frac{\mu_\fraki^2}{\mu_\frakj^2}\right)\pm\text{perms}.
 \end{aligned}
\end{equation}

\underline{For $\fraki = \frakj$}, we have \begin{equation}
	\begin{aligned}
	\mathcal{I}_{\left(\fraki=\frakj\right)}\left[\alpha^{L}\cdot\left\{ K\left(\alpha\right),E\left(\alpha\right)\right\} \right]_{L\geq0,L\neq\frac{1}{2}} & =  \frac{\pi\phantom{}_{3}F_{2}\left(\pm\frac{1}{2},\frac{1}{2},1+L;1,2+L;1\right)}{3\left(L+1\right)}\left[\left(\mu_\fraki^2\right)^{\frac{3}{2}}-\left(\mu_{\fraki-1}^2\right)^{\frac{3}{2}}\right] \\
 &  -\left[\left(\mu_{\fraki-1}^2\right)\left(\mu_\fraki^2\right)^{\frac{1}{2}}G_{L\pm}\left(\frac{\mu_{\fraki-1}^2}{\mu_\fraki^2}\right)-\left(\mu_{\fraki-1}^2\right)^{\fr{3}{2}}G_{L\pm}\left(1\right) \right],\\ \\
\mathcal{I}_{\left(\fraki=\frakj\right)}^{(\text{finite})}\left[\frac{K\left(\alpha\right)}{\alpha^{L}}\right]_{L\in\left\{ 1,2,3,\dots\right\} } & = \left[\left(\mu_\fraki^2\right)^{\frac{3}{2}}-\left(\mu_{\fraki-1}^2\right)^{\frac{3}{2}}\right]\frac{\Gamma^{2}\left(L+\frac{1}{2}\right)}{3\Gamma^{2}\left(L+1\right)} \\
 &  \phantom{}\times\phantom{}_{4}F_{3}\left(1,1,L+\frac{1}{2},L+\frac{1}{2};2,L+1,L+1;1\right) \\
 &  -\left[\left(\mu_{\fraki-1}^2\right)\left(\mu_\fraki^2\right)^{\frac{1}{2}}H_{L+}\left(\frac{\mu_{\fraki-1}^2}{\mu_\fraki^2}\right)-\left(\mu_{\fraki-1}^2\right)^{\fr{3}{2}} H_{L+}\left(1\right) \right],\\ \\
\mathcal{I}_{\left(\fraki=\frakj\right)}^{(\text{finite})}\left[\frac{E\left(\alpha\right)}{\alpha^{L}}\right]_{L\in\left\{ 1,2,3,\dots\right\} } & = \left[\left(\mu_\fraki^2\right)^{\frac{3}{2}}-\left(\mu_{\fraki-1}^2\right)^{\frac{3}{2}}\right]\frac{\left(1-2L\right)\Gamma^{2}\left(L-\frac{1}{2}\right)}{12\Gamma^{2}\left(L+1\right)} \\
 &   \phantom{}\times\phantom{}_{4}F_{3}\left(1,1,L-\frac{1}{2},L+\frac{1}{2};2,L+1,L+1;1\right) \\
 &   -\frac{\left(1-2L\right)}{4}\left[\left(\mu_{\fraki-1}^2\right)\left(\mu_\fraki^2\right)^{\frac{1}{2}}H_{L-}\left(\frac{\mu_{\fraki-1}^2}{\mu_\fraki^2}\right)-\left(\mu_{\fraki-1}^2\right)^{\fr{3}{2}} H_{L-}\left(1\right) \right],\\ \\
\mathcal{I}_{\left(\fraki=\frakj\right)}^{(\text{finite})}\left[\frac{K\left(\alpha\right)}{\alpha^{L}}\right]_{L\in\left\{ -\frac{1}{2},\frac{1}{2},\frac{3}{2},\dots\right\} } & = \frac{2\Gamma^{2}(L+1)}{9\Gamma^{2}(L+\frac{3}{2})}\left[\left(\mu_\fraki^2\right)^{\frac{3}{2}}-\left(\mu_{\fraki-1}^2\right)^{\frac{3}{2}}-\frac{3}{2}\left(\mu_{\fraki-1}^2\right)^{\frac{3}{2}}\log\left(\frac{\mu_\fraki^2}{\mu_{\fraki-1}^2}\right)\right] \\
 &   +\left[\left(\mu_\frakj^2\right)^{\frac{3}{2}}-\left(\mu_{\frakj-1}^2\right)^{\frac{3}{2}}\right]\frac{2\Gamma^{2}\left(L+2\right)}{15\Gamma^{2}\left(L+\frac{5}{2}\right)} \\
 &   \phantom{+}\times\phantom{}_{4}F_{3}\left(1,\frac{5}{2},L+2,L+2;\frac{7}{2},L+\frac{5}{2},L+\frac{5}{2};1\right) \\
 &   -\left[\left(\mu_{\fraki-1}^2\right)^{\frac{3}{2}}J_{L+}\left(\frac{\mu_{\fraki-1}^2}{\mu_\fraki^2}\right)-\left(\mu_{\fraki-1}^2\right)^{\frac{3}{2}}J_{L+}\left(1\right) \right],\\ \\
\mathcal{I}_{\left(\fraki=\frakj\right)}^{(\text{finite})}\left[\frac{E\left(\alpha\right)}{\alpha^{L}}\right]_{L\in\left\{ -\frac{1}{2},\frac{1}{2},\frac{3}{2},\dots\right\} } & = \frac{-L \,\Gamma^{2}(L)}{9\Gamma^{2}(L+\frac{3}{2})} \left[\left(\mu_\fraki^2\right)^{\frac{3}{2}}-\left(\mu_{\fraki-1}^2\right)^{\frac{3}{2}}-\frac{3}{2}\left(\mu_{\fraki-1}^2\right)^{\frac{3}{2}}\log\left(\frac{\mu_\fraki^2}{\mu_{\fraki-1}^2}\right)\right] \\
 &  -\left[\left(\mu_\fraki^2\right)^{\frac{3}{2}}-\left(\mu_{\fraki-1}^2\right)^{\frac{3}{2}}\right]\frac{\left(L+1\right)\Gamma^{2}\left(L+1\right)}{15\Gamma^{2}\left(L+\frac{5}{2}\right)} \\
 &  \phantom{+}\times\phantom{}_{4}F_{3}\left(1,\frac{5}{2},L+1,L+2;\frac{7}{2},L+\frac{5}{2},L+\frac{5}{2};1\right) \\
 &  + \left[ \frac{1}{2}\left(\mu_{\fraki-1}^2\right)^{\frac{3}{2}}J_{L-}\left(\frac{\mu_{\fraki-1}^2}{\mu_\fraki^2}\right)- \frac{1}{2}\left(\mu_{\fraki-1}^2\right)^{\frac{3}{2}}J_{L-}\left(1\right)\right].
	\end{aligned}
\end{equation}

\underline{Summary of $n \to n$ discretization integrals for \textit{even} $n$}: For the even sector of the $n \to n$ matrix elements, it will be useful to define the functions \begin{equation}
\begin{aligned}
P_{L}\left(x\right)&\equiv\begin{cases}
\frac{6x^{L+1}\atanh\sqrt{x} \, -2\left(L+1\right) \, x^{\frac{3}{2}}\beta \left(x,L,0\right)+\left(2L-1\right)\beta(x,\frac{3}{2}+L,0)}{3\left(L+1\right)\left(2L-1\right)} & x<1,\\
\frac{3\gamma_E+\log64+2\left(L+1\right)\psi\left(L\right)-\left(2L-1\right)\psi\left(\frac{3}{2}+L\right)}{3\left(L+1\right)\left(2L-1\right)} & x=1,
\end{cases} \\
Q_{L}\left(x\right)&\equiv\begin{cases}
-\frac{32}{21+28L} & x=0,\\
\frac{-9x\Phi\left(x,1,L+\frac{3}{4}\right)-24\left(L-1\right)x\phantom{}_{2}F_{1}\left(\frac{1}{4},1;\frac{5}{4};x\right)+4\left(2L+1\right)\left(-1+\phantom{}_{2}F_{1}\left(\frac{3}{4},1;\frac{7}{4};x\right)\right)}{9\left(L-1\right)\left(2L+1\right)x} & 0<x<1,\\
\frac{-8+18\gamma_E+3\pi-4L\left(4+3\pi\right)+54\log2+18\psi(L+\frac{3}{4})}{18\left(L-1\right)\left(2L+1\right)} & x=1,
\end{cases} \\
R_{L}\left(x\right)&\equiv\begin{cases} 
\frac{32}{5+20L} & x=0,\\
-\frac{9x\Phi\left(x,1,L+\frac{1}{4}\right)+4\left(3+6L+6x-6Lx-3\left(2L+1\right)\phantom{}_{2}F_{1}\left(\frac{1}{4},1;\frac{5}{4};x\right)+2\left(L-1\right)x^{2}\phantom{}_{2}F_{1}\left(\frac{3}{4},1;\frac{7}{4};x\right)\right)}{9\left(L-1\right)\left(2L+1\right)x} & 0<x<1,\\
-\frac{24+\pi-4L\pi-6H_{L-\frac{3}{4}}-18\log2}{6\left(L-1\right)\left(2L+1\right)} & x=1,
\end{cases}
\end{aligned}
\end{equation} where $\beta(z,a,b)$ is the incomplete beta function, $\gamma_E$ is the Euler constant, $\psi$ is the digamma function, $\Phi(z,s,a)$ is the Hurwitz-Lerch transcendent, and $H_k$ is the harmonic number.

With these definitions in mind, we have \underline{for $\fraki < \frakj$}: \begin{equation}
	\begin{aligned}
\mathcal{I}_{\left(\fraki<\frakj\right)}\left[\alpha^{L}\atanh\sqrt{\alpha}\right]_{L\geq-\frac{1}{2},L\neq\left\{ 0,\frac{1}{2}\right\} } = & -\left(\mu_{\frakj}^{2}\right)^{\frac{3}{2}}P_{L}\left(\frac{\mu_{\fraki}^{2}}{\mu_{\frakj}^{2}}\right)\pm\text{perms}, \\ \\
 \mathcal{I}_{\left(\fraki<\frakj\right)}^{(\text{finite})}\left[\frac{\atanh\sqrt{\alpha}}{\alpha^{L}}\right]_{L\in\left\{ \frac{3}{4},\frac{7}{4},\frac{11}{4},\dots\right\} } & = \left(\mu_{\fraki}^2\right)^{\frac{7}{4}}\left(\mu_{\frakj}^{2}\right)^{-\frac{1}{4}}Q_{L}\left(\frac{\mu_{\fraki}^2}{\mu_{\frakj}^{2}}\right)\pm\text{perms}, \\ \\
\mathcal{I}_{\left(\fraki<\frakj\right)}^{(\text{finite})}\left[\frac{\atanh\sqrt{\alpha}}{\alpha^{L}}\right]_{L\in\left\{ \frac{5}{4},\frac{9}{4},\frac{13}{4},\dots\right\} } & = \left(\mu_{\fraki}^{2}\right)^{\frac{5}{4}}\left(\mu_{\frakj}^{2}\right)^{\frac{1}{4}}R_{L}\left(\frac{\mu_{\fraki}^{2}}{\mu_{\frakj}^{2}}\right)\pm\text{perms},
	\end{aligned}
\end{equation} and for \underline{$\fraki = \frakj$}, \begin{equation}
	\begin{aligned}
		\mathcal{I}_{\left(\fraki=\frakj\right)}\left[\alpha^{L}\atanh\sqrt{\alpha}\right]_{L\geq-\frac{1}{2},L\neq\left\{ 0,\frac{1}{2}\right\} } &= \frac{\gamma_E+\log4+\psi\left(L+\frac{3}{2}\right)}{3\left(L+1\right)}\left[\left(\mu_{\fraki}^{2}\right)^{\frac{3}{2}}-\left(\mu_{\fraki-1}^{2}\right)^{\frac{3}{2}}\right]\\
 &  +\left[\left(\mu_{\fraki}^{2}\right)^{\frac{3}{2}}P_{L}\left(\frac{\mu_{\fraki-1}^{2}}{\mu_{\fraki}^{2}}\right)-\left(\mu_{\fraki-1}^{2}\right)^{\frac{3}{2}}P_{L}\left(1\right) \right],\\ \\
 		\mathcal{I}_{\left(\fraki=\frakj\right)}^{(\text{finite})}\left[\frac{\atanh\sqrt{\alpha}}{\alpha^{L}}\right]_{L\in\left\{ \frac{3}{4},\frac{7}{4},\frac{11}{4},\dots\right\} } & = \frac{\psi(\frac{7}{4})-\psi\left(L+\frac{3}{4}\right)}{3(1-L)}\left[\left(\mu_{\fraki}^{2}\right)^{\frac{3}{2}}-\left(\mu_{\fraki-1}^{2}\right)^{\frac{3}{2}}\right]\\
 & -\left[\left(\mu_{\fraki-1}^{2}\right)^{\frac{7}{4}}\left(\mu_{\fraki}^{2}\right)^{-\frac{1}{4}}Q_{L}\left(\frac{\mu_{\fraki-1}^{2}}{\mu_{\fraki}^{2}}\right)-\left(\mu_{\fraki-1}^{2}\right)^{\frac{3}{2}}Q_{L}\left(1\right) \right], \\ \\
 		\mathcal{I}_{\left(\fraki=\frakj\right)}^{(\text{finite})}\left[\frac{\atanh\sqrt{\alpha}}{\alpha^{L}}\right]_{L\in\left\{ \frac{5}{4},\frac{9}{4},\frac{13}{4},\dots\right\} } & =\frac{\psi(\frac{5}{4})-\psi(L+\frac{1}{4})}{3\left(1-L\right)}\left[\left(\mu_{\fraki}^{2}\right)^{\frac{3}{2}}-\left(\mu_{\fraki-1}^{2}\right)^{\frac{3}{2}}\right]\\
 &  -\left[\left(\mu_{\fraki-1}^{2}\right)^{\frac{5}{4}}\left(\mu_{\fraki}^{2}\right)^{\frac{1}{4}}R_L\left(\frac{\mu_{\fraki-1}^{2}}{\mu_{\fraki}^{2}}\right)-\left(\mu_{\fraki-1}^{2}\right)^{\frac{3}{2}} R_L\left(1\right) \right].
	\end{aligned}
\end{equation} These formulas exhaust all of the cases we encounter in the matrix elements. We have checked this expressions against matrix elements obtained using Fock space methods (outlined in appendix \ref{sec:FockSpace}). In practice, to implement these formulas, it is computationally fast to tabulate them over $(\fraki, \frakj)$ in the form of a ``reference'' matrix. Then, the final discretized matrix element can be obtained by summing over these matrices, weighted by the appropriate basis coefficients.

\section{Lightcone Fock Space Methods}
\label{sec:FockSpace}

In this section, we outline an alternative way to construct the basis and matrix elements using Fock space wavefunctions, rather than Fourier transforming position space correlators. Although this method is computationally slower than the methods given in appendix \ref{sec:MatrixElements}, it is nevertheless useful for two reasons. First, it provides a consistency check with the formulas in appendix \ref{sec:MatrixElements}. Those formulas have many moving parts, and finding agreement between these two methods is a highly nontrivial check. Second, it is often easier to gain intuition for the structure of matrix elements in Fock space (\emph{e.g.}, computing perturbative corrections is often conceptually and technically easier to see in Fock space).

The idea is to work directly with the free theory wavefunctions $F_{\cO}(p_1, \dots, p_n)$ of an operator, defined by the overlap \begin{equation}
	\langle \cO(0) | p_1, \dots , p_n \rangle \equiv F_{\cO}(p_1, \dots, p_n).
\end{equation} Note that our operators are Dirichlet, which in momentum space translates to overall factors of the momentum $p_{1-} \dots p_{n-}$ in the wavefunction. To tidy up some equations, we can pull out this overall factor of $p_{1-} \dots p_{n-}$ and define a `barred' wavefunction $\Fbar(p_1, \dots, p_n)$ such that \begin{equation}
	\Fnobar \equiv p_{1-} p_{2-} \dotsb p_{n-} \, \Fbar(p_1, \dots, p_n).
\end{equation} Then, any basis state can be expanded in terms of this wavefunction \begin{equation}
	\begin{aligned}
		\ket{\cO, \fraki} = &\frac{1}{\sqrt{2\pi}} \int_0^{\Lambda^2} \frac{d\mu^2}{ p_-^{|\bk_-|} \mu^{\frac{n-3}{2}-|\bk_\bot|}} b_{\fraki}(\mu) \frac{1}{n!}\int \frac{d^2 p_1 \dotsb d^2 p_n}{(2\pi)^{2n} 2p_{1-} \dotsb 2p_{n-}} (2\pi)^3 \delta^{(3)} \left( \sum_i p_i - p \right) \\
	&\times p_{1-} \dotsb p_{n-} \, \Fbar(p_1, \dots, p_n) \ket{p_1, \dots, p_n}, \label{eq:fockexpansion}
	\end{aligned}
\end{equation} where our normalization for the Fock space states is\footnote{We adhere to the conventions in \cite{Katz:2016hxp}.} \begin{equation}
	\langle p | q \rangle = 2 p_- (2\pi)^2 \delta^{(2)}(p-q).
\end{equation}Eq.~\eqref{eq:fockexpansion} leads to the inner product \begin{equation}
	\begin{aligned}
		&\frac{\langle \cO, \fraki | \cO', \frakj \rangle}{2p_- (2\pi)^2 \delta^{(2)}(\vec{p}-\vec{p}^{\,\prime})} = \frac{1}{2^n n!} \frac{1}{2\pi p^{|\bk_-|+|\bk_-'|}_-} \int_0^{\Lambda^2} \frac{d\mu^2}{\mu^{\frac{n-3}{2} + |\bk_\bot|}} \int_0^{\Lambda^2} \frac{d\mu^{\prime 2}}{\mu^{\prime \frac{n-3}{2} + |\bk_\bot'|}}  b_{\fraki}(\mu^2) b_{\frakj}(\mu^{\prime 2})  \\
		&\times (2\pi) \delta(\mu^2 - \mu^{\prime 2}) \int \frac{d^2 p_1 \dotsb d^2 p_n}{(2\pi)^{2n}} (2\pi)^3 \delta^{(3)}\left(\sum_i p_i - p \right) p_{1-} \dotsb p_{n-} {\Fbar}(p){ \Fbarp}(p). \label{eq:innerprodfock1}
	\end{aligned}
\end{equation} Using the equations of motion and the choice of our reference frame $p_\bot = \sum_{i}^n p_{i\bot} = 0$, the set of delta functions can be expanded out \begin{equation}
	\begin{aligned}
		\delta^{(3)}\left(\sum_i p_i - p \right) = \delta\left(\sum_i \frac{p_{i\bot}^2}{2p_{i-}}-\frac{\mu^2}{2p_-} \right)\delta\pth{\sum_i p_{i-}-p_-}\delta\pth{\sum_i p_{i\bot}}. \label{eq:fockdeltafcn}
	\end{aligned}
\end{equation} Here, it is useful to define the dimensionless variables \begin{equation}
	x_i \equiv \frac{p_{i-}}{p_-}, \quad\quad\quad y_i \equiv \frac{p_{i\bot}}{\mu},
\end{equation} such that the wavefunctions then have a scaling set by \begin{equation}
	F_{\cO}(p) = \mu^{|\bk_\bot|}p_-^{|\bk_-|} F_{\cO}(x,y).
\end{equation} Using this scaling, and the fact that the weight functions $b_{\fraki}(\mu)$, $b_{\frakj}(\mu)$ are defined to be orthonormal when integrated over $\mu^2$, the inner product factorizes into a piece that is diagonal with respect to $\fraki$, $\frakj$ and a piece that depends on $\cO$, $\cO'$: \begin{equation}
	\frac{\bra{\cO, \fraki} \cO', \frakj \rangle}{{2p_- (2\pi)^2 \delta^{(2)}(\vec{p}-\vec{p}^{\,\prime})}} = \delta_{\fraki \frakj} \cdot \cI_{\cO \cO'}.
\end{equation} To determine $\mathcal{I}_{\mathcal{O}\mathcal{O}'}$, we can choose integration variables defined by \begin{equation}
		\begin{aligned}
			x_1 &= (1-z_1)(1-z_2)(1-z_3) \dotsb (1-z_{n-1}), \\
			x_2 &= \quad\quad\,\,\, z_1(1-z_2)(1-z_3) \dotsb (1-z_{n-1}), \\
			x_3 &= \quad\quad\,\,\, \quad\quad\quad\,\, z_2(1-z_3) \dotsb (1-z_{n-1}), \\
			&\vdots \quad\quad\quad\quad\quad\quad\quad\quad\quad \ddots \\
			x_n &= \quad\quad\quad\quad\quad\quad\quad\quad\quad\quad\quad\quad\quad\quad z_{n-1}, \label{xtrans}
		\end{aligned}
	\end{equation} where the $z_i$ range from $[0, 1]$, and \begin{equation}
		\begin{aligned}
			y_1 &= -(y_2+ y_3 + \dotsb + y_n), \\
			y_2 &= \tilde{y}_1 \sqrt{z_1(1-z_1) \dotsb (1-z_{n-1})} - z_1(y_3 + \dotsb + y_n), \\
			y_3 &= \tilde{y}_2 \sqrt{z_2(1-z_2) \dotsb (1-z_{n-1})} - z_2(y_4 + \dotsb + y_n), \\
			& \, \, \, \vdots\\
			y_n &= \tilde{y}_{n-1} \sqrt{z_{n-1}(1-z_{n-1})}. \label{ytrans}
		\end{aligned}
	\end{equation} These variables nicely implement the constraints given in~\eqref{eq:fockdeltafcn}, such that the original set of delta functions simply reduce to a single delta function \begin{equation}
		\begin{aligned}
			\delta^{(3)}\left(\sum_i p_i - p \right) \to \frac{2}{\mu^3} \delta \left( \sum_i^{n-1} \tilde{y}_i^2 - 1 \right).
		\end{aligned}
	\end{equation} Introducing angular variables for the remaining $\tilde{y}$ variables to implement this constraint \begin{equation}
		\begin{aligned}
			\tilde{y}_1 &= \sin \theta_1 \sin \theta_2 \dotsb \sin \theta_{n-2}, \\
			\tilde{y}_2 &= \cos \theta_1 \sin \theta_2 \dotsb \sin \theta_{n-2}, \\
			\tilde{y}_3 &= \cos \theta_2 \sin \theta_3 \dotsb \sin \theta_{n-2}, \\
			&\, \, \, \vdots \\
			\tilde{y}_{n-1} &= \cos \theta_{n-2}, \label{thetatrans}
		\end{aligned}
	\end{equation} where $\theta_i \in [0, \pi]$ for $i = 1, \dots, n-3$ and $\theta_{n-2} \in [0, 2\pi]$, we find that the inner product becomes \begin{equation}
		\boxed{
		\begin{aligned}
			\frac{\bra{\cO, \fraki} \cO', \frakj \rangle}{{2p_- (2\pi)^2 \delta^{(2)}(\vec{p}-\vec{p}^{\,\prime})}} &=  \delta_{\fraki \frakj} \cdot \mathcal{I}_{\mathcal{O}\mathcal{O}'}, \\
			\mathcal{I}_{\mathcal{O}\mathcal{O}'} =  \frac{1}{n! 2^n (2\pi)^{2n-3}} &\int dz_1 \dotsb dz_{n-1} \left( \prod_i z_i^{\frac{3}{2}} (1-z_i)^{\frac{5}{2}i-1} \right) \\
			&\times \int d\theta_1 \dotsb d\theta_{n-2} \left( \prod_j \sin^{j-1} \theta_j \right) \Fbar(z,\theta) \Fbarp(z,\theta).
		\end{aligned}
		} \label{eq:innerFock}
	\end{equation}\eqref{eq:innerprodfock1} gives the inner product of two operators in our basis in terms of their momentum space wavefunctions. One can use this expression to construct the basis states by tabulating a list of Dirichlet monomials at and below a given $\Delta_{\textrm{max}}$ and their associated momentum space wavefunctions, and then compute the Gram matrix using~\eqref{eq:innerFock} between different monomials.

 Now, let us turn to the matrix elements. First, let us start with kinetic term matrix elements corresponding to the free part of the Hamiltonian. This part of the Hamiltonian can be expressed in terms of raising and lowering operators as\footnote{See \cite{Katz:2016hxp} for details about constructing the Hamiltonian in terms of creation and annihilation operators.} \begin{equation}
	P_+^{(\textrm{CFT})} = \int \frac{d^2 p}{(2\pi)^2} a^\dagger_p a_p \frac{p_\bot^2}{2p_-}. \label{eq:kineticP}
\end{equation} Note that this term preserves particle number, so that we consider sectors with differing particle number separately. Inserting~\eqref{eq:kineticP} in between two basis states and using the coordinate transformations in~\eqref{xtrans},~\eqref{ytrans}, and~\eqref{thetatrans} we find that \begin{equation}
	\begin{aligned}
		\boxed{\frac{\bra{\cO, \fraki} P_+^{(\textrm{CFT})} | \cO', \frakj \rangle}{{2p_- (2\pi)^2 \delta^{(2)}(\vec{p}-\vec{p}^{\,\prime})}} =  \left( \frac{\mu_{\fraki+1}^2 + \mu_{\fraki}^2}{2} \right) \delta_{\fraki \frakj} \cdot \mathcal{I}_{ \mathcal{O}\mathcal{O}'}.}
	\end{aligned}
\end{equation}

Next, the mass term gives rise to a correction \begin{equation}
	\begin{aligned}
		\delta P_+^{(m)} &= \int \frac{d^2 p}{(2\pi)^2} a^\dagger_p a_p \frac{m^2}{2p_-}.
	\end{aligned}
\end{equation} Like the kinetic term, this term preserves particle number. We can use the same coordinate transformations as in the inner product and kinetic terms to arrive at \begin{equation}
\boxed{
\begin{aligned}
	\frac{\bra{\cO, \fraki} \delta P_+^{(m)} | \cO', \frakj \rangle}{{2p_- (2\pi)^2 \delta^{(2)}(\vec{p}-\vec{p}^{\,\prime})}} = \delta_{\fraki \frakj} &\frac{m^2(1+\delta_{n,2})}{(n-1)! 2^n (2\pi)^{2n-3}} \int dz_1 \dotsb dz_{n-1} \left( \prod_i z_i^{\frac{3}{2}} (1-z_i)^{\frac{5}{2}i-1}  \right) \left( \frac{1}{z_{n-1}} \right) \\
	&\times \int d\theta_1 \dotsb d\theta_{n-2} \left( \prod_j \sin^{j-1} \theta_j \right) \Fbar(z,\theta) \Fbarp(z,\theta).
\end{aligned} }
\end{equation}

Next, the quartic interaction in the Hamiltonian gives the terms \begin{equation}
	\begin{aligned}
		\delta P_+^{(g)} = \frac{g}{24} \int \frac{d^2p d^2 q d^2 k}{(2\pi)^6 \sqrt{8 p_- q_- k_-}} \left( \frac{4 a^\dagger_p a^\dagger_q a^\dagger_k a_{p+q+k}}{\sqrt{2(p_- + q_- + k_-)}} + \textrm{ h.c. } + \frac{6 a^\dagger_p a^\dagger_q a_k a_{p+q-k}}{\sqrt{2(p_- + q_- - k_-)}} \right). \label{quarticdef}
	\end{aligned}
\end{equation} As explained in appendix \ref{sec:MatrixElements}, we can divide this into an $n \to n$ particle number preserving piece $\delta P_+^{(n \to n)}$ and an $n \to n+2$ piece $\delta P_+^{(n \to n+2)}$. Recalling the definition for the matrix element in~\eqref{eq:matrixelemOOpdef}, the $n \to n+2$ matrix element (before discretization) takes the form \begin{equation}
	\begin{aligned}
		\Mcal^{(n\to n+2)}_{\Ocal \Ocal'}(\mu, \mu') &= \frac{g}{6(n-1)!} \int \frac{d^2 p_1 \dotsb d^2 p_n}{(2\pi)^{2n} 2p_{1-} \dotsb 2p_{n-}} (2\pi)^3 \delta^3 \left(\sum_i p_i - p \right) p_{1-} \dotsb p_{n-} \bar{F}_{\Ocal}(p) \\
		&\times \int \frac{d^2p_1' \dotsb d^2p_{n+2}'}{(2\pi)^{2n+4} 2p_{1-}' \dotsb 2p'_{n+2-}} (2\pi)^3 \delta^3\left(\sum_i p'_i - p' \right) p'_{1-} \dotsb p'_{n+2-} \bar{F}_{\Ocal'}(p') \\
		&\times 2p_{2-} (2\pi)^2 \delta^2(p_2 - p'_4) \dotsb 2 p_{n-} (2\pi)^2 \delta^2 (p_n - p'_{n+2}).
	\end{aligned}
\end{equation}

 It is useful to change variables separately for the both the primed and unprimed variables. That is, we take~\eqref{xtrans} and~\eqref{ytrans} for the unprimed variables and \begin{equation}
		\begin{aligned}
			x_1' &= (1-z'_1)(1-z'_2)(1-z'_3) \dotsb (1-z'_{n+1}), \\
			x_2' &= \quad\quad\,\,\, z'_1(1-z'_2)(1-z'_3) \dotsb (1-z'_{n+1}), \\
			x_3' &= \quad\quad\,\,\, \quad\quad\quad\,\, z'_2(1-z_3') \dotsb (1-z_{n+1}'), \\
			&\vdots \quad\quad\quad\quad\quad\quad\quad\quad\quad \ddots \\
			x_{n+2}' &= \quad\quad\quad\quad\quad\quad\quad\quad\quad\quad\quad\quad\quad\quad z_{n+1}' \label{xtransprimed}
		\end{aligned}
	\end{equation} for the primed coordinates and analogously for~\eqref{ytrans}. We then find \begin{equation}
		\begin{aligned}
			&\Mcal^{(n\to n+2)}_{\Ocal \Ocal'}(\mu, \mu') = \frac{g n \sqrt{(n+1)(n+2)}}{24 \pi} \frac{1}{\mu'} \alpha^{\frac{n-3}{4}} \int dz_1 \dotsb dz_{n-1}   d\tilde{y}_1 \dotsb d\tilde{y}_{n-1}  \\
			&\times  \left( \prod_{i=1}^{n-1} z_i^{\frac{3}{2}} (1-z_i)^{\frac{5}{2}i+1} \right)  \delta\left(\sum_i^{n-1} \tilde{y}^2_{n-1} -1  \right) \bar{F}_{\Ocal}(z,\tilde{y}) \int dz'_1 dz'_2  d\tilde{y}'_1 d\tilde{y}_2'\, {z'_1}^{\frac{1}{2}} (1-z'_1)^{\frac{1}{2}} {z'_2}^{\frac{1}{2}} (1-z'_2)^2 \\
			&\times \delta\left(\tilde{y}_1^{\prime 2} + \tilde{y}_2^{\prime 2} + \alpha \sum_{i=1}^{n-1} \tilde{y}_i^2 - 1 \right) \bar{F}_{\Ocal'}(z', \tilde{y},\tilde{y}'),
		\end{aligned}
	\end{equation} where $\alpha \equiv \mu^2/\mu^{\prime 2}$. The first delta function constrains the $n-1$ $\tilde{y}$'s, which correspond to the variables of the spectator particles, to a sphere of radius 1. The other delta function for the interacting particles constrains $\tilde{y}'$ to a sphere of radius $1-\alpha$, which makes manifest the restriction $\alpha < 1$ (\emph{i.e.}, $\mu < \mu'$). Parameterizing these two spheres with angular variables for the spectators and interacting particles we obtain\footnote{Note that the special case of $1\to 3$ is given by \begin{equation}
		\Mcal^{(1 \to 3)}_{\cO'}(\mu') = \frac{g}{3\pi^3 2^7} \int dz_1' dz_2' d\theta' z_1^{\frac{1}{2}} (1-z_1')^{\half} z_2^{\prime \half} (1-z_2')^2 \bar{F}_{\cO'}(z',\theta').
	\end{equation}} \begin{equation}
	\boxed{
	\begin{aligned}
			&\Mcal^{(n\to n+2)}_{\Ocal \Ocal'}(\mu, \mu') = \frac{g}{(n-1)! 3 \pi^{2n} 2^{3n+4}} \frac{1}{\mu'} \alpha^{\frac{n-3}{4}}   \\
			&\times \int dz_1 \dotsb dz_{n-1} dz'_1 dz'_2   \left( \prod_{i=1}^{n-1} z_i^{\frac{3}{2}} (1-z_i)^{\frac{5}{2}i+1} \right) {z'_1}^{\frac{1}{2}} (1-z'_1)^{\frac{1}{2}} {z'_2}^{\frac{1}{2}} (1-z'_2)^2  \\
			&\times \int d\theta_1 \dotsb d\theta_{n-2} d\theta' \left(\prod_j \sin^{j-1} \theta_j \right) \bar{F}_{\Ocal}(z,\theta) \bar{F}_{\Ocal}(z',\theta',\theta,\alpha).
		\end{aligned} \label{eq:ntonp2Fockfinal}
	}
	\end{equation} 

Finally, we turn to the $n$-to-$n$ part of the quartic interaction. It takes the form \begin{equation}
	\begin{aligned}
		\Mcal_{\Ocal \Ocal'}^{(n\to n)}(\mu, \mu') = \frac{g n(n-1)}{4}& \frac{1}{n!} \int \frac{d^2p_1 \dotsb d^2 p_n}{(2\pi)^{2n} 2p_{1-} \dotsb 2p_{n-}} (2\pi)^3 \delta^3\left(\sum_i p_i - p \right) p_{1-} \dotsb p_{n-} \bar{F}_{\Ocal}(p) \\
		&\times \int \frac{d^2 p'_1 \dotsb d^2 p'_n}{(2\pi)^{2n} 2p'_{1-} \dotsb 2p'_{n-}} (2\pi)^3 \delta^3\left(\sum_i p'_i - p' \right) p'_{1-} \dotsb p'_{n-} \bar{F}_{\Ocal}(p') \\
		&\times 2p_{3-}(2\pi)^2 \delta^2(p_3 - p_3') \dotsb 2p_{n-} (2\pi)^2 \delta^2(p_n -p'_n).
	\end{aligned}
\end{equation} Performing the coordinate transforms in~\eqref{xtrans} and \eqref{ytrans} for both the primed and unprimed coordinates, we find \begin{equation}
	\begin{aligned}
		\Mcal_{\Ocal \Ocal'}^{(n\to n)}(\mu, \mu') =& \frac{g}{(n-2)! \pi^{2n-2} 2^{3n}}\int dz_1  d\tilde{y}_1 dz'_1 d\tilde{y}'_1 \sqrt{z_1 (1-z_1) z'_1 (1-z'_1)} \\
		&\times \delta\left(\sum_i \tilde{y}_i^2 - 1 \right) \delta\left({\tilde{y}_1}^{\prime 2} +  \alpha \sum_{i =2}\tilde{y}_i^2  -1\right) \\
		&\times \int dz_2 \dotsb dz_{n-1} d\tilde{y}_2 \dotsb d\tilde{y}_{n-1} \left( \prod_{i>1} z_i^{\frac{3}{2}} (1-z_i)^{\frac{5i-3}{2}} \right) F_{\Ocal}(z,\tilde{y}) F_{\Ocal}(z',\tilde{y}').
	\end{aligned}
\end{equation} We can use the delta functions to perform the integration over the $\tilde{y}$ coordinates of the interacting particles. Note that they impose the constraints \begin{equation}
	\tilde{y}_1 = \pm\sqrt{1-\tilde{r}^2}, \quad\quad\quad\quad\quad \tilde{y}_1^{\prime} = \pm \sqrt{ 1 - \alpha \tilde{r}^2},
\end{equation} where \begin{equation}
	\tilde{r}^2 \equiv \tilde{y}_2^2 - \tilde{y}_3^2 \dotsb - \tilde{y}_{n-1}^2.
\end{equation} Note that when $\alpha =1$, the two constraints coincide, and the range of integration $\tilde{r}$ is taken to be between $[0,1]$. Similarly, when $\alpha < 1 $, the reality condition on $\tilde{y}_1$ requires $\tilde{r} \in [0,1]$, which automatically satisfies the constraint on $\tilde{y}'_1$. However, when $\alpha > 1$, the reality condition on $\tilde{y}'_1$ provides a stronger constraint and requires $\tilde{r} \in [0, \alpha^{-1/2} ]$.

Defining spherical coordinates for the remaining spectators \begin{equation}
	\begin{aligned}
		\tilde{y}_2 &= \tilde{r} \sin \theta_1 \sin \theta_2 \dotsb \sin \theta_{n-3}, \\
		\tilde{y}_3 &= \tilde{r} \cos \theta_1 \sin \theta_2 \dotsb \sin \theta_{n-3}, \\
		\tilde{y}_4 &= \tilde{r} \cos \theta_2 \sin \theta_3 \dotsb \sin \theta_{n-3}, \\
		&\, \, \, \vdots \\
		\tilde{y}_{n-1} &= \tilde{r} \cos \theta_{n-3},
	\end{aligned}
\end{equation} and defining \begin{equation}
	\bar{F}_{\Ocal \pm} \equiv \bar{F}_{\Ocal}(\tilde{y}_1 = \pm \sqrt{1-\tilde{r}^2}),\quad\quad\quad\quad \bar{F}_{\Ocal' \pm} \equiv \bar{F}_{\Ocal'}(\tilde{y}_1 = \pm \sqrt{1-\alpha \tilde{r}^2}),
\end{equation} the matrix element can be summarized as \begin{equation}
	\boxed{
	\begin{aligned}
			&\Mcal^{(n\to n)}_{\Ocal \Ocal'}(\mu, \mu') = \frac{g}{(n-2)! \pi^{2n-2} 2^{3n+2}} \frac{1}{\mu'} \alpha^{\frac{n-3}{4}}  \\
			&\times \int dz_1 \dotsb dz_{n-1} dz'_1 \sqrt{z_1(1-z_1)z'_1(1-z'_1)} \left(\prod_{i > 1} z_i^{\frac{3}{2}} (1-z_i)^{\frac{5i-3}{2}} \right) \\
			&\times \int_0^{\min \left(1,\alpha^{-1/2} \right)} d\tilde{r} \int d\theta_1 \dotsb d\theta_{n-3} \left( \prod_j \sin^{j-1} \theta_j \right) \frac{\tilde{r}^{n-3}}{\sqrt{(1-\tilde{r}^2)(1-\alpha \tilde{r}^2)}} \\
			&\times \left(\sum_{\pm} \Fbar(z,\tilde{r},\theta) \right)\left(\sum_{\pm} \Fbarp(z,\sqrt{\alpha} \tilde{r}, \theta) \right).
	\end{aligned} \label{eq:ntonFockfinal}
	}
\end{equation} The discretization procedure for the matrix elements~\eqref{eq:ntonp2Fockfinal} and~\eqref{eq:ntonFockfinal} can be carried out using the methods given in \ref{sec:discretizingme}. Note that~\eqref{eq:ntonFockfinal} already suggests the simplifications in the hypergeometric functions we saw in appendix \ref{subsec:muDiscretization}. If we imagine expanding $\Fbar$, $\Fbarp$ in monomials, then the $\tilde{r}$ integral will schematically give rise to elliptic functions and $\atanh$ functions, depending on $n$ and the monomial power of $\tilde{r}$.

\section{State-Dependence in ET and LC Quantization}
\label{app:StateDependence}

In this appendix, we discuss in more detail the origin of state-dependence in UV divergences. Specifically, we consider the leading correction in ET quantization to the mass of an $n$-particle state with general total momentum $\vec{p}_\tot$ due to the sunset diagram. Given a UV cutoff $\Lambda$ on the total invariant mass of the intermediate state, we then demonstrate that the effective cutoff seen by the loop momenta depends on the momenta of the spectator particles, as well as the total momentum of the external state. Finally, we obtain the effective cutoff for LC quantization by taking the infinite momentum limit $|\vec{p}_\tot| \ra \infty$, which behaves qualitatively differently than ET quantization at finite total momentum.

In ET quantization, let's consider some initial $n$-particle state $|p_1,\ldots,p_n\>$ in the free massive theory with total momentum $\vec{p}_\tot$, which we can choose to point in the $x$-direction,
\be
\vec{p}_\tot \equiv \sum_{i=1}^n \vec{p}_i = (p_\tot,0).
\ee
If we add the interaction $g\phi^4$, at $O(g^2)$ in perturbation theory this $n$-particle state receives a correction due to mixing with $(n+2)$-particle states, represented by the sunset diagram in figure~\ref{fig:Sunset}(b). Of course, this state also receives a correction from $(n+4)$-particle states, corresponding to a ``backwards'' sunset diagram. Here we'll only focus on the $(n+2)$-particle contribution, as this is the one that survives in the large momentum limit, but both have the same qualitative behavior for the resulting state-dependence.

In the sunset diagram, one of the particles (which we have labeled by momentum $p_i$), splits into three intermediate particles, with individual momenta $\ell_1$, $\ell_2$, $\ell_3$. Of course, the spatial momentum is conserved in this process,
\be
\vec{p}_i = \vec{\ell_1} + \vec{\ell}_2 + \vec{\ell}_3,
\ee
while the energy is not. We can parametrize the total energy of these loop momenta with their invariant mass $\mu_\Loop^2$,
\be
\mu_\Loop^2 \equiv (\ell_1 + \ell_2 + \ell_3)^2,
\ee
such that the energy is 
\be
E_\Loop = \sqrt{|\vec{p}_i|^2 + \mu_\Loop^2}.
\ee
The remaining $n-1$ of the particles are ``spectators'', with energy fixed by their incoming momentum,
\be
E_j = \sqrt{|\vec{p}_j|^2 + m^2}.
\ee
The total energy of the intermediate $(n+2)$-particle state is thus
\be
E_\tot = \sqrt{|\vec{p}_i|^2 + \mu_\Loop^2} + \sum_{j\neq i} \sqrt{|\vec{p}_j|^2 + m^2},
\ee
and the total invariant mass is
\be
\mu_\tot^2 = \left( \sqrt{|\vec{p}_i|^2 + \mu_\Loop^2} + \sum_{j\neq i} \sqrt{|\vec{p}_j|^2 + m^2} \right)^2 - |\vec{p}_\tot|^2.
\label{eq:MuTot}
\ee

When imposing a UV cutoff $\Lambda$ on the intermediate state, we specifically place a bound on the total invariant mass,
\be
\mu_\tot^2 \leq \Lambda^2.
\ee
We can discover the resulting effective cutoff $\Lambda_\Loop$ seen by the loop momenta by inverting eq.~\eqref{eq:MuTot}, to determine $\mu_\Loop^2$ as a function of $\mu_\tot^2$,
\be
\mu_\Loop^2 = \left( \sqrt{|\vec{p}_\tot|^2 + \mu_\tot^2} - \sum_{j\neq i} \sqrt{|\vec{p}_j|^2 + m^2} \right)^2 - |\vec{p}_i|^2.
\ee
We therefore obtain the following relation between $\Lambda_\Loop$ and the UV cutoff $\Lambda$: 
\be
\Lambda_\Loop^2 = \left( \sqrt{|\vec{p}_\tot|^2 + \Lambda^2} - \sum_{j\neq i} \sqrt{|\vec{p}_j|^2 + m^2} \right)^2 - |\vec{p}_i|^2.
\label{eq:LambdaLoop}
\ee
The effective cutoff is clearly \emph{state-dependent}, with its relation to the universal cutoff $\Lambda$ dependent on the momenta of all particles in the external $n$-particle state.

We are particularly interested in studying this effective cutoff in two limits. The first is the standard center-of-mass frame where $\vec{p}_\tot = 0$, in which case we have
\be
\Lambda_\Loop^2 = \left( \Lambda - \sum_{j\neq i} \sqrt{|\vec{p}_j|^2 + m^2} \right)^2 - |\vec{p}_i|^2 \qquad (|\vec{p}_\tot| \ra 0).
\ee
In this case, the state-dependence is simply a finite additive shift, such that in the limit $\Lambda \ra \infty$ we have
\be
\boxed{\Lambda_\Loop \approx \Lambda - O(\Lambda^0) \qquad (|\vec{p}_\tot| \ra 0).}
\label{eq:LambdaLoopET}
\ee
This additive shift is unsurprising, as some of the ``energy budget'' is taken up by the finite energies of the spectator particles, leading to a lower effective cutoff for the loop momenta.

The second limit we would like to study is the infinite momentum frame, which corresponds to working in LC quantization~\cite{Dirac:1949cp,Weinberg:1966jm,Bardakci:1969dv}. Specifically, if we choose $\vec{p}_\tot$ to point in the $x$-direction, we want to take the infinite momentum limit with the ratio of the $x$-component of each particle's momentum to the total momentum fixed,
\be
|\vec{p}_\tot| \ra \infty, \quad x_j \equiv \fr{p_j^x}{p_\tot} \, \textrm{ fixed}.
\ee
In this limit, we then obtain the qualitatively different relation between $\Lambda_\Loop$ and $\Lambda$:
\be
\Lambda_\Loop^2 = x_i\left( \Lambda^2 - \sum_{j\neq i} \fr{p_{j\perp}^2 + m^2}{x_j} \right) - p_{i\perp}^2 \qquad (|\vec{p}_\tot| \ra \infty),
\ee
where $p_{j\perp} \equiv p_j^y$ is simply the $y$-component of each momentum. As we can see, the state-dependence is now a \emph{multiplicative} factor, in addition to the additive shift, such that at large cutoff we obtain
\be
\boxed{\Lambda_\Loop \approx \sqrt{x_i}\Lambda - O\Big(\fr{1}{\Lambda}\Big) \qquad (|\vec{p}_\tot| \ra \infty).}
\label{eq:LambdaLoopLC}
\ee

In LC quantization, the effective cutoff for the loop momenta is thus suppressed by the momentum fraction $x_i$ carried by the loop. This is quite different from the behavior in ET quantization, where the effective cutoff is approximately the same as the total cutoff, up to a finite difference due to the spectator energies.

The origin of this multiplicative factor is perhaps simplest to understand directly in LC quantization (rather than by taking the infinite momentum limit of ET quantization). In this case, the LC energy of the three particles in the sunset diagram is
\be
p_{\Loop+} \equiv \ell_{1+} + \ell_{2+} + \ell_{3+} = \fr{p_{i\perp}^2 + \mu_\Loop^2}{2p_{i-}}.
\ee
Given a cutoff $\Lambda$ on the total invariant mass, we thus obtain the following bound on the LC energy available to the loop momenta,
\be
p_{\Loop+} \leq \fr{\Lambda^2}{2p_{\tot-}} - \sum_{j\neq i} \fr{p_{j\perp}^2 + m^2}{2p_{j-}}.
\label{eq:PplusCutoff}
\ee
Converting this to a bound on $\mu_\Loop^2$, we find
\be
\mu_\Loop^2 \leq 2p_{i-} \left( \fr{\Lambda^2}{2p_{\tot-}} - \sum_{j\neq i} \fr{p_{j\perp}^2 + m^2}{2p_{j-}} \right) - p_{i\perp}^2,
\label{eq:LCMuCutoff}
\ee
exactly matching the expression obtained by taking the infinite momentum limit of~\eqref{eq:LambdaLoop}.

\begin{figure}[t!]
\begin{center}
\includegraphics[width=0.65\textwidth]{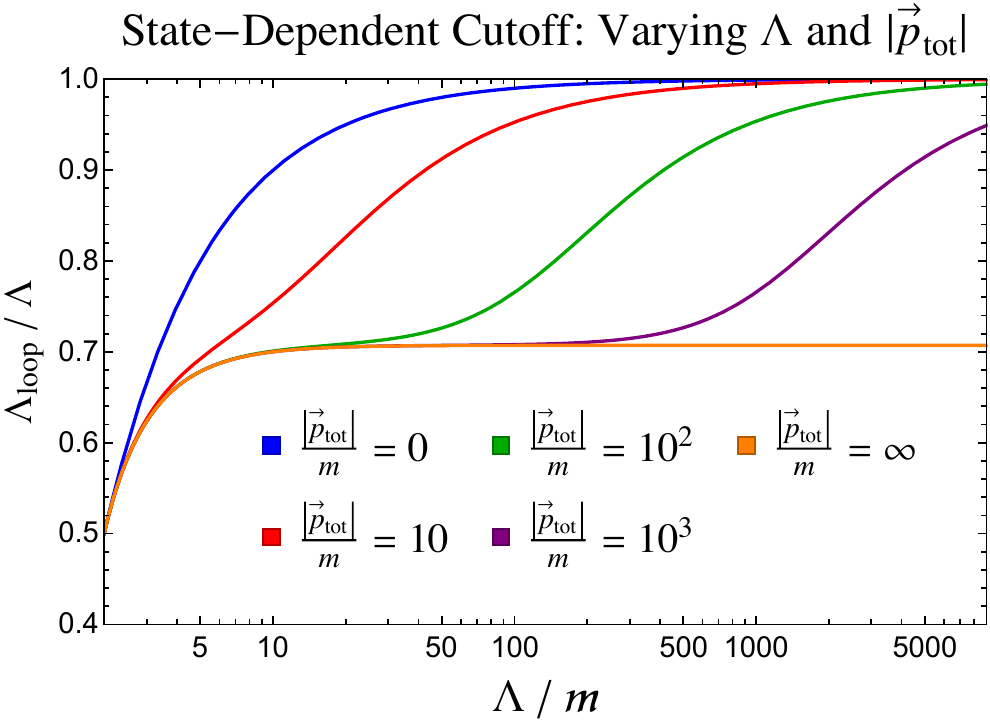}
\caption{Effective cutoff $\Lambda_\Loop$ (in units of the UV cutoff $\Lambda$) as a function of $\Lambda$ for different values of $|\vec{p}_\tot|$, when the external state is the lowest two-particle state. For any finite value of $|\vec{p}_\tot|$, the effective cutoff eventually reaches $\Lambda$, but for $|\vec{p}_\tot| \ra \infty$, the effective cutoff only reaches $\fr{1}{\sqrt{2}}\Lambda$.}
\label{fig:LambdaLoop} 
\end{center}
\end{figure}

To better understand the transition between ET and LC behavior as $|\vec{p}_\tot|$ increases, let's consider the case where the external state is the lowest $n$-particle state, where all particles have $\vec{p}_j = \fr{1}{n} \vec{p}_\tot$. In this case, eq.~\eqref{eq:LambdaLoop} takes the form
\be
\Lambda_\Loop^2 = \Lambda^2 + m^2(n-1)^2 + \fr{2(n-1)}{n} \bigg( |\vec{p}_\tot|^2 - \sqrt{(|\vec{p}_\tot|^2+\Lambda^2)(|\vec{p}_\tot|^2+n^2m^2)} \bigg).
\ee
Figure~\ref{fig:LambdaLoop} shows $\Lambda_\Loop$ as a function of $\Lambda$ for the case $n=2$, with a variety of values for $|\vec{p}_\tot|$. As we can see, for $|\vec{p}_\tot| = 0$, the effective cutoff quickly asymptotes to $\Lambda$ for large cutoffs, as the finite additive shift becomes negligible. However, as we increase $|\vec{p}_\tot|$, we see that $\Lambda_\Loop$ first approaches the reduced value of $\fr{1}{\sqrt{2}}\Lambda$ (set by the momentum fraction $x_i = \half$ carried by the sunset diagram) before eventually reaching $\Lambda$ once the cutoff becomes much larger than the total momentum. As $|\vec{p}_\tot| \ra \infty$, this transition is removed, such that the effective cutoff remains fixed at the reduced value of $\sqrt{x_i}\Lambda$.

This multiplicative factor leads to very different behavior in the structure of state-dependent counterterms between ET and LC quantization. Given a UV divergence of $O(\Lambda^n)$, we see from eq.~\eqref{eq:LambdaLoopET} that in ET quantization the state-dependence only arises at $O(\Lambda^{n-1})$. In other words, state-dependent counterterms necessarily have a lower power of $\Lambda$ than the leading local, state-independent counterterms. For example, in 3d $\phi^4$-theory the logarithmic divergence in the bare mass results in no finite state-dependence for ET quantization, as discussed in~\cite{EliasMiro:2020uvk}. However, in LC quantization, the state-dependence occurs at $O(\Lambda^n)$, such that state-dependent counterterms have the same power of $\Lambda$ as the leading local counterterms. Though we have been specifically discussing the UV divergence in the mass due to the sunset diagram, note that this analysis generalizes to any UV divergence in deformations of free field theory.

\begin{figure}[t!]
\begin{center}
\includegraphics[width=0.38\textwidth]{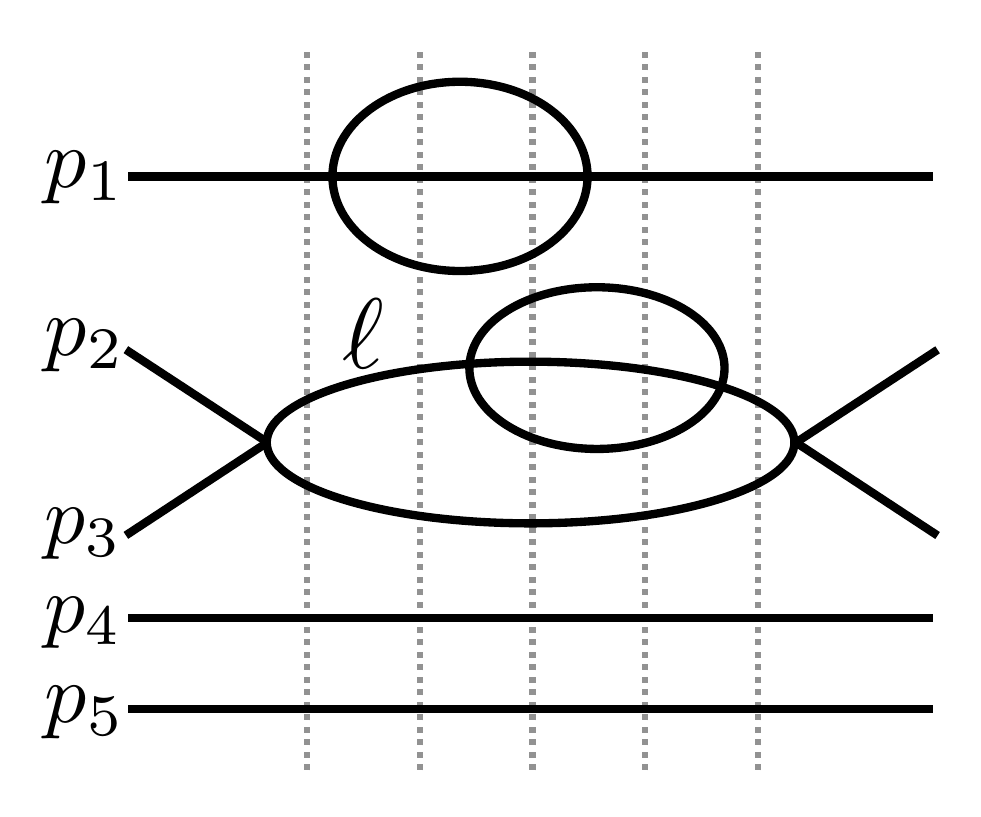}
\caption{Example of an $O(g^6)$ contribution in perturbation theory involving multiple sunset subdiagrams. The finite state-dependence for all such higher order processes is set only by the momentum fraction carried by the individual sunset diagrams (for this example, $\fr{p_{1-}}{p_{\tot-}}$ and $\fr{\ell_-}{p_{\tot-}}$), just like for second-order processes.}
\label{fig:HigherOrderSunsets} 
\end{center}
\end{figure}

So far, this discussion has focused specifically on state-dependence that arises at second order in the coupling $g$. At higher orders in perturbation theory, one can encounter much more complicated processes, such as the example shown in figure~\ref{fig:HigherOrderSunsets}, containing one or more sunsets as subdiagrams. Note that this figure is \emph{not} a Feynman diagram, but rather a representation of an $O(g^6)$ contribution to Hamiltonian eigenstates in old-fashioned perturbation theory. The relative horizontal positions of interaction vertices thus indicate the order in which the insertions of the Hamiltonian act and determine which intermediate states contribute to this process.\footnote{Concretely, this example is a $5 \to 5 \to 7 \to 9 \to 7 \to 5 \to 5$-particle process, indicated schematically by the vertical dashed lines between interaction vertices.}

If we repeat the above analysis for the two overlapping sunset diagrams in this example, we see that the \emph{additive} shifts for the loop cutoffs are much more complicated than the $O(g^2)$ case from above. However, the \emph{multiplicative} factors $\Lambda_\Loop^2 \to x_i \Lambda^2$ (where $x_i$ is the momentum fraction carried by each individual sunset diagram) follow simply from LC kinematics, as we saw in eq.~\eqref{eq:LCMuCutoff}, and are therefore insensitive to any other part of the diagram. Because each sunset diagram is logarithmically divergent, the only finite state-dependence comes from the multiplicative factors. Concretely, this means that sunset subdiagrams of higher order processes, even those in parallel with other sunsets, all lead to the same finite factors of $\log x_i$ as in second-order processes. This observation is crucial to ensure that our $O(g^2)$ state-dependent counterterm is sufficient to remove state-dependent UV sensitivities at all higher orders in $g$.

\section{UV and IR Cutoffs in Truncation}
\label{app:LambdaEffective}

As discussed in appendix~\ref{sec:Basis}, each CFT primary operator constructs a continuum of basis states, parametrized by their invariant mass $\mu^2$. To obtain a finite-dimensional basis, we must therefore discretize the parameter $\mu^2$ in some way. Regardless of the choice of discretization scheme, this inherently leads to a UV cutoff $\Lambda_\UV$ and IR cutoff $\Lambda_\IR$, roughly corresponding to the largest and smallest values of $\mu$, respectively.

In this work, we have chosen to keep $\Lambda_\IR$ fixed at some small scale and increase the size of our basis by increasing $\Lambda_\UV$. However, one could naively consider doing the opposite, keeping $\Lambda_\UV$ fixed at some large value and adding more states to the basis by decreasing $\Lambda_\IR$. However, as we demonstrate in this appendix, there is actually a lower bound to the size of $\Lambda_\IR$ we can safely consider for a given value of $\Dmax$. In short, at any truncation level $\Dmax$, there is an \emph{effective UV cutoff} $\Lambda_\eff$ which is set by the \emph{IR scale} $\Lambda_\IR$, with the following schematic relation,
\be
\Lambda_\eff(\Dmax) \sim \Lambda_\IR \Dmax^\alpha,
\ee
for some parameter $\alpha$. If we set the IR cutoff $\Lambda_\IR$ to be too small, the effective UV cutoff $\Lambda_\eff$ then becomes comparable to the parameters in the theory (such as $g$ or $m$), leading to errors in our truncation results. By contrast, if we hold $\Lambda_\IR$ fixed and increase $\Lambda_\UV$, we simply find that results become insensitive to $\Lambda_\UV$ once it passes $\Lambda_\eff$.\footnote{Note that this effective cutoff $\Lambda_\eff$ is not directly related to the state-dependence discussed in appendix~\ref{app:StateDependence}. State-dependent UV divergences arise from the relation between a cutoff on the total mass of intermediate states and the resulting cutoff on subsets of particles, and are present even in continuum QFT. The effective cutoff we are discussing now is specifically a consequence of truncating the set of primary operators in intermediate states.}

To see the emergence of this $\Lambda_\IR$-dependent effective cutoff, let's compare the leading correction to the mass of the one-particle state and the lowest two-particle state, due to the sunset diagram. Figure~\ref{fig:MassShiftIR} shows both mass shifts as a function of $\Lambda_\IR$, for $\Dmax=16$ and $\Lambda_\UV/m=100$, compared to the theoretical prediction
\be
\left.\de m^2\right|_{O(g^2)} = - \fr{g^2}{96\pi^2} \log \left( \fr{(\fr{\Lambda_\UV}{m}+1)^2}{8(\fr{\Lambda_\UV}{m}-1)} \right).
\ee

\begin{figure}[t!]
\begin{center}
\includegraphics[width=0.7\textwidth]{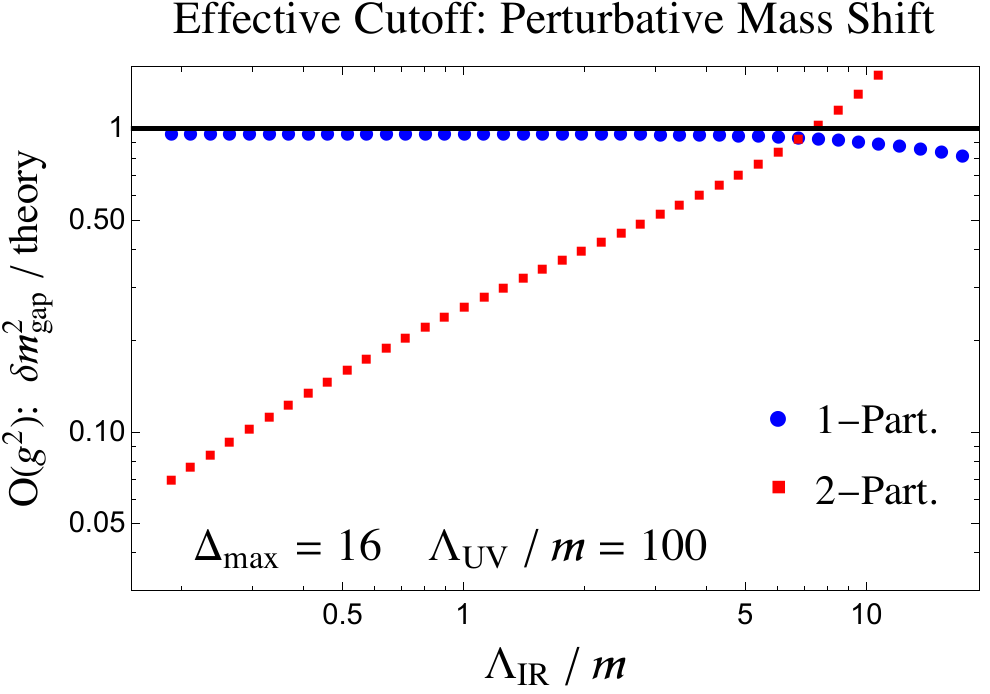}
\caption{The $O(g^2)$ correction to the one-particle state (blue) and lowest two-particle state (red), normalized by the theoretical prediction set by $\Lambda_\UV$ (black line), as a function of $\Lambda_\IR$, for $\Dmax=16$, $\Lambda_\UV/m=100$, and $r=0.8$. The two-particle shift continually decreases as $\Lambda_\IR$ decreases, indicating that it sees a separate cutoff than $\Lambda_\UV$.}
\label{fig:MassShiftIR} 
\end{center}
\end{figure}

As we decrease $\Lambda_\IR$ (\emph{i.e.}, add more states to the basis), the shift for the one-particle state (blue) quickly approaches the theoretical prediction, as we would naively expect. However, the shift for the lowest two-particle state (red) starts out comparable to the theoretical prediction, but continually decreases as we decrease $\Lambda_\IR$. This indicates that the sum over intermediate states for the two-particle mass shift is seeing a \emph{different} cutoff than the fixed $\Lambda_\UV$, one which depends on $\Lambda_\IR$.

Note that this particular quantity (the second-order mass shift) is very UV-sensitive, which makes it a useful probe of the effective cutoff seen by intermediate states. However, IR observables (\emph{e.g.}, spectral densities) are insensitive to the precise UV cutoff, so long as it is far above the other mass scales of the theory. We therefore do not see significant deviations in these observables as we vary $\Lambda_\IR$, such as in figure~\ref{fig:SDVaryIR}. 

We can see this effective cutoff even more explicitly by looking at the sum over intermediate mass eigenstates that determines the mass shift. From old-fashioned perturbation theory, the second-order mass shift for a state $|\psi\>$ is simply
\be
\de m_\psi^2 = \left(\fr{g}{4!}\right)^2 \sum_i \fr{|\<\psi|\phi^4|\mu_i\>|^2}{m_\psi^2 - \mu_i^2},
\ee
where the intermediate states $|\mu_i\>$ are three-particle mass eigenstates for the one-particle shift, and four-particle states for the two-particle shift. Let's focus specifically on the matrix elements in the numerator, removing the difference in masses from the denominator to obtain the simpler sum
\be
\sum_i |\<\psi|\phi^4|\mu_i\>|^2 \sim \Lambda^2.
\label{eq:MESum}
\ee
Because the mass shift is logarithmically divergent, we expect this sum to be quadratically divergent.

\begin{figure}[t!]
\begin{center}
\includegraphics[width=0.96\textwidth]{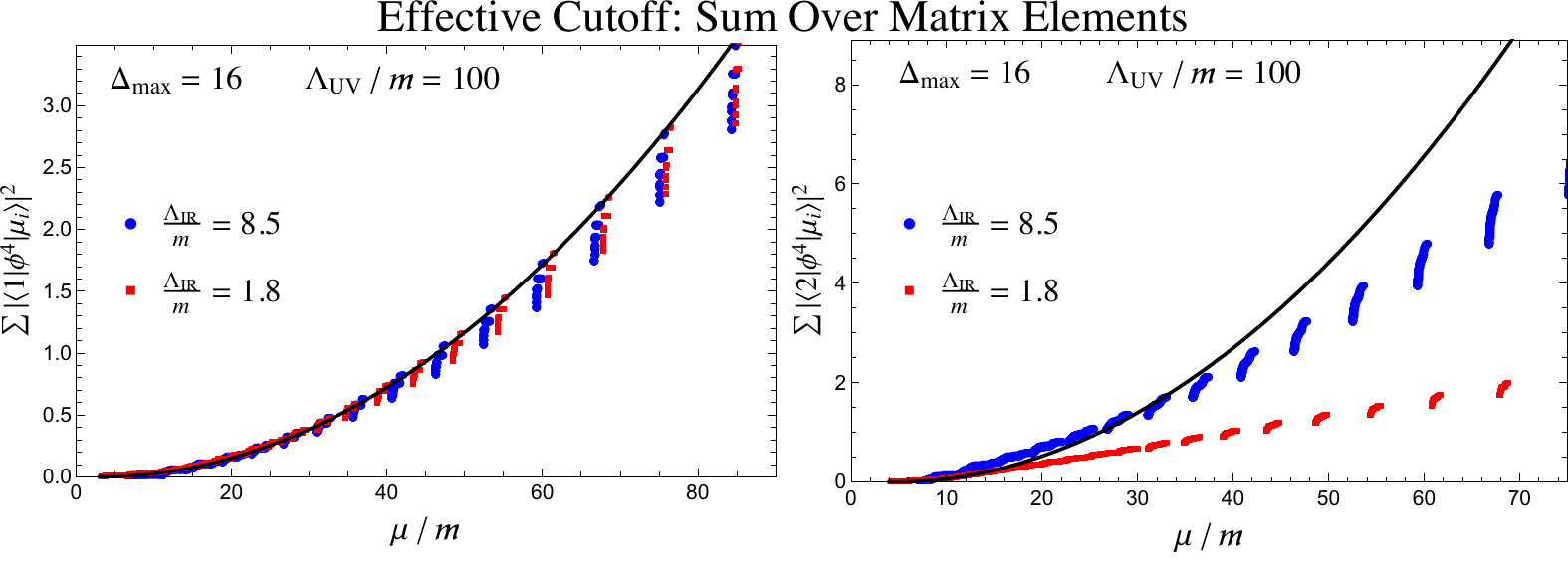}
\caption{Sum over intermediate matrix elements for the one-particle state (left) and lowest two-particle state (right) as a function of the intermediate invariant mass $\mu$, for $\Lambda_\IR / m = 8.5$ (blue) and $\Lambda_\IR / m = 1.8$ (red). Both plots were made with $\Dmax=16$, $\Lambda_\UV / m = 100$, and $r=0.8$, and are compared to the theoretical prediction (black line).}
\label{fig:MESum} 
\end{center}
\end{figure}

Figure~\ref{fig:MESum} shows this cumulative sum over matrix elements as a function of the invariant mass of the intermediate states, for the one-particle state (left) and lowest two-particle state (right), with $\Lambda_\IR / m = 8.5$ (blue) and $\Lambda_\IR / m = 1.8$ (red), $\Dmax=16$, and $\Lambda_\UV/m=100$. As we can see, the one-particle matrix elements match the theoretical prediction (black line) across the full range of $\mu$, and decreasing $\Lambda_\IR$ simply increases the density of intermediate states in the IR. However, the two-particle matrix elements initially agree with the theoretical prediction in the IR, but then deviate as we move to higher values of $\mu$. In particular, the data for the lower value of $\Lambda_\IR$ begins to flatten out near $\fr{\mu}{m} \sim 20$, indicating that while the naive UV cutoff is $\Lambda_\UV$, in practice this sum sees a much lower effective cutoff. If we continue to decrease $\Lambda_\IR$, this deviation occurs for continually lower values of $\mu$. 

It therefore appears that $\Lambda_\IR$ indirectly sets an effective UV cutoff for the intermediate states, but why? As we'll now argue, this effective cutoff arises due to our truncation of the CFT scaling dimension of the intermediate basis states. To see this, note that the sum over matrix elements~\eqref{eq:MESum} is approximately equivalent to computing a free field theory four-point function
\be
\sum_i |\<\Ocal|\phi^4|\mu_i\>|^2 \quad \Rightarrow \quad \<\Ocal \phi^4 \phi^4 \Ocal\>,
\ee
where $\Ocal=\phi$ for the one-particle matrix elements and $\Ocal$ corresponds to any two-particle operator (such as $\phi^2$) for the two-particle matrix elements.

In particular, by computing the UV divergence due to the sunset diagram, we are essentially computing the contribution of $\phi^2$ in the OPE $\phi^4 \times \phi^4$,
\be
\<\Ocal(x_1) \phi^4(x_2) \phi^4(x_3) \Ocal(x_4)\> \supset \fr{1}{x_{23}^3} \<\Ocal(x_1) \phi^2(x_2) \Ocal(x_4)\>.
\ee
More precisely, we are computing the Fourier transform of this contribution to momentum space, but it will be simpler to study this behavior in position space. From this perspective, we are inserting a complete set of intermediate states from the OPE $\Ocal \times \phi^4$ to obtain an OPE singularity in another channel, which we can represent by the familiar bootstrap diagram in figure~\ref{fig:Crossing}. As is well-known, this singularity is specifically reproduced by the infinite sum of high-dimension primary operators in the cross-channel~\cite{Fitzpatrick:2012yx,Komargodski:2012ek,Qiao:2017xif,Mukhametzhanov:2018zja}. If we truncate this sum and only keep primary operators below $\Dmax$, this limits our resolution of the OPE singularity.

\begin{figure}[t!]
\begin{center}
\includegraphics[width=0.65\textwidth]{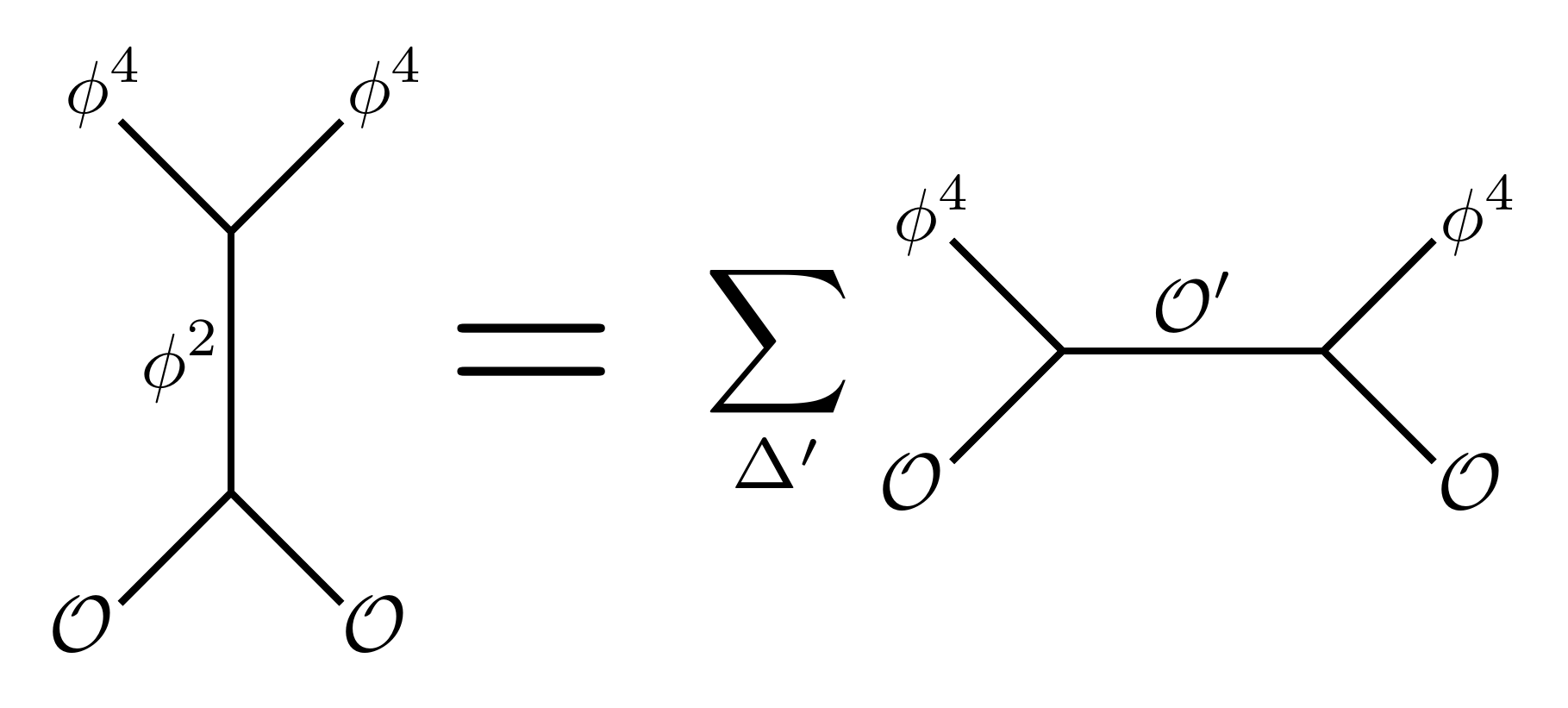}
\caption{Schematic bootstrap diagram showing the relation between the UV divergence at second-order in perturbation theory and the sum over intermediate states. The UV divergence comes from the contribution of $\phi^2$ to the OPE $\phi^4 \times \phi^4$ (left), and is reproduced by the infinite sum of primary operators in the $\Ocal \times \phi^4$ OPE (right). Truncating the sum by $\Dmax$ sets an effective UV cutoff for the OPE singularity.}
\label{fig:Crossing} 
\end{center}
\end{figure}

We can phrase this statement most concretely in terms of the familiar conformally-invariant cross-ratios,
\be
u \equiv \fr{x_{12}^2 x_{34}^2}{x_{13}^2 x_{24}^2}, \qquad v \equiv \fr{x_{14}^2 x_{23}^2}{x_{13}^2 x_{24}^2}.
\ee
An OPE singularity in the $\phi^4 \times \phi^4$ channel corresponds to the limit $v \ra 0$. From the large-$\De$ behavior of conformal blocks~\cite{Qiao:2017xif}, we expect the following bound on our resolution in $v$ due to truncation:
\be
v \gtrsim \fr{1}{\Dmax^2}.
\ee
Due to the nature of crossing, the resulting cutoff on the \emph{short-distance} resolution in $x_{23}^2$ (\emph{i.e.}, the $v\ra 0$ limit) is set by the \emph{long-distance} resolution in $x_{12}^2$ of the conformal blocks for the intermediate primary operators (\emph{i.e.}, the $u \ra 1$ limit of the other channel), which is set by $\Lambda_\IR$. We therefore expect
\be
x_{12}^2 \lesssim \fr{1}{\LIR^2} \quad \Rightarrow \quad x_{23}^2 \gtrsim \fr{1}{\Lambda_\eff^2} \sim \fr{1}{\LIR^2 \Dmax^2} \quad \Rightarrow \quad \Lambda_\eff \sim \Lambda_\IR \Dmax.
\ee
Given our limited range of values for $\Dmax$, we are currently unable to confirm the precise power $\alpha=1$ from our numerical data. It would be useful in future work to make this relation more precise and experimentally confirm the scaling of $\Lambda_\eff$ with $\Dmax$. For the present work, though, it is sufficient to keep $\Lambda_\IR$ fixed such that the effective cutoff does not significantly affect our truncation results.

However, there remains one important question: why doesn't the one-particle shift have the same effective cutoff? From figure~\ref{fig:MassShiftIR}, it is clear that the UV divergence for the one-particle mass is controlled by $\Lambda_\UV$ and is insensitive to $\Lambda_\IR$. This is because the case $\Ocal=\phi$ is unique. In taking the OPE $\phi \times \phi^4$, we only obtain a \emph{single} three-particle operator,
\be
\phi(x_1) \times \phi^4(x_2) \supset \fr{1}{x_{12}} \phi^3(x_2) + \phi^5(x_2) + \ldots,
\ee
In LC quantization, the contributions from all five-particle operators vanish, leaving only the contribution from $\phi^3$. For this specific case, the $\phi^2$ contribution in the cross-channel is therefore reproduced by a single operator, such that we do not obtain a $\Dmax$-dependent effective cutoff.

For all higher-particle operators, such as $\Ocal=\phi^2$, we instead obtain an infinite set of intermediate $n+2$-particle states in the OPE,
\be
\phi^2(x_1) \times \phi^4(x_2) \supset \fr{1}{x_{12}} \phi^4(x_2) + \fr{x_{12}^\mu}{x_{12}} \phi^3\p_\mu\phi(x_2) + \ldots,
\ee
resulting in the emergent cutoff $\Lambda_\eff$ when we truncate this sum.

\section{State-Dependent Counterterm Example}
\label{app:CountertermExample}

 In this section, we explicate the details of our counterterm prescription by means of an example. Let us quickly recap the overall steps: \begin{enumerate}
	\item Diagonalize the free, massive Hamiltonian ($g = 0$) and determine the mass eigenstates $\ket{\mu_i}$.
	\item Compute the perturbative $\cO(g^2)$ mass shift $\delta \mu_i^2|_{(n+2,g^2)}$ for each mass eigenstate, coming from the  $n \to n +2$ interaction.
	\item In the basis of mass eigenstates, the state-dependent counterterm is then given by a diagonal matrix with entries $- \delta \mu_i^2 |_{(n+2, g^2)}$. By applying a rotation, we can compute it in terms of the original primary basis.
	\item Finally, we also introduce a local mass shift $\delta \mathcal{L} = \half c_L g^2 \phi^2$, where for a given $\Delta_{\textrm{max}}$, the coefficient $c_L$ is chosen to optimize the rate of convergence (see section \ref{sec:Results}).
\end{enumerate}

In order to illustrate these steps, let us consider a reduced basis at $\Delta_{\textrm{max}} = 8$. At this truncation, the $\mathbb{Z}_2$-odd basis consists of a one-particle state, five three-particle states, and a single five-particle state. The state-dependent counterterm will therefore have a piece that comes from the mixing between one- and three-particle states, as well as a piece due to the mixing between three- and five-particle states. To keep this toy example as simple as possible, let us set $\fraki_{\textrm{max}} = 1$. To implement step 1, we set $g = 0$, so that the Hamiltonian is comprised of the CFT piece and the mass term,
{\small \begin{equation}
	\Mcal_{\Ocal\Ocal'} = \Lambda_{\textrm{UV}}^2 
\left(\begin{array}{c|ccccc|c}
 0 &  0 & 0 & 0 & 0 & 0 & 0 \\ \hline
 0 & 0.5 & 0 & 0 & 0 & 0 & 0 \\
 0 & 0 & 0.5 & 0 & 0 & 0 & 0 \\
 0 & 0 & 0 & 0.5 & 0 & 0 & 0 \\
 0 & 0 & 0 & 0 & 0.5 & 0 & 0 \\
 0 & 0 & 0 & 0 & 0 & 0.5 & 0 \\ \hline
 0 & 0 & 0 & 0 & 0 & 0 & 0.5 \\
\end{array} \right) + m^2 
\left(\begin{array}{c|ccccc|c}
 1 &  0 & 0 & 0 & 0 & 0 & 0 \\ \hline
 0 & 13 & -4.502 & 0 & -2.287 & 0 & 0 \\
 0 & -4.502 & 21. & 0 & -2.574 & 0 & 0 \\
 0 & 0 & 0 & 17 & 0 & 3.162 & 0 \\
 0 & -2.287 & -2.574 & 0 & 25 & 0 & 0 \\
 0 & 0 & 0 & 3.162 & 0 & 17.5 & 0 \\ \hline
 0 & 0 & 0 & 0 & 0 & 0 & 38.333 \\
\end{array} \right),
\label{eq:FreeHamiltonianExample}
\end{equation}}
where we have omitted any $\fraki$, $\frakj$ labels, since $\fraki_{\textrm{max}} = 1$. For clarity, we have delineated the different particle sectors. We can now diagonalize this Hamiltonian to obtain a set of mass eigenstates $\ket{\mu_i}$. For each eigenstate, we can then compute the mass shift due to the $n \to n+2$ interaction from old-fashioned perturbation theory, \begin{equation}
	\delta \mu_i^2|_{(n+2,g^2)} =\left( \frac{g}{4!} \right)^2 \sum_k \frac{|\bra{\mu_i} \phi^{4}\ket{\mu_k}|^2}{\mu_i^2-\mu_k^2}, \label{eq:exampleoldfashioned}
\end{equation} where $\bra{\mu_i}$ is the $n$-particle mass eigenstate and $\ket{\mu_k}$ is the $(n+2)$-particle mass eigenstate. Here, due to the $\Delta_{\textrm{max}}$ we have chosen, the states that recieve a correction due to the $n \to n+2$ interaction are only the one- and three-particle states, as there are no seven-particle states to contribute to the five-particle state.  As matrices (or in this case, as row and column vectors), the $n \to n+2$ interaction term takes the form \begin{equation}
	\begin{aligned}
		(\phi^4)^{1 \to 3} &= \Lambda_{\textrm{UV}} \left(
\begin{array}{ccccc}
 0.0155 & -0.0095 & 0 & -0.0008 & 0 \\
\end{array}
\right),\quad\quad\quad
(\phi^4)^{3\to 5} &= \Lambda_{\textrm{UV}} \left(
\begin{array}{c}
 0.0559 \\
 -0.0094 \\
 0 \\
 0 \\
 0 \\
\end{array}
\right),
	\end{aligned}
\end{equation} where the factor of $\Lambda_{\textrm{UV}} = 0.5$ comes from the discretization integral. Plugging this into~\eqref{eq:exampleoldfashioned}, we obtain the resulting mass shifts \begin{equation}
	\delta \mu_1^2 \approx -g^2 \cdot 5.1512 \times 10^{-6}, \quad\quad\quad \delta \mu_3^2 \approx -g^2 \begin{pmatrix}
		0.0000183 \\
		0 \\
		0 \\
		0.0000181 \\
		2.76788 \times 10^{-7}
	\end{pmatrix}.
\end{equation} Note that to obtain our final counterterm, we must rotate these quantities to the space of CFT basis states, rather the space of mass eigenstates. This can be accomplished by noting that diagonalizing the Hamiltonian gives us a unitary transformation $U$ that relates the basis of CFT eigenstates to the mass eigenstates. The rotation back to the CFT basis is then given by $U D U^{\dagger}$, where $D$ is a diagonal matrix whose entries are the mass shifts.\footnote{More explicitly, we can write the free theory Hamiltonian as the matrix $M$, such that \begin{equation}
	M=A+B,
\end{equation} where $A$ corresponds to the CFT Hamiltonian and $B$ corresponds to the mass term. When we diagonalize $M$, this gives us the unitary matrix $U$ that relates the eigenstates $|\mu_i\>$ of $M$ to the basis of CFT states $|\Ocal\>$, which are eigenstates of $A$,
\be
|\mu_i\> = U |\Ocal\>.
\ee
The state-dependent counterterm $D$ is diagonal in the massive basis,
\be
\<\mu_i|D|\mu_j\> = -\de \mu_i^2 \, \de_{ij},
\ee
and we can rotate this matrix to the CFT basis by evaluating $U D U^\dagger$.} Performing this rotation and accounting for the minus sign in step 3, we finally obtain the state-dependent counterterm for this example\begin{equation}
	\begin{aligned}
		\Mcal^{(\textrm{state-dep.)}}_{\Ocal\Ocal'} = g^2 \left( \begin{array}{c|ccccc|c}
			5.151 \times 10^{-6} & 0 & 0 & 0 & 0 & 0 & 0 \\ \hline
			0 & 1.826 \times 10^{-5} & 2.587 \times 10^{-8} & 0 & 1.730 \times 10^{-7} & 0 & 0 \\
			0 & 2.587 \times 10^{-8} & 1.477 \times 10^{-5} & 0 & 6.979 \times 10^{-6} & 0 & 0 \\
			0 & 0 & 0 & 0 & 0 & 0 & 0 \\
			0 & 1.730 \times 10^{-7} & 6.979 \times 10^{-6} & 0 & 3.639 \times 10^{-6} & 0 & 0 \\
			0 & 0 & 0 & 0 & 0 & 0 & 0 \\ \hline
			0 & 0 & 0 & 0 & 0 & 0 & 0 \\
		\end{array} \right).
	\end{aligned}
    \label{eq:CountertermExample}
\end{equation} Finally, we add the local mass shift $\delta \mathcal{L} = \half c_L g^2 \phi^2$, which just shifts the coefficient of the mass term matrix elements in~\eqref{eq:FreeHamiltonianExample}. We can then diagonalize the resulting matrix of \eqref{eq:FreeHamiltonianExample} (with the $c_L$ shift) plus \eqref{eq:CountertermExample} to determine the spectrum and mass eigenstates as a function of $g$.

Finally, let us comment on a technical issue that can arise in constructing the state-dependent counterterm. In principle, the $O(g^2)$ mass shift for every mass eigenstate $|\mu_i\>$ should be negative, since it is due to mixing with higher particle states. However, at finite $\Delta_{\textrm{max}}$ and $\fraki_{\textrm{max}}$, the mass shifts computed from~\eqref{eq:exampleoldfashioned} can actually be positive for a few states. This typically occurs only for the heaviest $n$-particle states, if there are not enough $(n+2)$-particle states with higher mass to ensure that their overall mass shift is negative. However, as we take $\Delta_{\textrm{max}} \to \infty$, we expect these spurious positive shifts to disappear as we add more $(n+2)$-particle states to the basis.

\begin{figure}[t!]
\begin{center}
\includegraphics[width=\textwidth]{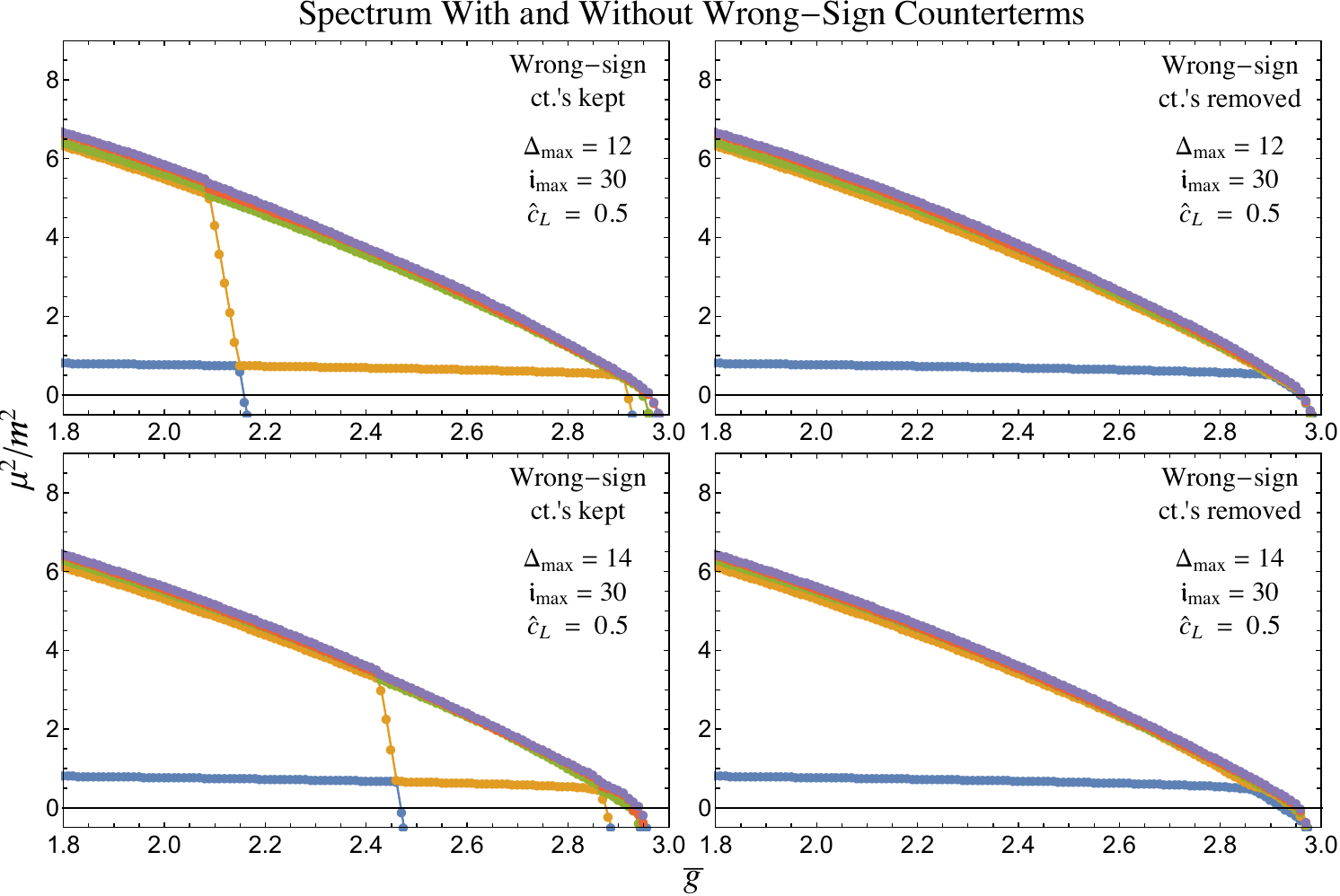}
\caption{The first five eigenvalues in the $\mathbb{Z}_2$-odd sector, where the wrong-sign entries in the counterterm have not been removed (left side) and where they have been removed (right side). Note the relatively small truncation cutoff $\Dmax=12$ chosen in the top row of plots. Increasing to $\Dmax=14$ (bottom row), the crashing behavior moves to higher values of $\bar{g}$.}
\label{fig:CrashingEigenvalues} 
\end{center}
\end{figure}

Naively, we should just cancel these positive mass shifts (like all other $O(g^2)$ corrections) with our state-dependent counterterm. However, in practice we have found that canceling these ``wrong-sign'' mass shifts leads to errors in the resulting spectrum, where these high-mass eigenvalues rapidly become negative, ``crashing'' through the low-mass spectrum. This behavior is shown in the top left plot of figure \ref{fig:CrashingEigenvalues}, where we see one such crash in the odd sector around $\bar{g} \approx 2.1$. In order to fix this, we simply identify all entries in our state-dependent counterterm with the ``wrong'' sign (in the massive basis) and set those entries to zero. This yields the top right plot of figure \ref{fig:CrashingEigenvalues}, where the crashing behavior has disappeared. Note that these wrong-sign entries are a set of measure zero of the full basis and only affect the highest mass states, such that removing them has no significant effect on the low-mass states (apart from the absence of crashes).

Note the relatively small values of $\Dmax$ and $\imax$ in these plots. As $\Dmax$ is increased, the crashing behavior is pushed to higher values of $\bar{g}$ (bottom left plot). For most cases of interest where $\Dmax$ and $\imax$ are increased even further, the spurious mass shifts do not have any noticeable effect on the spectrum. Nevertheless, in implementing our counterterm procedure we always set to zero any entries with the wrong sign, to ensure no crashing behavior.

\section{Varying Truncation Parameters}
\label{app:VaryParameters}

This appendix is an addendum to section~\ref{sec:Results}. In that section, we used LCT to compute the spectrum, eigenvalue ratios, and nonperturbative spectral densities of $\phi^4$ theory. Unless otherwise indicated, the results presented there were obtained with a maximum truncation level of $\Dmax=16$, the IR cutoff set to be $\LIR=0.5$, and the coefficient of the local counterterm in our Hamiltonian~(\ref{eq:HplusCT}) set to $\hat{c}_L=1$ (recall the normalization of $\hat{c}_L$ from eq.~(\ref{eq:cLhat})). In this appendix, we vary these parameters in order to study their effects on our results. Our broad observation is that varying these parameters only leads to changes in some scheme-dependent overall constants, with no significant effect on physical observables.

Let us begin by revisiting the ratios  $R_{2:1} \equiv  \frac{1}{4}\frac{\mu_{2,\text{-part.}}^2}{\mu_{1\text{-part.}}^2}$ and $R_{3:1} \equiv  \frac{1}{9}\frac{\mu_{3\text{-part.}}^2}{\mu_{1\text{-part.}}^2}$ plotted in Figure~\ref{fig:SpectrumRatios}. In Figure~\ref{fig:VaryDeltaLIR}, we recompute these ratios for different values of $\Dmax$ at fixed $\LIR=0.5$ (left column) and then for different values of $\LIR$ at fixed $\Dmax=16$ (right column). As expected, we see that increasing $\Dmax$ and decreasing $\LIR$ bring both $R_{2:1}$ and $R_{3:1}$ closer to their theoretical prediction of 1, although for the range of $\Dmax$ and $\LIR$ shown the relative variation is quite small. At the same time, at a fixed $\Dmax$, we see that there is a cost to decreasing $\LIR$, because the ratios deviate from 1 at larger values of $\fr{\mgap}{g/4\pi}$. In the main text, we chose $\LIR=0.5$ as a happy medium, however we could have selected a different value in the range $0.25 \lesssim \LIR \lesssim 0.75$ without significantly affecting the resulting spectrum.

\begin{figure}[t!]
\begin{center}
\includegraphics[width=1\textwidth]{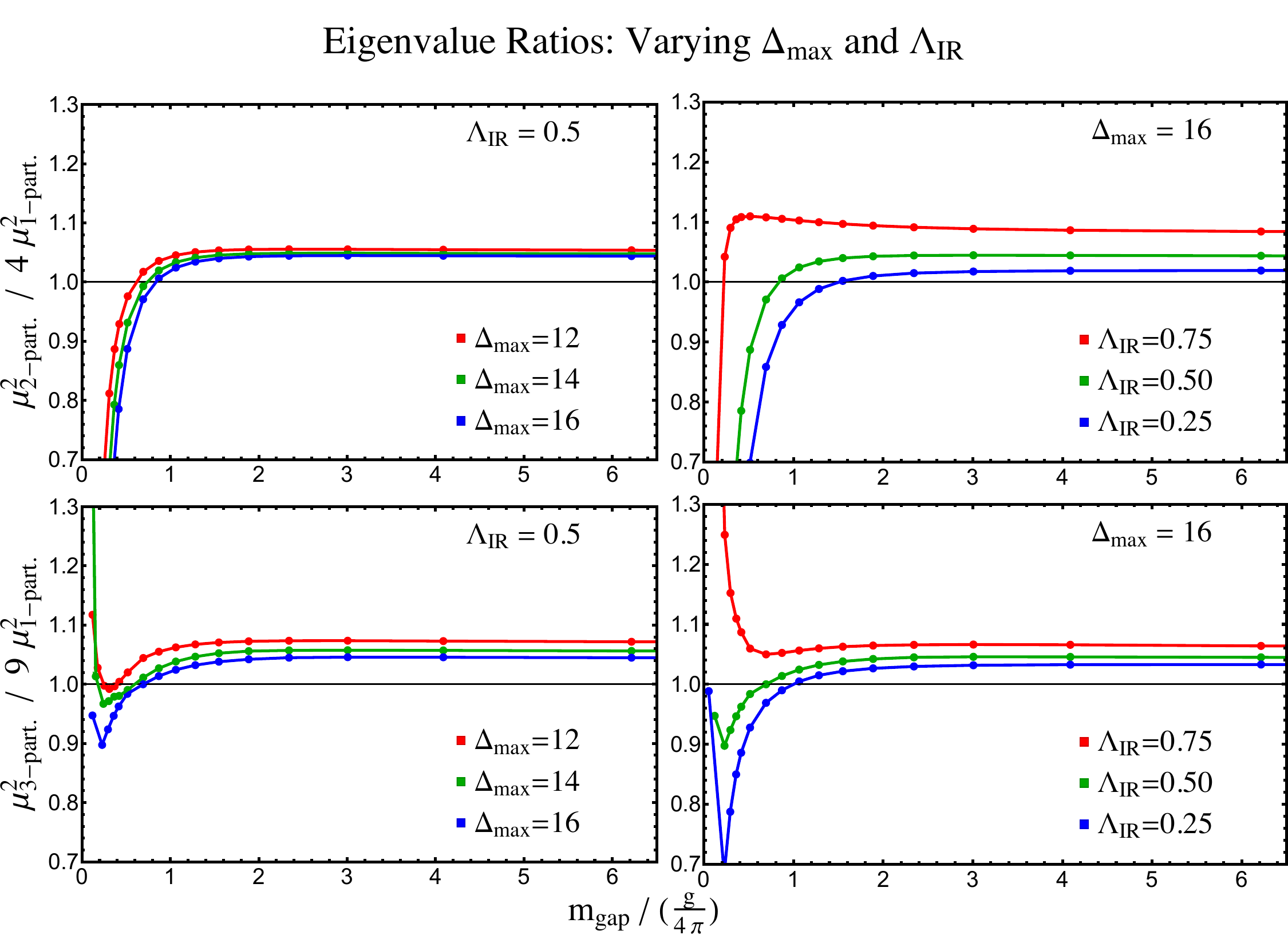}
\caption{The ratios $R_{2:1} \equiv  \frac{1}{4}\frac{\mu_{2,\text{-part.}}^2}{\mu_{1\text{-part.}}^2}$ (top row) and $R_{3:1} \equiv  \frac{1}{9}\frac{\mu_{3\text{-part.}}^2}{\mu_{1\text{-part.}}^2}$ (bottom row) computed for different $\Dmax$ at $\LIR=0.5$ (left column) and for different $\LIR$ at $\Dmax=16$ (right column).}
\label{fig:VaryDeltaLIR} 
\end{center}
\end{figure}

\begin{figure}[t!]
\begin{center}
\includegraphics[width=1\textwidth]{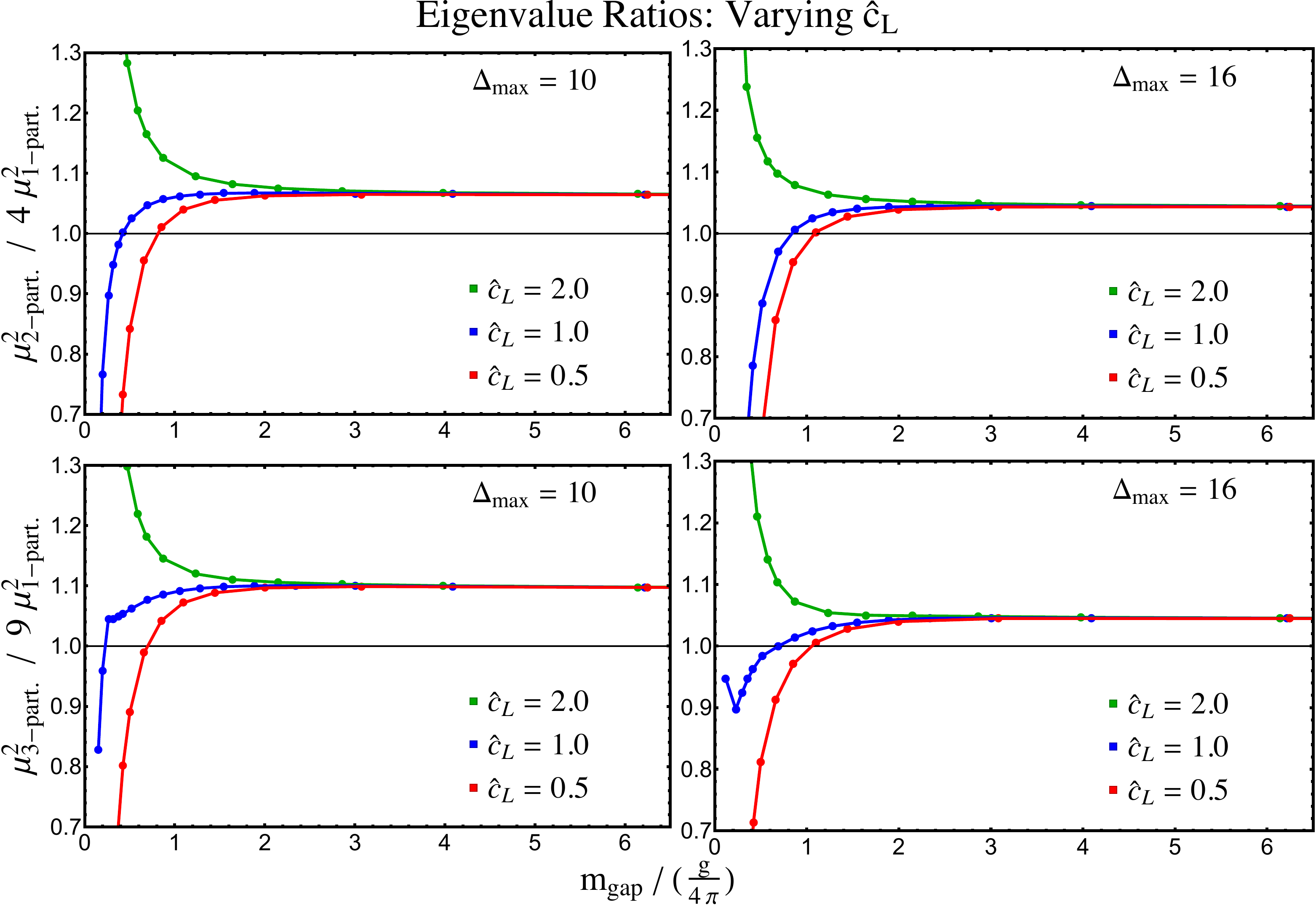}
\caption{The ratios $R_{2:1} \equiv  \frac{1}{4}\frac{\mu_{2,\text{-part.}}^2}{\mu_{1\text{-part.}}^2}$ (top row) and $R_{3:1} \equiv  \frac{1}{9}\frac{\mu_{3\text{-part.}}^2}{\mu_{1\text{-part.}}^2}$ (bottom row) computed for different $\hat{c}_L$ at $\Dmax=10$ (left column) and at $\Dmax=16$ (right column).}
\label{fig:VarycL} 
\end{center}
\end{figure}

Next, let us consider the effect of $\hat{c}_L$ on the ratios $R_{2:1}$ and $R_{3:1}$. Figure~\ref{fig:VarycL} shows our results for $R_{2:1}$ and $R_{3:1}$ for different values of $\hat{c}_L$ at $\Dmax=10$ (left column) and at $\Dmax=16$ (right column). We see that varying $\hat{c}_L$ has almost no effect until $\fr{\mgap}{g/4\pi} \approx 2$, below which the results for different $\hat{c}_L$ begin to `fan out'. At these values of $\Dmax$, it is clear from the figure that we do not want to choose $\hat{c}_L$ too far outside of the range $0.5 \lesssim \hat{c}_L \lesssim 2.0$ shown, as this leads to significant truncation effects. In the main text, we chose $\hat{c}_L=1$ as a generic value in this range. In the $\Dmax\rightarrow \infty$ limit, we expect that there should be some threshold value $\hat{c}_{L,\min}$ such that for $\hat{c}_L < \hat{c}_{L,\min}$ the mass gap does not close for any real coupling $g$. So long as we are above this threshold, varying $\hat{c}_L$ should have no effect on the resulting spectrum, as changing $\hat{c}_L$ is just a redefinition of the physical mass. Because we find that the gap closes for all values $0.5 \lesssim \hat{c}_L \lesssim 2.0$, our results indicate that the threshold value $\hat{c}_{L,\min} \lesssim 0.5$. As we go from $\Dmax=10$ to $\Dmax=16$, we see that overall the ratios move closer to 1, and moreover the width of the fan becomes slightly narrower, which indicates that our range of reliable $\hat{c}_L$ increases with $\Dmax$. Nevertheless, $\Dmax=16$ is still quite low, and it is not yet clear what will happen for larger $\Dmax$. In particular, it would be useful to see whether as $\Dmax \to \infty$ we encounter a finite $\hat{c}_{L,\min}$ below which the gap does not close.

Now, let us turn to the effect of $\LIR$ and $\hat{c}_L$ on spectral densities. We start with Figure~\ref{fig:SDVaryIR}, which shows the effect of varying $\LIR$ on various spectral densities computed in section~\ref{sec:Results}. The top row shows the $\phi^2$ and $\phi^4$ integrated spectral densities (which should be compared to Figure~\ref{fig:SDStrong}). We see that the $\phi^2$ spectral density does not noticeably change when we vary $\LIR$, whereas the $\phi^4$ spectral density seems to change by a multiplicative factor. We confirm the latter observation in the bottom left plot, where allowing for an overall coefficient collapses the $\phi^4$ spectral densities at different $\LIR$ back onto a single curve. This is consistent with the fact that the definition of $\phi^4$ is dependent on our regularization scheme, as mentioned in section~\ref{subsec:Results-SD}. An overall scaling clearly does not change our universality results (see Figure~\ref{fig:Universality}), which only hold up to multiplicative factors anyway. In the bottom right of Figure~\ref{fig:SDVaryIR}, we additionally check that varying $\LIR$ also does not change the spectral density of the stress tensor trace. Following the procedure set forth in section~\ref{subsubsec:trace}, for each $\LIR$ we compute $\delta_T$ from the one-particle entry of the state-dependent counterterm and then compute the $T_{\,\,\mu}^\mu$ spectral density. We see that at the critical point, the $T_{\,\,\mu}^\mu$ spectral density vanishes in the IR for all three choices of $\LIR$ (compare with Figure~\ref{fig:StressTensorTrace}). In these plots, we have also included the integrated spectral density of $\tfrac{g}{4!}\phi^4$ for comparison (with $\LIR$ labeled along each curve). Recall that $\tfrac{g}{4!}\phi^4$ is one of two contributions to $T_{\,\,\mu}^\mu$, which does not vanish on its own and must cancel against the $\phi^2$ contribution in order for the trace to vanish.

\begin{figure}[t!]
\begin{center}
\includegraphics[width=1\textwidth]{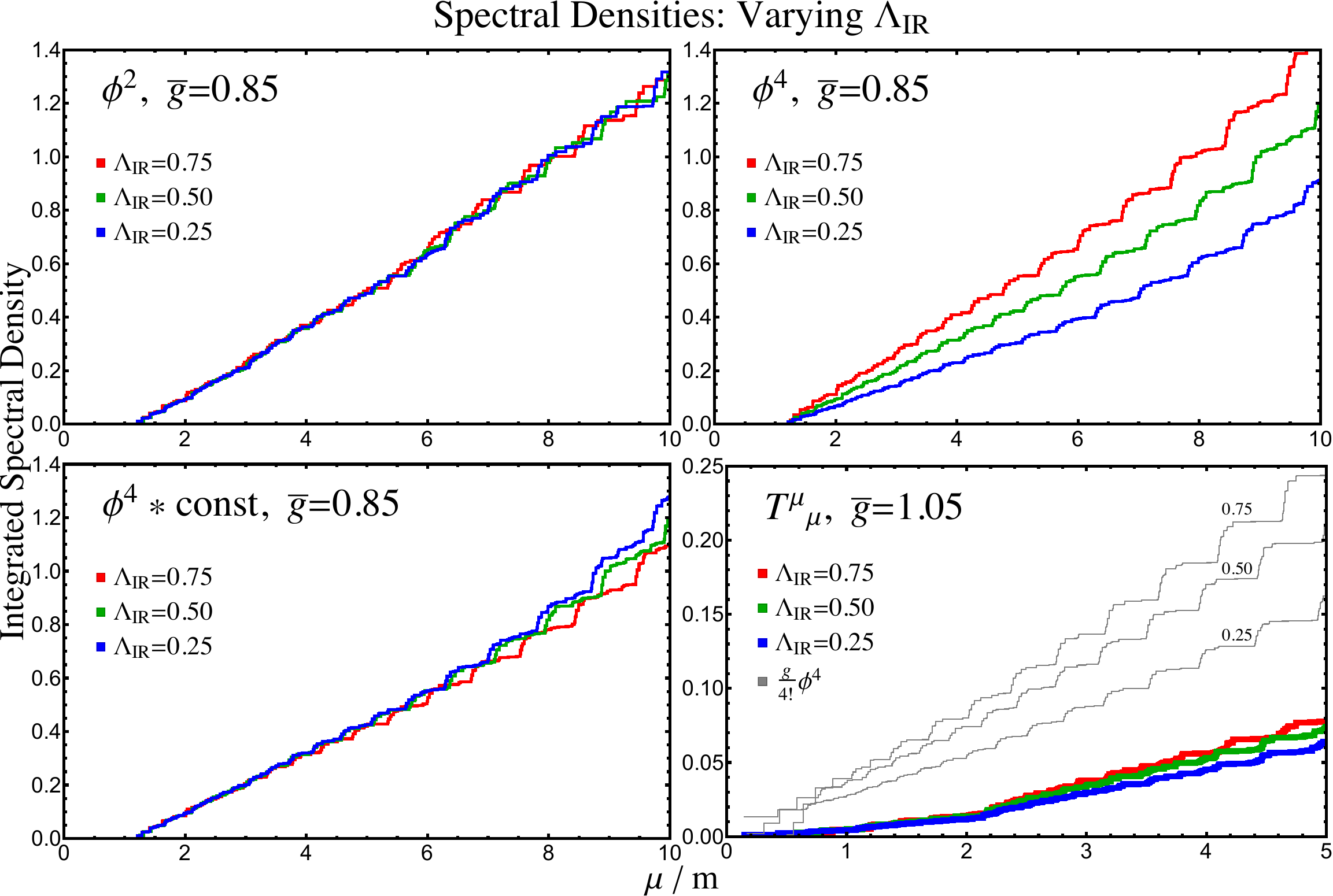}
\caption{Integrated spectral densities computed using different $\LIR$ at $\Dmax=16$ (and $\hat{c}_L=1$). At bottom left, we have rescaled the $\LIR=0.25, 0.75$ results by overall coefficients compared to top right. }
\label{fig:SDVaryIR} 
\end{center}
\end{figure}

\begin{figure}[t!]
\begin{center}
\includegraphics[width=1\textwidth]{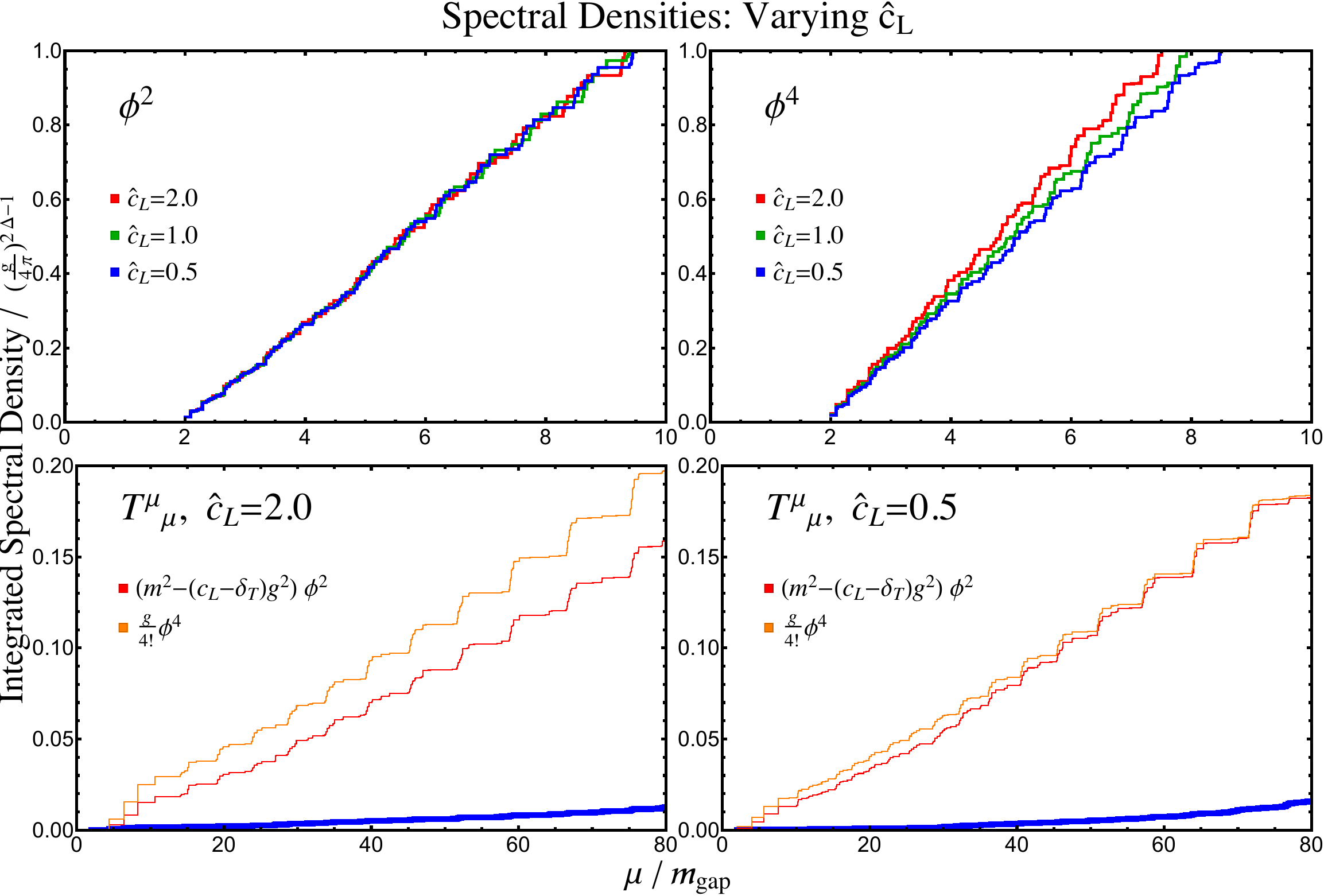}
\caption{Integrated spectral densities of $\phi^2$ and $\phi^4$ at $\fr{\mgap}{g/4\pi} = 1.0$ (top row) and of $T_{\,\,\mu}^\mu$ at  $\fr{\mgap}{g/4\pi} \approx 0.07$ (bottom row) with $\Dmax=16$ and different $\hat{c}_L$. All integrated spectral densities are expressed in units of the coupling $\fr{g}{4\pi}$, and remain consistent as we vary $c_L$.}
\label{fig:SDVarycL} 
\end{center}
\end{figure}

Finally, let us consider the effect of varying $\hat{c}_L$ on spectral densities. The story is similar to varying $\LIR$ in that $\phi^2$ and $T^\mu_{\,\,\mu}$ are insensitive to $\hat{c}_L$, and $\phi^4$ changes only by an overall constant. One difference, though, is that when varying $\LIR$, we held $\bar{g}$ fixed, because this was essentially equivalent to holding $\fr{\mgap}{g/4\pi}$ fixed. When we vary $\hat{c}_L$, on the other hand, $\mgap$ changes significantly (in units of the bare mass), and thus to compare spectral densities, we directly hold $\fr{\mgap}{g/4\pi}$ fixed when plotting results for different $\hat{c}_L$. For instance, figure~\ref{fig:SDVarycL} shows the integrated spectral densities of $\phi^2$ and $\phi^4$ computed at  $\fr{\mgap}{g/4\pi} = 1.0$ (top row) and of $T_{\,\,\mu}^\mu$ computed at  $\fr{\mgap}{g/4\pi} \approx 0.07$ (bottom row).\footnote{In practice, it is computationally expensive to keep the precise value of $\fr{\mgap}{g/4\pi}$ fixed near the critical point. The bottom row of figure~\ref{fig:SDVarycL} was computed at $\fr{\mgap}{g/4\pi} = 0.077$ for $\hat{c}_L=2.0$ and $\fr{\mgap}{g/4\pi} = 0.062$ for $\hat{c}_L=0.5$, compared to the value $\fr{\mgap}{g/4\pi} = 0.074$ in figure~\ref{fig:StressTensorTrace}.} Note that the horizontal axis is now $\mu / \mgap$.\footnote{Since we are working exclusively in the even-particle sector, in figure~\ref{fig:SDVarycL} we define $\mgap=\tfrac{1}{2}\mu_{2\text{-part.}}$.}

Relatedly, the overall scale of the spectral densities (in units of the bare mass) also changes as we vary $\hat{c}_L$. We have therefore normalized each spectral density by the appropriate power of the coupling $\fr{g}{4\pi}$. As we can see, the properly normalized spectral density for $\phi^2$ is remarkably insensitive to $\hat{c}_L$, while $\phi^4$ varies slightly by an overall constant, again due to its dependence on the regularization scheme. Crucially, this overall coefficient has no effect on the observation of IR universality near the critical point. Meanwhile, in the bottom row we verify that the spectral density of $T^\mu_{\,\,\mu}$ is insensitive to $\hat{c}_L$ and vanishes in the IR.

To summarize, our main takeaway from the analyses in this appendix is that varying $\LIR$ and $\hat{c}_L$ does not significantly affect our main results. In particular, as long as we keep these parameters within a reasonable range (approximately $0.25 \lesssim \LIR \lesssim 0.75$ and $0.5 \lesssim \hat{c}_L \lesssim 2.0$), eigenvalue ratios are largely unaffected, and integrated spectral densities are modified by at most an overall multiplicative constant. We therefore see that the parameters $\LIR = 0.5$ and $\hat{c}_L = 1.0$ used in the main text were not fine-tuned choices, but rather representative values from a range of consistent parameters.

\newpage

\bibliographystyle{utphys}
\bibliography{Refs3D}

\end{document}